\documentclass[a4paper,10pt,dvipsnames]{article}
\usepackage[utf8]{inputenc}
\usepackage[english]{babel}
\usepackage{amsthm}
\usepackage{amsfonts}
\usepackage{amssymb}	
\usepackage{amsmath}
\allowdisplaybreaks
\usepackage{chemarr}
\usepackage{slashed}
\usepackage{enumerate}
\usepackage{graphicx}
\usepackage[colorlinks=true, allcolors=blue]{hyperref}
\usepackage{systeme}
\usepackage{dsfont}
\usepackage{relsize}
\usepackage[margin=0.90in]{geometry}
\usepackage{float}
\usepackage{tikz}
\usepackage[labelformat=simple]{subcaption}

\usepackage{bmpsize}
\usepackage{epstopdf}
\usepackage{bbm}
\usepackage{etoolbox}
\usepackage{xcolor}
\usepackage{titling}
\newcommand*\samethanks[1][\value{footnote}]{\footnotemark[#1]}
\usepackage{blindtext,graphicx}
\usepackage[absolute]{textpos}
\usepackage[hang,flushmargin]{footmisc}

\usepackage{xfrac}
\usepackage{nicefrac}
\usepackage{esvect}
\usepackage{cases}
\usepackage{empheq}
\usepackage{stmaryrd}
\usepackage{cancel}
\usepackage[linguistics]{forest}
\usepackage{url}
\usepackage[font=small]{caption}
\usepackage{array}
\usepackage{dsfont}
\usepackage{bbold}
\usepackage[most]{tcolorbox}
\usepackage{dashrule}
\usepackage{pifont}
\newcommand{\cmark}{\ding{51}}%
\newcommand{\xmark}{\ding{55}}%
\usepackage{fancybox}
\usepackage{fontawesome}
\usepackage{setspace}
\usepackage[locale=US,exponent-product=\cdot,per-mode=fraction]{siunitx}
\newcommand{\C}{\mathbb{C}}
\newcommand{\R}{\mathbb{R}}
\newcommand{\F}{\mathcal{F}}
\renewcommand{\.}{\hspace*{0.07em}}
\definecolor{myblue}{RGB}{80,191,227}
\definecolor{myblue2}{RGB}{80,141,227}
\definecolor{myblue3}{rgb}{0.0, 0.44, 1.0}
\definecolor{myblue4}{rgb}{0.01, 0.28, 1.0}
\usepackage[outermarks]{titlesec}

\titleformat{\paragraph}
  {\normalfont\bfseries}{\theparagraph}{1em}{}

\titlespacing*{\paragraph}{0pt}{3.25ex plus 1ex minus .2ex}{4mm}
\newcommand{\corrections}[1]{{\color{black}#1}}
\newcommand{\cor}[1]{{\color{black}#1}}

\newcommand{\cort}[1]{{\color{black}#1}}
\newcommand{\corf}[1]{{\color{black}#1}}
\newcommand{\corc}[1]{{\color{black}#1}}
\newcommand{\cors}[1]{{\color{black}#1}}
\newcommand{\remove}[1]{{\color{brown}#1}}
\newcommand{\coran}[1]{{\color{black}#1}}
\newcommand{\coranR}[1]{{\color{black}#1}}
\newcommand{\angcor}[1]{{\color{black}#1}}
\usepackage{soul}
\usepackage[colorinlistoftodos]{todonotes}
\usepackage{mathabx}
\newcommand{\xra}[1]{\overset{#1}{\rightsquigarrow}}

\usepackage[giveninits=true,sortcites=true,date=year,maxbibnames=99,doi=false,isbn=false,url=false,eprint=false]{biblatex}
\renewbibmacro{in:}{}
\DeclareFieldFormat{pages}{#1}
\AtEveryBibitem{%
\clearlist{language}%
}
\usepackage{csquotes}

\usepackage{setspace}
\makeatletter
\newcommand{\thickhline}{%
\noalign {\ifnum 0=`}\fi \hrule height 1pt
\futurelet \reserved@a \@xhline
}
\newcolumntype{"}{@{\hskip\tabcolsep\vrule width 1pt\hskip\tabcolsep}}
\makeatother
\newtheorem{theorem}{Theorem}[section]

\theoremstyle{definition}

\newtheorem{remark}[theorem]{Remark}
\makeatletter
\newcommand{\pushright}[1]{\ifmeasuring@#1\else\omit\hfill$\displaystyle#1$\fi\ignorespaces}
\newcommand{\pushleft}[1]{\ifmeasuring@#1\else\omit$\displaystyle#1$\hfill\fi\ignorespaces}
\makeatother
\title{\coran{From frequency-dependent models to frequency-independent enriched continua for mechanical metamaterials}}
\author{
Gianluca Rizzi\.\thanks{Faculty of Architecture and Civil Engineering, TU Dortmund, August-Schmidt-Str. 8, 44227 Dortmund, Germany},
\quad
Marco Valerio d'Agostino\.\thanks{GEOMAS, INSA-Lyon, Université de Lyon, 20 avenue Albert Einstein, 69621, Villeurbanne Cedex, France},
\quad
Jendrik Voss\.\samethanks[1],
\quad
Davide Bernardini\,\thanks{Department of Structural and Geotechnical Engineering, Sapienza University of Rome, Rome, Italy},
\\[2mm]
Patrizio Neff\.\thanks{Head of Chair for Nonlinear Analysis and Modelling, Fakultät für Mathematik, Universität Duisburg-Essen, \\ \indent \,\,\,\, Thea-Leymann-Straße 9, 45127 Essen, Germany},
\quad and \quad
Angela Madeo\.\samethanks[1]
}

\thanksmarkseries{arabic}
\date{\today}
\begin{document}
\maketitle

\begin{abstract}
Mechanical metamaterials have recently gathered increasing attention for their uncommon mechanical responses enabling unprecedented applications for elastic wave control.
Many research efforts are driven towards the conception of always new metamaterials’ unit cells that, due to local resonance or Bragg-Scattering phenomena, may produce unorthodox macroscopic responses such as band-gaps, cloaking, focusing, channeling, negative refraction, etc.
To model the mechanical response of large samples made up of these base unit cells, so-called homogenization or upscaling techniques come into play trying to establish an equivalent continuum model describing these macroscopic metamaterials’ characteristics.
A rather common approach is to assume \textit{a priori} that the target continuum model is a classical linear Cauchy continuum featuring the macroscopic displacement as the only kinematical field.
This implies that the parameters of such continuum models (density and/or elasticity tensors) must be considered to be frequency-dependent to capture the complex response of the considered mechanical systems in the frequency domain.
These frequency-dependent models can be useful to describe some of the aforementioned macroscopic metamaterials’ properties, yet, they suffer some drawbacks such as featuring negative masses and/or elastic coefficients in some frequency ranges which are close to resonance frequencies of the underlying microstructure.
This implies that the considered Cauchy continuum is not positive-definite for all the considered frequencies.
In this paper, we present a procedure, based on the definition of extra kinematical variables (with respect to displacement alone) and through the use of the inverse Fourier transform in time, to convert a frequency-dependent model into an enriched continuum model of the micromorphic type.
All the parameters of the associated enriched model are constant (i.e., frequency-independent) and the model itself remains positive-definite for all the considered frequency ranges.
The response of the frequency-dependent model and the associated micromorphic model coincide in the frequency domain, in particular when looking at the dispersion curves.
\angcor{Moreover, the micromorphic (frequency-independent) model results to be well defined  both in time- and in the frequency-domain, while the Cauchy (frequency-dependent) model can only exist in the frequency domain.}
This paper aims to build a bridge between the upscaling techniques usually found in the literature and our persuasion that macroscopic continua of the micromorphic type should be used to model metamaterials’ response at the macroscopic scale.
\end{abstract}
\textbf{Keywords}: metamaterials, inertia-augmented, dispersion curves, band-gap, enriched continua, frequency-dependent model, generalized continua, dynamic homogenization, Galilean invariance, inverse partial Fourier transform, frequency domain.
%
%
%
%
{
  \hypersetup{linkcolor=black}
  \tableofcontents{}
}
\section{Introduction}
\label{sec:intro}

\subsection{A \textit{material} from an engineering point of view}
Using the word material, we are often referring to a solid substance (e.g., sandstone, marble, steel, iron, etc.) or a fluid substance (e.g., water, oil, etc.) with characteristic macroscopic properties making it easily recognizable for us.
For example, marble is known for its aptitude to be worked in resistant slabs of beautiful colors that are often used to pave internal and external surfaces, steel for its stiffness which makes it irreplaceable for our civil and aeronautic structures, water for its transparency and purity and, like all other fluids, for its habit of taking the form of the recipient that contains it.
All these macroscopic characteristics are certainly conferred by a specific organization of small particles of different sorts (atoms) that are arranged together in different ways (molecules).
In other words, each material can be seen to have a specific discrete (or heterogeneous) nature as soon as we look at it ``close enough''.
However, knowing all details of this underlying heterogeneity often adds little value to our macroscopic observation of the material itself.
For example, knowing how the molecules of a block of marble are made up of silica, oxygen and other atoms and how different molecules are distributed into the marble block does not really help us if we just want to cut large slices out of the block and polish them to pave our living room.
It is exactly the ability of our thought to focus uniquely on these macroscopic materials' properties that allowed the modern scientific method to produce systematic technological and cultural advancement:
when Archimedes explained why certain solids float and others sink (Archimedes' principle), he did not focus his attention on the fact that both water and the solid are made of molecules, but only on the overall interaction forces between the two materials at the macroscopic scale.
On the same line, all the progress achieved in the last centuries enabling the efficient design of civil and aeronautical structures heavily relied on the ability to focus attention only on the relevant macroscopic materials properties (stiffness, mass, etc).

There is mostly unanimous agreement in the scientific community about the fact that engineers must take advantage of this macroscopic way of ``observing'' materials to design a building or an aircraft.
Today, while living in the era of high computational performances and artificial intelligence, we should not renounce to our critical thinking by stating that we should compute the dimensions of a dam by accounting for all the mutual interactions of water and concrete molecules.
Instead, we should most willingly focus our efforts to use these new tools to optimize the dam's shape, mass distribution, etc. so as to achieve the same result (building a dam) by using less material in view of sustainable construction.
It is even very likely that using the new computational capabilities to build a dam starting from atoms, while forgetting the achievements of classical continuum mechanics, would not provide a result that is as reliable as the classical one.
Such new computational tools can certainly push forward the achievements of classical mechanics by enabling the exploration of more sustainable structures in a way that could not be possible otherwise.

Scientists and engineers mostly agree on the general view given here about the ``macroscopic observation of materials''\footnote{ It is clear that physicists and chemists would be more interested in the discrete nature of matter, given the smaller scales at which occur the phenomena they are interested in.}.
When an engineer talks about a given \angcor{elastic} material (steel, concrete, etc.), the most relevant quantities to him are often the Young modulus, the Poisson ratio, and the apparent mass.
As a matter of fact, fixing specific values for these quantities is in some sense equivalent to choosing a specific material: the value of these quantities can be calculated once and for all (for example with mechanical tests in a laboratory) and subsequently used to design structures made up of the chosen materials.
It would be hard to find an engineer stating that the elastic modulus of steel can vary depending on the intensity of the applied load, as long as the material remains in the linear elastic regime.
To be more precise, we can briefly recall what is done in classical elasticity to describe the mechanical response of large blocks of a homogeneous material (i.e., a material in which we neglect its discrete structure).
In classical linear elasticity, a displacement field $u(x,t)$ is introduced describing the motion of a material point $x$ from the reference configuration to the current one. 
Each material point indeed represents a small homogeneous volume of matter that can interact with the adjacent elementary volumes in a way that is specific to each material.
From the study of the equilibrium of this continuous system, one can obtain a PDE governing its motion in which the only unknown is the macroscopic displacement $u$ as we can see in eq.(\ref{eq:equi_equa_Cauchy}).
Once the values of the mass density and the elastic coefficients of the desired material are chosen, solving this PDE will give information about the response (displacement and deformation) of the material under the application of a given external load (see Section \ref{sec:introCauchy} for a brief summary).

\subsection{Metamaterials and how to model them at the engineering scale}
\begin{figure}[!ht]
\centering
\includegraphics[width=0.75\textwidth]{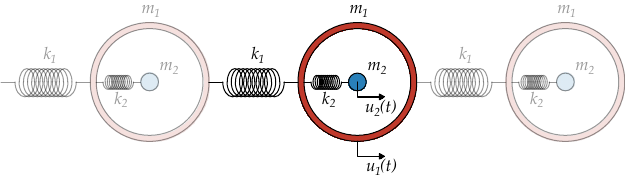}
\caption{Schematic representation of a metamaterial's ``unit-cell'' usually found in literature: its repetitions in space along one direction give rise to a 1D macroscopic metamaterial.}
\label{fig:2}
\end{figure}

In the last two or three decades, the classical concept of ``material'' has been revolutionized by the design and realization of materials whose heterogeneous nature can have visible effects at the macroscopic scale.
In particular, scientists and engineers purposely created materials with architectured microstructures in which the vibration of the microscopic components has an important effect at the macroscopic scale.
These materials are often called metamaterials in the sense that their mechanical response goes beyond (from the Greek ``meta'' = beyond) the one usually shown by the more classical materials that we are used to know.
The exotic dynamic metamaterials' responses at the macroscopic scale are indeed triggered by special vibration mechanisms taking place at the level of the architectured microstructure.
We are standing in front of something that was never observed before: the motion of the material constituents at lower scales has a non-negligible impact on the mechanical response at the macroscopic-scale.
More than this: the \angcor{overall} properties at the macroscopic-scale are almost completely determined by the motion of metamaterials' microstructure, at least for certain frequency ranges.
Typical examples are metamaterials exhibiting band-gaps (frequency ranges in which elastic waves cannot propagate) \cite{bilal2018architected,celli2019bandgap,liu2000locally,wang2014harnessing,el2018discrete,koutsianitis2019conventional,goh2019inverse,zhu2015study,fedele2023effective}, cloaking (elastic waves proceed unperturbed even if hitting the metamaterial) \cite{buckmann2015mechanical,misseroni2016cymatics,rossi2020numerical,misseroni2019omnidirectional,norris2014active}, focusing (a diffused incident wave is focused in a ray while passing inside the metamaterial) \cite{cummer2016controlling,guenneau2007acoustic,}, channeling (elastic waves take patterns with specific orientations while passing into the metamaterial) \cite{kaina2017slow,tallarico2017edge,bordiga2019prestress,wang2018channeled,miniaci2019valley}, negative refraction (waves are reflected in unusual way when hitting an interface) \cite{willis2016negative,bordiga2019prestress,zhu2015study,srivastava2016metamaterial,lustig2019anomalous,morini2019negative}, and many others.
Let us now repeat the same reasoning done in the case of classical elasticity 
when a material point $x$ does not represent a homogeneous elementary volume, but indeed represents a two-mass system of the type presented in Fig.~\ref{fig:2}.\footnote{The considerations drawn here are of general nature and are not bound to specific mass/spring microstructures. The mass/spring example proposed here is aimed at exposing the main concepts in one of the forms usually found in the literature, but the same considerations could be repeated for any heterogeneous material in which microscopic-motions have a non-negligible macroscopic effect.}

Metamaterials are often designed starting from a periodic repetition of building blocks like those in Fig.~\ref{fig:2}: the wanted exotic properties (e.g., macroscopic stopping of wave propagation) are obtained exploiting local resonances of the internal mass that starts vibrating at the microscopic-level and ``traps'' the energy of the propagating wave (the macroscopic effect of this microscopic-energy trapping is that no macroscopic propagation can be observed in the macroscopic metamaterial's block).
A common approach to describe the response of such metamaterials' blocks at the macroscopic scale is to use classical Cauchy elasticity 
This hypothesis implies that the ``unit-cell'' in Fig.~\ref{fig:2} is treated as a homogeneous ``black box'' that has an ``effective mass'' and a macroscopic displacement u. This ``effective mass'' is introduced as a suitable combination of the two original masses and depends on frequency\footnote{This dependence on the frequency of the ``effective mass'' can be found explicitly via identification between the dispersion relation for the considered unit-cell and a classical single mass-spring unit-cell \cite{shen2018analysis,huang2009negative}.
However, it can be intuitively understood that if we try to replace a two-mass (2DOF) system with a ``black box'' having only the displacement of the outer mass as a single DOF, the latter system implicitly requires the assumption that the average effective mass of the ``black box'' changes for different ways of vibrating of the inner mass.
This results in a frequency dependence of the effective mass.
In particular, when the internal mass strongly oscillates (for frequencies close to its own resonance frequency) in counter-phase with the displacement of the outer mass, the effective mass may become negative.}: as soon as the frequency approaches the resonance frequency of the internal mass, the effective mass may become negative giving rise to evanescent waves and thus band-gap behaviors \cite{huang2009negative,liu2000locally,liu2005analytic,milton2007modifications,shen2018analysis,faraci2023twoscale}.
An equivalent way to look at the same problem is to consider that the elastic stiffness of the spring (instead of the mass) is frequency-dependent so that a negative macroscopic elastic stiffness can be observed in frequency ranges where band-gaps occur \cite{liu2000locally,fang2006ultrasonic,seo2012acoustic}.

\begin{figure}[!ht]
\centering
\includegraphics[width=0.60\textwidth]{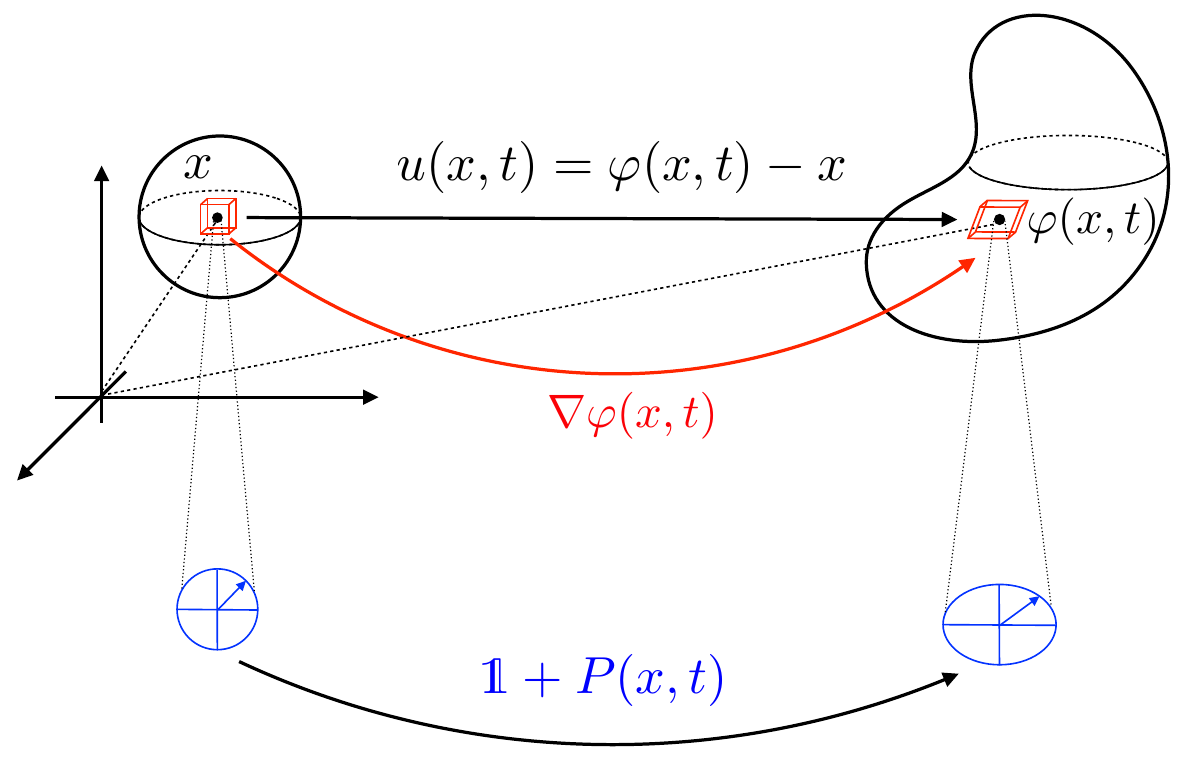}
\caption{
Schematic representation of the kinematics of a micromorphic continuum featuring an additional second-order tensor $P$ (the microdistortion) with respect to the macroscopic displacement $u$.
This extended kinematics allows us to describe affine microscopic-motions.
}
\label{fig:3}
\end{figure}

An approach of this type leads to a PDE of the same type as that of classical mechanics 
(\angcor{see} Section~\ref{sec:introCauchy}) where now the density and/or elastic moduli are not material constants \cors{anymore}, but depend on frequency \cite{willis1981variational,willis2011effective,willis2012construction,willis2016negative,milton2007modifications,nemat2011homogenization,srivastava2017evanescent}.
While this methodology can give important insight into macroscopic metamaterials' response (description of dispersion, band gaps, etc.) it has the main drawback that the definition of ``\angcor{engineering} material'', as we know \angcor{it}, 
results to be strongly perturbed.
Indeed, using this methodology we are implicitly assuming that the (meta-) material's macroscopic properties depend on the type (here the frequency) of the externally applied load.
In other words, we cannot give a finite set of constant parameters allowing us to describe the metamaterial's elastic response under any applied external load.
Moreover, when the motion of the internal mass becomes particularly important (local resonance/band-gap), the macroscopic mass density and elastic stiffness may counterintuitively become negative.
This is fundamentally related to the fact that we decided \textit{a priori} to neglect the presence of an additional degree of freedom, although its macroscopic effect is not negligible (see also \cite{huang2009negative}).
A solution to this drawback can be searched by considering continuous models allowing the presence of additional kinematical fields (in addition to the displacement) also at the macroscopic level.
This naturally leads to the introduction of the so-called micromorphic continuum models, whose extended kinematics classically features an additional second-order tensor $P(x,t)$ (called the microdistortion) with respect to the simple displacement field (Fig.~\ref{fig:3} and \cite{mindlin1964micro,eringen_mechanics_1968,eringen2012microcontinuum,neff2014unifying,madeo2015wave,ghiba2015relaxed,madeo2015band}).
In this way, the overall macroscopic response results to be simplified with respect to considering a detailed periodic juxtaposition of unit cells of the type in Fig.~\ref{fig:2} and one can arrive at a model featuring constant (i.e., frequency-independent) elastic parameters, while allowing the description of the main macroscopic metamaterials' characteristics (dispersion, band-gaps, etc.).

It is known that in the literature one can find extended continuum models featuring enriched constitutive behaviors while keeping the macroscopic displacement $u(x,t)$ as the only macroscopic kinematical field.
This is the case for, e.g., so-called second gradient continua \cite{barbagallo2021model,barbagallo2017modelling,madeo2014towards,madeo2013continuum,madeo2012second,sciarra2008variational,madeo2008variational,askes2011gradient,askes2009gradient,askes2006gradient,auffray2015analytical,germain2020method}, or also so-called Willis materials \cite{muhlestein2017experimental,willis1981variational}.
Even if these models may help to account for some effects of metamaterials' underlying microstructure (namely dispersion of the acoustic curves), they suffer drawbacks similar to those discussed before.
More specifically, while these models may describe some dispersion (also with constant, frequency-independent coefficients), it is impossible to describe optic curves (and thus band-gaps), without considering counterintuitive properties such as negative mass or stiffness.

In the last decades, many homogenization techniques have emerged trying to establish how to derive suitable macroscopic PDEs for mechanical metamaterials starting from specific microscopic unit-cells (upscaling techniques) \cite{craster2010high,boutin2014large,willis2011effective,willis2012construction,allaire1992homogenization,andrianov2008higher,auriault2012long,bensoussan2011asymptotic,boutin2014large,chen2001dispersive,marigo2016two,touboul2020effective}.
However, since the target macroscopic model is generally chosen \textit{a priori} to depend only on the displacement field, the associated parameters (mass and/or stiffness) turn \cors{out} to be frequency-dependent and may become negative for frequencies approaching the resonance frequency of the internal mass.

Recently, so-called computational homogenization techniques have been proposed that complement these upscaling techniques to include the possibility of letting enriched continua of the micromorphic type emerge at the macroscopic scale \cite{liu2021computational,sridhar2016homogenization}.

From our viewpoint, the micromorphic continuum structure can be postulated directly at the macroscopic scale (without trying to obtain it from a specific microstructure) and the metamaterials' properties specific to each metamaterial can be retrieved in a second instance by means of an inverse approach allowing to identify the micromorphic parameters.
This avoids complex descriptions and hypotheses that have to be drawn at the microscopic-scale to achieve the desired upscaling and allows one to focus attention on the macroscopic metamaterials' properties that one wants to exploit at the engineering scale.
However, it is not the aim of this paper to discuss how the specific examples presented here could be approached by reproducing the corresponding dispersion curves directly at the macroscopic scale (without starting from the specific microstructures) by using, e.g., the so-called relaxed micromorphic model. We address the interested reader to some of our previous papers for more details in this sense \cite{voss2022modeling,neff2020identification,d2020effective,d2017panorama,barbagallo2017transparent,rizzi2021exploring,rizzi2022boundary,rizzi2022metamaterial,ramirez2023multi,demore2022unfolding,rizzi2022towards,}.
\angcor{Instead, the present paper will thoroughly show a detailed procedure that allows to construct a micromorphic-type (frequency-independent) continuum model starting from a given frequency-dependent Cauchy model. It will be shown that the two models are fully equivalent in the frequency domain, while only the micromorphic model results to be well defined both in the frequency and in the time domain thanks to the time-Fourier transform.}

\subsection{Reconciling Cauchy frequency-dependent models and micromorphic frequency-independent models for mechanical metamaterials}
In the present paper, we explicitly show how specific frequency-dependent Cauchy models in the frequency domain can be transformed into their frequency-independent micromorphic counterparts in the time domain. 

To this aim, we propose a detailed procedure allowing us to pass from a frequency-dependent model to an associated micromorphic one by making use of suitable changes of variables and inverse partial Fourier transform.
Similar arguments can also be found in \cite{bellis2019simulating} for a specific 1D case.

The proposed procedure, which is able to ``transform'' a frequency-dependent model into its frequency-independent micromorphic counterpart must include mechanical consistency checks to ensure that the obtained micromorphic model is physically meaningful.
These consistency checks are:
\begin{enumerate}[i)]
    \item existence of an action functional from which the partial differential equilibrium equations in strong form can be obtained via a least-action principle together with consistent boundary conditions;
    \item \coran{positive definiteness;}
    \item conservation of total energy;
    \item Galilean invariance.
    \end{enumerate}
These consistency checks strongly reduce the set of possible micromorphic models that can be considered to be physically meaningful.

We show that when considering the associated micromorphic model, no elastic parameter depends on frequency anymore.
In this respect, the micromorphic model restores the classical notion of ``material'' also when metamaterials are considered: the micromorphic coefficients can be fixed once and for all for each metamaterial and all frequencies will describe its response notwithstanding the nature of the externally applied load.\footnote{This implies also that while the frequency-dependent model can only be used formally in the time-harmonic regime, the associated micromorphic one will not be limited to this special case anymore.}

The procedure proposed here aims at building a bridge between the frequency-dependent models usually found in the literature and our claim according to which macroscopic continua of the micromorphic type should be used to describe metamaterials’ response at the macroscopic scale.

%
%
%
%
%
%
\section{Classical elasticity: a summary on the Cauchy continuum model in the time domain and the frequency domain}
\label{sec:introCauchy}
Since this is widely used throughout the paper, we recall here some well-known features of classical elasticity.
\coran{Specifically, we present the process of 
transforming
the dynamic equations of the linear elasticity problem
from the time domain to the frequency domain.} 
Subsequently, we will perform the dispersion analysis
through two different approaches: a rigorous mathematical procedure involving the space-time-Fourier transform (or the space-Fourier transform if beginning with the associated frequency-dependent model), and a second approach employing the commonly found technique in the literature known as the ``plane wave ansatz" for the displacement field $u$.

\bigskip

The equilibrium equations for a classical linear elastic Cauchy continuum are
\begin{equation}
\rho \, \ddot{u}
=
\text{Div} \, \sigma 
\, ,
\label{eq:equi_equa_Cauchy}
\end{equation}
where $u$ is the displacement field, $\rho$ is the (constant) density, and $\sigma$ is  the symmetric Cauchy force-stress tensor.
The most general \coranR{linear elastic constitutive law} and the isotropic one are
\begin{equation}
\sigma \coloneqq \mathbb{C} \, \text{sym}\nabla u
\quad
\xrightleftharpoons[\text{general}]{\text{isotropic}}
\quad
\sigma \coloneqq 2\mu \, \text{sym}\nabla u + \lambda \, \text{tr} \left( \nabla u\right)\mathds{1} \, ,
\label{eq:sigma_Cauchy}
\end{equation}
where $\mathbb{C}$ is the classical 4$^{\text{th}}$ order elasticity tensor, and $\lambda$ and $\mu$ are the Lamé constants.

When restricted to the 2D case, eq.(\ref{eq:equi_equa_Cauchy}) remains formally the same, but it is intended that the displacement has only two non-zero components ($u_1$,$u_2$) which only depend on the coordinates in the plane ($x_1$,$x_2$).
%
%
%
%
\subsection{Dispersion analysis through the space-time-Fourier transform}
\label{dispersion_Fourier}

The dispersion analysis addresses a highly specific problem associated
to the physical relevant dynamic initial value problem for the linear elasticity system. In fact, the problem of primary physical interest takes the form:
\begin{equation}
    \begin{aligned}
    &\textrm{given  }\ensuremath{(f,g,h,u_{0},v_{0})}\textrm{ in a }\\
    &\textrm{suitable space of functions,}\\
    &\textrm{find }u\textrm{ solving}
\end{aligned}
    \hspace{2cm}
    \left.
         \begin{aligned}
              \rho\,\ddot{u}-\text{Div}\left[\C \, \text{sym}\nabla u\right] & =f & \textrm{in} & \; \Omega\times(0,T]
              \\
              \left.u\right|_{\Gamma_{D}} & =g & \textrm{on} &  \;  \Gamma_{D}\times[0,T]
              \\
              \left.\sigma\,n\right|_{\Gamma_{N}} & =h & \textrm{on} &  \;  \Gamma_{N}\times[0,T]
              \\
              u(0,\cdot)=u_{0},\quad\dot{u}(0,\cdot) & =v_{0} & \textrm{in} &  \;  \Omega\times\left\{ 0\right\} 
         \end{aligned}
    \right\}
\end{equation}
where $\Omega$ represents a domain in $\mathbb{R}^3$ with a boundary $\partial\Omega$ that is divided into two complementary parts: $\Gamma_N$ and $\Gamma_D$, on which Neumann and Dirichlet boundary conditions can be respectively applied. The initial data are denoted by $u_0$ and $v_0$. In contrast, the dispersion analysis exclusively addresses the bulk problem across the full space-time $\mathbb{R}^3 \times \mathbb{R}$ \coran{and, for this reason, can only be representative of the response of infinite media}. As a result, the information conveyed by the dispersion relations is limited, as it overlooks various aspects that characterize the specific problem under consideration, such as boundary and initial conditions. Consequently, as we will elaborate further, there exist multiple non-equivalent (they can show different behaviors such as differences in aspects like the conservation of total energy) models in the time domain yielding the same frequency-dependent model (through the time-Fourier transform). 

\bigskip

In current literature, the use of the Fourier transform is often replaced by the adoption of the plane-wave ansatz, i.e. considering $u(x,t)=\psi\,e^{i(\langle x , q \rangle-\omega\,t)}$ where $\psi\in\R^3$ \angcor{(or its partial representations as $u(x,t)=\varphi(x) \, \, e^{- \, i \, \omega \, t}$ or $u(x,t)=\phi(t)\, \,e^{i\langle x , q \rangle}$)}. Although this latter approach is formally equivalent to the space-time-Fourier approach when dealing with the dispersion analysis, it presents several inconveniences\footnote{
For instance, in the expression $\psi \, e^{i(\langle x , q \rangle-\omega \, t)}$, the Fourier variables $(q,\omega)$ are introduced alongside the space-time variables $(x,t)$, leading to a lack of clear separation between these two domains. The Fourier transform resolves this issue, enabling the proper introduction of functions that depend solely on one of the four pairs of variables: $(x,t)$, $(x,\omega)$, $(q,t)$, and $(q,\omega)$.
Moreover, the primary advantage of employing the Fourier transform lies in the flexibility it offers, as it allows to consider a broader range of functions beyond the specific structure of monochromatic plane waves.}. In this paper, we adopt the Fourier transform formalism if not differently specified. 
The space-time-Fourier transform 
$
      \mathcal{F}_{x,t}:L^2(\R^3_x\times\R_t)
      \xrightarrow[\hphantom{r}]{}
      L^2(\R^3_q\times\R_\omega),
$
is introduced such that
\begin{equation}
      u(x,t)
      \xmapsto[\hphantom{r}]{}
      \widehat{u}(q,\omega)
      \coloneqq
      \mathcal{F}_{x,t}[u](q,\omega)
      \coloneqq
      \frac{1}{(2\pi)^2}
      \int_{t\in\R} \int_{x\in\R^3} u(x,t)\,e^{i (\langle q , x \rangle - \omega \, t ) }
      \, dx \, 
      \,dt.
\end{equation}
The new variables $\omega$ and $q=[k_1,k_2,k_3]^{\rm T}$, whose norm is denoted by $\vert q\vert=k=\sqrt{k_1^2+k_2^2+k_2^3}$, are respectively the frequency and  the wavevector (and $k$ the wavenumber).

Applying $\mathcal{F}_{x,t}$ to the bulk equation \eqref{eq:equi_equa_Cauchy}
we obtain
\begin{equation}
    \mathcal{F}_{x,t}\big[\rho\,\ddot{u}-\text{Div}\left[\C\text{sym}\nabla u\right]\big]
    =
    - \, \rho \, \omega^2 \,  \widehat u(q,\omega) + \big( \C \, \text{sym} (\widehat u(q,\omega)\otimes q) \big) \, q
    =
    0.
    \label{eigenproblem}
\end{equation}
Let us define the linear operator $\mathbb A(\omega,q,\rho,\C):\R^3\to\R^3$ as
\begin{equation}
    \mathbb A(\omega,q,\rho,\C)
    \,
    \widehat u
    \coloneqq
    - \, \rho \, \omega^2 \,  \widehat u + \big( \C \, \text{sym} (\widehat u\otimes q) \big) \, q.
\end{equation}
The matrix representation of $\mathbb A(\omega,q,\rho,\C)$ is known as \textbf{dispersion matrix}\footnote{If we define a linear map $\mathbb D :\R^3\to\R^3$ via $\mathbb D\,\widehat u=\mathbb C\,\text{sym}\,(\widehat u\otimes q)\,q$, the matrix representation of $\mathbb D$ is known as the \textbf{acoustic tensor}.}.
As it is well known, in order to obtain non-trivial solutions of $\mathbb A(\omega,q,\rho,\C)
    \,
    \widehat u(q,\omega)=0$, it is necessary to search for values of $\omega$ (as functions of $q$) such that:
\begin{equation}
    \det
    \,
     \mathbb A(\omega,q,\rho,\C)
    =
    0.
    \label{eigenproblem_3}
\end{equation}
\coran{Considering $\rho$ and $\C$ to be known (fixing the material), the roots $\omega=\omega(q)$ (or $q=q(\omega)$) of eq.\eqref{eigenproblem_3} are known as dispersion curves of the material. }
\begin{figure}[!ht]
\centering
\includegraphics[width=0.49\textwidth]{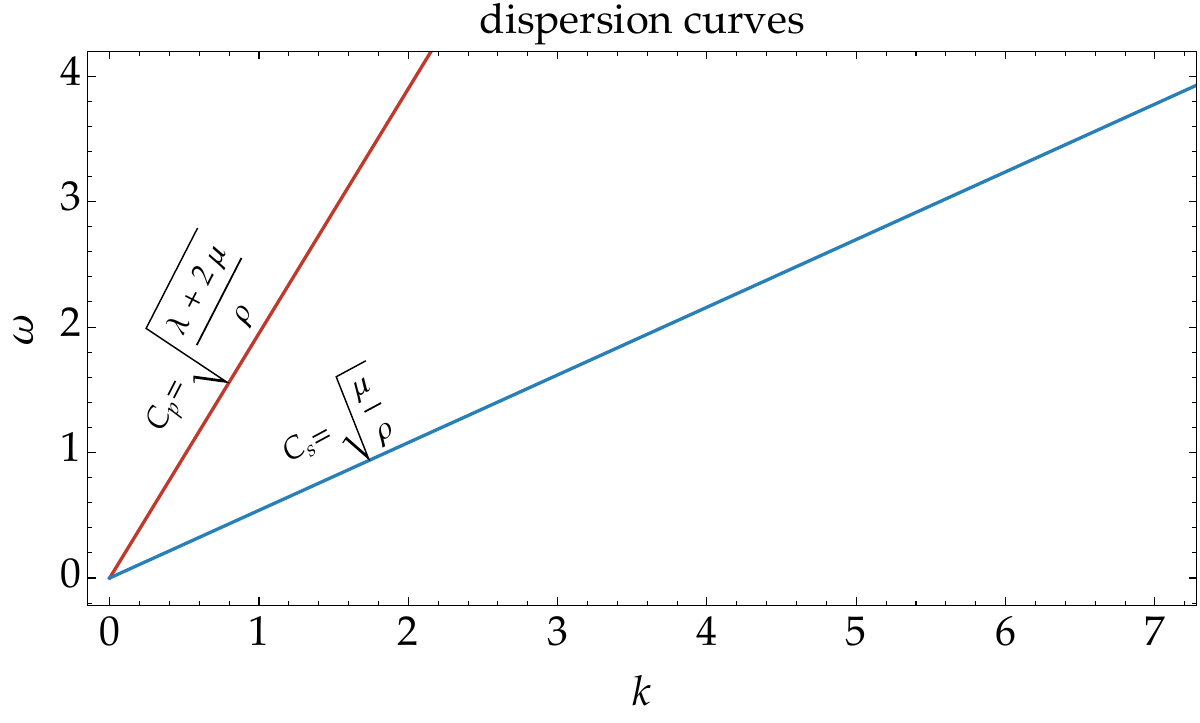}
\caption{Dispersion curves for a classical 2D isotropic Cauchy model in which the following values for the parameters have been used: $\rho=900$ kg/m$^3$, $\lambda=2898$ Pa, $\mu=262$ Pa. Given that the sole kinematic variable at play is the displacement field $u$, the resultant dispersion curves are exclusively acoustic. Furthermore, the linearity exhibited by the relationships $\omega_{\text p}(k)$ and $\omega_{\text s}(k)$ implies that the model is non-dispersive.}
\label{fig:classic_Cauchy_disp_curve}
\end{figure}
%
%
%
%
\subsection{Dispersion analysis through the space-time plane wave ansatz}
\label{sec:disp_relations}

The algebraic problems in \eqref{eigenproblem_3} can also be  derived introducing the monochromatic plane-wave ansatz for the displacement field $u$ i.e. setting
\begin{align}
u(x,t)
=
\psi \, e^{i(\langle q,x\rangle-\omega \, t)}
\label{eq:time_harmo_ansatz}
\end{align}
where $\psi\in \mathbb{R}^3$ is the amplitude vector. 
Substituting the ansatz (\ref{eq:time_harmo_ansatz}) in the equilibrium equations (\ref{eq:equi_equa_Cauchy}), we formally obtain 
the same family of algebraic problems, namely $ \mathbb A(\omega,q,\rho,\C)\,\psi=0$
\coran{which thus implies eq.\eqref{eigenproblem_3} for the search of dispersion curves.}
%
%
%
%
\subsection{Linear elasticity in the frequency domain (dispersion curves through the subsequent application of time and space-Fourier transform)}
\label{sec:Cle-thr}

Let us consider the linear elastic problem \corc{in the full space-time}
\begin{equation}\label{general_problem}
     \rho
     \,
     \ddot{u}
     -
     \text{Div}[\mathbb{C}\,\text{sym}\nabla u]
     =
     0
\end{equation}
where $u:\R^3_x\times\R_t\to\R^3$ is the displacement field. To derive a solution for eq.\eqref{general_problem}, we can break it down into a parameterized collection of simpler problems using the time-Fourier transform $\mathcal{F}_t$. This transform is defined as follows:
\[
      \mathcal{F}_t:L^2(\R_x\times\R_t)
      \xrightarrow[\hphantom{r}]{}
      L^2(\R_x\times\R_\omega),
      \qquad
      u(x,t)
      \xmapsto[\hphantom{r}]{}
      \widehat{u}(x,\omega)
      \coloneqq
      \mathcal{F}_t[u](x,\omega)
      \coloneqq
      \frac{1}{\sqrt{2\pi}}
      \int_{t\in\R} u(x,t)\,e^{-\,i\,\omega\,t}\,dt.
\]
Throughout this section and the remainder of the paper, whenever we need to emphasize the distinction between the time domain $\R$ in the domain of definition of the time-Fourier transform and the $\R$ in the codomain, we will denote them as $\R_t$ and $\R_\omega$, respectively. The same will be done for the $\R^3_x$ and $\R^3_q$ for the space-Fourier transform. The advantage of the time-Fourier transform is related to the fact that it ``converts" (as a consequence of the integration by parts)
\footnote{Indeed, \begin{align}
\mathcal{F}_t[\dot u](x,\omega)
&=
\int_\R \dot u(x,t)\,e^{-\,i\,\omega\,t}\,dt
=
\int_\R \Big[\frac{d}{dt}\Big(u(x,t)\,e^{-\,i\,\omega\,t}\Big)-u(x,t)\,\frac{d}{dt}\,e^{-\,i\,\omega\,t}\Big] \,dt
=
\overbrace{\left.u(x,t)\,e^{-\,i\,\omega\,t}\right|_{-\infty}^{+\infty}}^{=\,0\;\text{because}\,u\in L^2}-\int_\R -\,i\,\omega\,u(x,t)\,e^{-\,i\,\omega\,t} \,dt
\nonumber
\\
&=
i\,\omega\int_\R u(x,t)\,e^{-\,i\,\omega\,t} \,dt
=
i\,\omega\F_t[u](x,\omega).
\end{align}} 
derivatives in polynomial factors, i.e. for example 
\[
      \ddot{u}(x,t)
      \xmapsto[\hphantom{r}]{}
      \mathcal{F}_t[\ddot{u}](x,\omega)
      =
      -\,\omega^2\,\mathcal{F}_t[u](x,\omega)
      =
      -\,\omega^2\,\widehat{u}(x,\omega).
\]
To make the notation lighter from now on, we will simply write $\widehat u$ instead of $\widehat u(x, \omega)$ when this does not create confusion. In this way, applying $\mathcal{F}_t$ to eq.\eqref{general_problem} leads to a family (parameterized by $\omega$) of PDEs 
\begin{equation}
     -\,
     \rho
     \,
     \omega^2
     \,
     \widehat{u}
     -
     \text{Div}[\mathbb{C}\,\text{sym}\nabla \widehat u]
     =
     0.
     \label{freq_dep_problems}
\end{equation}
\coran{Eq.s \eqref{freq_dep_problems} are referred to as the ``linear elasticity problem in the frequency domain" and are often the starting point for frequency-dependent models usually found in the literature to describe metamaterial responses when letting $\rho$ or $\C$ to be frequency-dependent. Concerning the dispersion analysis associated to eq.\eqref{freq_dep_problems}, we }
explicitly remark that we can obtain the family of algebraic problems $\det \, \mathbb A(\omega,q,\rho,\C)=0$, \coran{equivalent to eq.\eqref{eigenproblem_3}}, also considering the model in the frequency domain  \eqref{freq_dep_problems} and \coran{subsequently} applying the space-Fourier transform 
\begin{equation}
      \mathcal{F}_{x}:L^2(\R^3_x\times\R_\omega)
      \xrightarrow[\hphantom{r}]{}
      L^2(\R^3_q\times\R_\omega),
      \qquad
      \widehat u(x,\omega)
      \xmapsto[\hphantom{r}]{}
      \widehat{u}(q,\omega)
      \coloneqq
      \mathcal{F}_{x}[ \widehat u](q,\omega)
      \coloneqq
      \frac{1}{(2\pi)^{\frac{3}{2}}}
      \int_{x\in\R^3} \widehat u(x,\omega)\,e^{i \, \langle q , x \rangle }
      \, dx \, 
\end{equation}
to\footnote{By an abuse of notation, we employ the same symbol, $\widehat u $ \coran{without specifying its argument}, to represent the three images $\mathcal{F}_{x,t}[u]$, $\mathcal{F}_{x}[u]$ and $\mathcal{F}_{t}[u]$, \coran{when no confusion can arise}. 
} $\widehat u(x,\omega)$. In other words, we have that 
$
      \mathcal{F}_{x,t}
      =
      \mathcal{F}_{x}\circ \mathcal{F}_{t}
      =
      \mathcal{F}_{t} \circ \mathcal{F}_{x}.
$
\coran{The same family of algebraic problems can be also derived via the space-plane-wave ansatz for the displacement $\widehat u(x,\omega)$.}
\coran{\subsubsection{Dispersion curves for 2D isotropic Cauchy media}}
When the accounted medium is isotropic, i.e.
$
\C
\,
\text{sym}(\widehat u \, \otimes \, q)
=
2 \, \mu \, \text{sym}(\widehat u \, \otimes \, q) 
+  \lambda \, \langle\mathds{1},\text{sym}(\widehat u \, \otimes \, q)\rangle \, \mathds{1},
$
remarking that
$
\langle\mathds{1},\text{sym}(\widehat u \, \otimes \, q)\rangle
=
\text{tr}\,(\text{sym}(\widehat u \, \otimes \, q))
=
\text{tr}\,(\widehat u \, \otimes \, q)\,
=
\langle \widehat u , q \rangle,
$
we see that
$
 \text{tr}\,(\widehat u \, \otimes \, q) \, \mathds{1} \, q
 =
 \langle
 \widehat u , q 
 \rangle
 \,q
 =
 (q\otimes q) \, \widehat u, 
$
and
\begin{align}
\text{sym}(\widehat u \, \otimes \, q) \, q
&
=
\frac{1}{2}\Big(\widehat u \otimes q + q \otimes \widehat u \Big) \, q
=
\underbrace{\text{tr}\,(q\otimes q)}_{k^2} \, \widehat u
+
\langle
\widehat u , q 
\rangle
\,q
=
k^2 \, \widehat u
+
(q\otimes q) \, \widehat u.
\end{align}
\cors{Equation \eqref{eigenproblem_3} then simplifies as}
\begin{equation}
\det
\bigg(
-
\,
\rho  \, \omega^2 \, \mathds{1} 
+
\mu \, k^2 \, \mathds{1} + (\mu+\lambda) \, q\otimes q \bigg)
=
0,
\qquad
\qquad
q=(k_1,k_2,k_3).
\end{equation}
Considering now the 2D isotropic case, we obtain
\begin{align}
    \det
    \begin{pmatrix}
        - \, \rho \, \omega^{2} \,
        +
        \mu\,k^{2}+(\mu+\lambda)\,k_{1}^{2} 
        & (\mu+\lambda)\,k_{1}\,k_{2}
        \\
        (\mu+\lambda)\,k_{1}\,k_{2} 
        & -\,\rho\,\omega^{2} \, + \mu \, k^{2}+(\mu+\lambda)\,k_{2}^{2}
    \end{pmatrix}
    &
    \\[3mm]
    =
    \mu \, (\lambda + 2 \mu) 
    \,
    k^4
    - 
    \rho \, \omega^2 \, k^2 \, (\lambda + 3 \mu)  
    + 
    \rho^2 \, \omega^4
    &
    =0, \nonumber
\end{align}
i.e. eq.(\ref{eigenproblem_3}) gives rise to a polynomial \angcor{that is bi-quadratic both in $k^2$ and $\omega^2$}.
The polynomial (\ref{eigenproblem_3}) can be equivalently solved in terms
of both $\omega(k)$ and $k(\omega)$ giving equivalent but inverse relations.
The roots of eq.(\ref{eigenproblem_3}) in terms of $k(\omega)$ are:
\begin{align}
&\corrections{k_{\rm p}}
\coloneqq \frac{1}{C_{\rm p}} \, \omega
=\pm \sqrt{\frac{\rho}{\lambda+2\mu}} \, \omega
\, ,
&&\corrections{k_{\rm s}}
\coloneqq \frac{1}{C_{\rm s}} \, \omega
=\pm \sqrt{\frac{\rho}{\mu}} \, \omega
\, ,
\label{eq:disp_relations_Cau_express}
\end{align}
where $\corrections{k_{\rm p}}$ is a solution associated with the propagation of pressure waves, while $\corrections{k_{\rm s}}$ with the propagation of shear waves. 
In Fig.~\ref{fig:classic_Cauchy_disp_curve} it is possible to see the plot of the dispersion relations (\ref{eq:disp_relations_Cau_express}), i.e., the dispersion curves of an isotropic Cauchy continuum, for specific values of the parameters. \coran{The equivalent but inverse relations $\omega(k)$ have a more complex expression and will not be shown here.}
Here $\corrections{C_{\rm p}}$ and $\corrections{C_{\rm s}}$ are the speed of propagation of pressure and shear waves, respectively.
It is highlighted that, since we have chosen an isotropic constitutive law, the dispersion relations depend just on the wavenumber $k$ and not on the direction of propagation since the response of the material must be the same regardless the direction.
%
%
%
%
\begin{tcolorbox}[colback=myblue4!15!white,
                  colframe=myblue4!50!white, 
                  coltitle=black,
                  breakable,
                  pad at break=1mm,
                  title={\centering \textbf{Summary:} \coran{Equivalent techniques for the dispersion analysis for classical linear elasticity} }]
\textbf{1. Through the space-time-Fourier transform:}
\\[1em]
Applying $\mathcal{F}_{x,t}$ to $\rho \, \ddot{u}
=
\text{Div} \, \sigma $ we obtain the family of algebraic problems $\mathbb A(\omega,q,\rho,\C) \, \widehat u=0$. To obtain non trivial solutions, we need to look for the couples $(\omega,q)$ \coran{such that the characteristic polynomial is vanishing}:
\[
\det \mathbb A(\omega,q,\rho,\C)
    =
    0.
\]
The roots $k(\omega)$ of the characteristic polynomial, for the 2D-isotropic case, give the dispersion curves: 
\begin{align}
&\corrections{k_{\rm p}}
\coloneqq \frac{1}{C_{\rm p}} \, \omega
=\pm \sqrt{\frac{\rho}{\lambda+2\mu}} \, \omega
\, ,
&&\corrections{k_{\rm s}}
\coloneqq \frac{1}{C_{\rm s}} \, \omega
=\pm \sqrt{\frac{\rho}{\mu}} \, \omega
\, . \nonumber
\end{align}
\textbf{2. Through the space-time plane wave ansatz:}
\\[1em]
Setting
$
u(x,t)
=
\psi \, e^{i(\langle q,x\rangle-\omega \, t)}
$,
inserting it into the bulk equation $\rho \, \ddot{u}=\text{Div} \, \sigma $ we obtain the same family of algebraic problems 
$
\mathbb A(\omega,q,\rho,\C)\,\psi
=0
\;
\Longrightarrow
\;
\det \mathbb A(\omega,q,\rho,\C)=0
$,
\coran{whose roots are the dispersion curves.}
The constant amplitude vector $\psi$ takes formally the role \coran{that $\widehat u$ had} in the space-time-Fourier approach.
\\[1em]
\textbf{3. Through the space-Fourier transform of the elasticity model in the frequency domain:}
\\[1em]
Starting from linear elasticity written in the frequency domain 
\begin{equation}
-\,
     \rho
     \,
     \omega^2
     \,
     \widehat{u}
     -
     \text{Div}[\mathbb{C}\,\text{sym}\nabla \widehat u]
     =
     0
     \label{ddd}
\end{equation}
the dispersion relations
$
\mathbb A(\omega,q,\rho,\C)\,\widehat u
=0
\;
\Longrightarrow
\;
\det \mathbb A(\omega,q,\rho,\C)=0
$
can be also obtained by applying the space-Fourier transform $\mathcal F_x$ to eq.\eqref{ddd}. 
\\[1em]
\textbf{4. Through the space-plane wave ansatz applied to the elasticity model in the frequency domain:}
\\[1em]
Setting 
$
    \widehat u(x,\omega)=\angcor{\psi(\omega)}\,e^{i \, \langle x, q\rangle}
$
and inserting it into 
     $-\,
     \rho
     \,
     \omega^2
     \,
     \psi
     -
     \text{Div}[\mathbb{C}\,\text{sym}\nabla \psi]=0$ we obtain again the dispersion relations from
\[
\mathbb A(\omega,q,\rho,\C)\,\psi
=0
\qquad
\Longrightarrow
\qquad
\det \mathbb A(\omega,q,\rho,\C)=0.
\]
\coran{\textbf{Conclusion:} These techniques for dispersion analysis are equivalent when the goal is to derive the dispersion curves.}
\end{tcolorbox}

\subsubsection{Passing from the frequency domain to the time domain through inverse time-Fourier transform}\label{sec:new_ang}

Since it will be at the basis of the new procedure presented in this paper to transform a frequency dependent model into an enriched model, we briefly illustrate here how it is possible to pass from the frequency domain to the time domain by making use of the inverse time-Fourier transform for classical linear elasticity.
Starting from 
\begin{equation}
     -\,
     \rho
     \,
     \omega^2
     \,
     \widehat{u}
     -
     \text{Div}[\mathbb{C}\,\text{sym}\nabla \widehat u]
     =
     0
\end{equation}
and remarking that $\F_t^{-1}[-\,\omega^2\,\widehat{u}]=\ddot{u}$, applying $\F_t^{-1}$ to both sides of the previous equation we obtain
\begin{align*}
     -\,
     \rho
     \,
     \omega^2
     \,
     \widehat{u}
     -
     \text{Div}[\mathbb{C}\,\text{sym}\nabla \widehat u]
     =
     0
     \quad
     &
     \xRightarrow{\F_t^{-1}}
     \quad
     \F_t^{-1}\big[ -
     \rho
     \,
     \omega^2
     \,
     \widehat{u}
     -
     \text{Div}[\mathbb{C}\,\text{sym}\nabla \widehat u] \big] = \F_t^{-1}[0]
     \\
     &
     \xLeftrightarrow{\hphantom{\F_t^{-1}}}
     \quad
     \rho  \,
     \ddot{u}
     -
     \text{Div}[\mathbb{C}\,\text{sym}\nabla u] =0.
\end{align*}

\begin{tcolorbox}[colback=BlueGreen!15!white,colframe=BlueGreen!50!white,coltitle=black,title={\centering \textbf{Summary:} linear elasticity in the frequency domain \coran{and its time domain counterpart}}]
\textbf{Linear elasticity in the frequency domain:}
\\[1em]
$
- \,
\rho
\,\omega^2 \, \widehat u
=
\text{Div} \left[ \mathbb{C} \, \text{sym}\nabla\widehat u\right]
$
\\[1em]
\textbf{Time domain counterpart:}
\\[1em]
The action functional associated to the  model in the time domain obtained from the model in the frequency domain through the inverse time-Fourier transform is
\\[1em]
$
\mathcal{A}
=
\displaystyle\iint\displaylimits_{\Omega\times [0,T]}
\underbrace{
\frac{1}{2}
\, \rho \, \langle \dot{u} , \dot{u} \rangle
}_{\substack{\text{K - kinetic}\vphantom{\int}\\\text{energy density}}}
-
\;\;
\underbrace{
\frac{1}{2} \, \langle \mathbb{C} \, \text{sym}\nabla u, \text{sym}\nabla u \rangle
}_{\text{W - strain energy density}}
\mathrm{d}x\.\mathrm{d}t\,.
$
\\[1em]
The equilibrium equations in $\Omega$ are:
$
\quad
\rho \, \ddot{u}
- \, \text{Div} \left[ \mathbb{C} \, \text{sym}\nabla u\right]
=0
\, ,
$
\\[1em]
and the Neumann boundary conditions on $\partial\Omega\times[0,T]$ are:
$
\quad
(\mathbb{C} \, \text{sym}\nabla u
) \, n = 0 
$.
\\[1em]
\textbf{Consistency checks of the model in the time domain:}
\\[1em]
\text{positive-definiteness} \!\!\!\quad\textcolor{Green}{\cmark}\quad (but does not allow for band-gaps)
\text{energy conservation}\quad\textcolor{Green}{\cmark}
\text{infinitesimal Galilean invariance \eqref{eq:ap_Galilean}}\quad\textcolor{Green}{\cmark}
\text{extended infinitesimal Galilean invariance \eqref{eq:ap_GalileanExtended}}\quad\textcolor{Green}{\cmark}
\end{tcolorbox}
%
%
%
%
\section{A simple Cauchy model with frequency-dependent density and its enriched frequency-independent counterpart}
\label{sec:rho_freq}
In the frequency domain, we have the option to tackle a fresh set of problems, wherein the inertia and elastic tensors are functions of the frequency\footnote{\angcor{Frequency-dependent densities (elastic tensors) result from the attempt to reduce the degrees of freedom of systems with complex kinematics by means of upscaling (homogenization)
procedures (see e.g. \cite{huang2009negative} and references there cited). For example, each unit cell of the 1D system presented in Fig.1 would naturally have 2 degrees of freedom. Instead, the classical approach is to replace the true unit cell with a « homogenized » cell which only considers the displacement of an « equivalent » single mass as the only degree of freedom. Consequently, this hypothesis requires adjustments in the average effective mass (elastic tensor) to accommodate different modes of vibration of the inner spring-mass system. These adjustments result in frequency-dependent homogenized material properties}} $\omega$. This entails considering the parametrized family of PDE systems
%
\begin{equation}\label{freq_dep_problems_general}
     -\,
     \widetilde \rho(\omega)
     \,
     \omega^2
     \,
     \widehat{u}
     -
     \text{Div}[\widetilde{\mathbb{C}}(\omega)\,\text{sym}\nabla \widehat u]
     =
     0,
     \qquad\qquad
     \forall\omega\in\R
     ,
\end{equation}
where
$
     \widetilde \rho:\text{Dom} \, \widetilde \rho\subset\R\xrightarrow[\hphantom{r}]{}\R
$
 and 
$
     \widetilde{\mathbb{C}}:\text{Dom} \, \widetilde \C\subset\R\xrightarrow[\hphantom{r}]{}
     \mathrm{Sym}^{\,+}\!\left(\mathrm{Sym}(3),\mathrm{Sym}(3)\right)
$
and where $\mathrm{Sym}^{\,+}\!\left(\mathrm{Sym}(3),\mathrm{Sym}(3)\right)$ is the space of positive definite fourth order elasticity tensors.

\begin{figure}[!ht]
\centering
\includegraphics[width=0.49\textwidth]{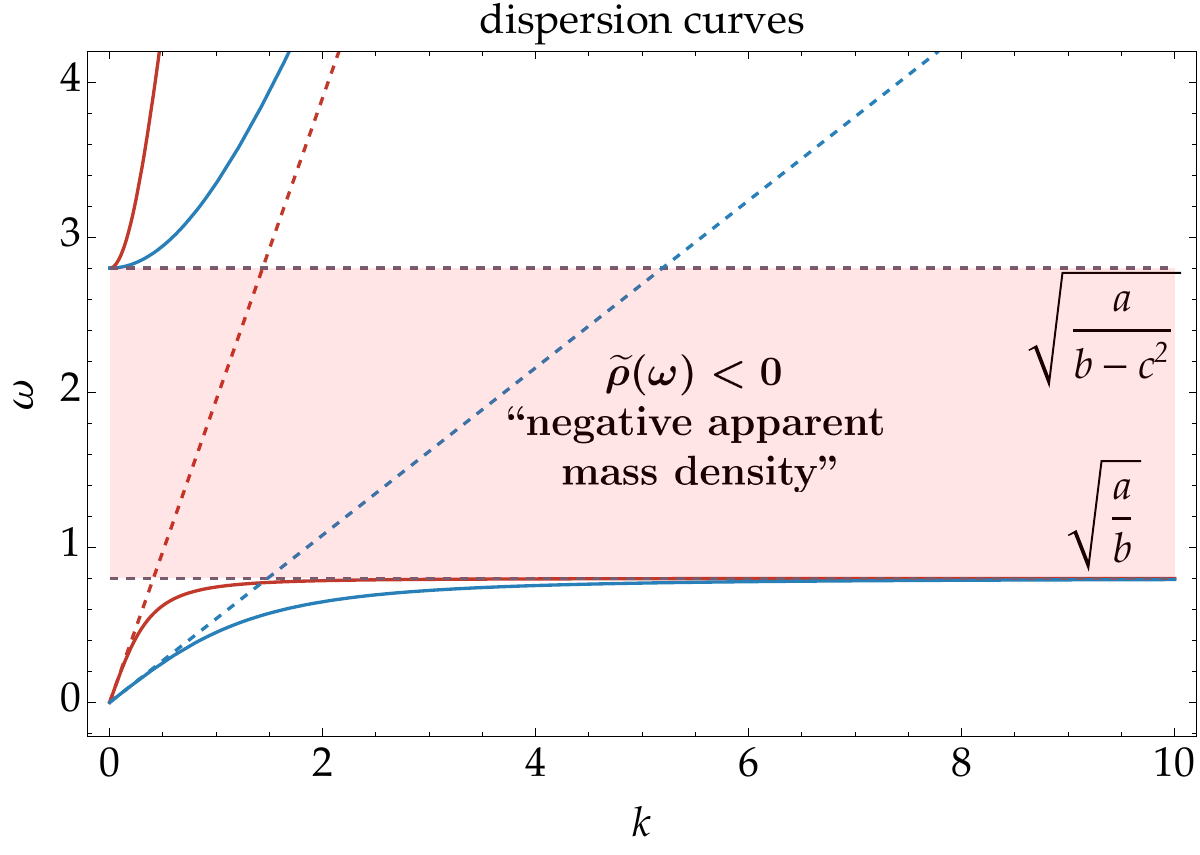}
\caption{
Dispersion curves for the 
\coran{$\widetilde\rho(\omega)$ model}
in which the following values for the parameters have been used: $\rho=900$ kg/m$^3$, $\lambda=2898$ Pa, $\mu=262$ Pa, $a=0.697\;\si{1\per s\squared}$, $b=1.089$, and $c=1$.
The curves for the frequency-dependent model and the corresponding enriched model (Fig.~\ref{fig:disp_curves_freq_depe_negative}) coincide.
However, while the enriched model remains \cors{positive-definite} in the band-gap region, the frequency-dependent one does not.
In this plot we also report the dispersion curves for the classical Cauchy model (dashed lines) already presented in Fig.~\ref{fig:classic_Cauchy_disp_curve}.
In correspondence of the band-gap interval $(\sqrt{\nicefrac{a}{b}},\sqrt{\nicefrac{a}{(b-c^2)}})$ the frequency-dependent mass density function $\widetilde \rho$ attains negative values.}
\label{fig:disp_curves_freq_depe_negative_Cau}
\end{figure}

In this section, we will study the case in which only the density $\widetilde\rho$ is a function of the frequency, reserving more general cases for subsequent paragraphs. Therefore, we  consider the parametrized family of PDE systems in the frequency domain,
\begin{equation}
-
\,
\widetilde{\rho}(\omega) \.\omega^2\.\widehat u
=
\text{Div} \left[ \mathbb{C} \, \text{sym}\nabla\widehat u\right]
\qquad
\text{with}
\qquad
\widetilde{\rho}(\omega)
=
\rho\left( 1+\frac{c^2 \, \omega^2}{a - b \, \omega^2
}
\right) 
\quad
\text{and}
\quad
\text{Dom} \, \widetilde \rho =  \R \, \setminus \left\{\pm\sqrt{\frac{a}{b}}\right\}
\, ,
\label{eq:poly_rho_freq_depe_negative}
\end{equation}
where $b$ and $c$ are dimensionless coefficients, and $a$ has the dimension of s$^{-2}$.
Note that $\displaystyle\lim_{\omega \to 0}\widetilde{\rho}(\omega)=\rho$, which means that the density approaches the classical value in the long-wave limit.
\angcor{It must be underlined that the choice of the function $\widetilde \rho(\omega)$ in eq.(\ref{eq:poly_rho_freq_depe_negative}) cannot be completely arbitrary and that expressions of this type can be often found in the literature when considering ``unit cells'' like those in Fig.~\ref{fig:2} \cite{milton2012metamaterial,huang2009negative,shen2018analysis}.}
The procedure to derive the dispersion curves for the frequency-dependent model is formally the same as the one presented in Section \ref{sec:introCauchy} and is summarized in Appendix \ref{disp_analysis_app} for this particular case.
The assumed dependence of the inertia $\rho$ on the frequency $\omega$ implies that the determinant of the associated acoustic tensor will now be a rational fraction with respect to $\omega$.
When solved in terms of $k$, the dispersion relations for the frequency-dependent density model (for an isotropic medium) read
\begin{align}
    k_{\rm p}&=\sqrt{\frac{\rho \, \omega ^2 \left(a+\omega ^2 \left(c^2-b\right)\right)}{(\lambda + 2 \mu) \left(a-b \, \omega ^2\right)}}=\omega\sqrt{\frac{\rho}{\lambda + 2 \mu}\left( 1+\frac{c^2 \, \omega^2}{a - b \, \omega^2} \right)}\, ,
    \label{eq:disp_relations_freq_depe}
    \\
    k_{\rm s}&=\sqrt{\frac{\rho \, \omega ^2 \left(a+\omega ^2 \left(c^2-b\right)\right)}{\mu \left(a-b \, \omega ^2\right)}}=\omega\sqrt{\frac{\rho}{\mu}\left( 1+\frac{c^2 \, \omega^2}{a - b \, \omega^2} \right)}
    \,,\notag
\end{align}
where 
$ k_{\rm p}$ and $k_{\rm s}$ represent the wavenumber for pressure and shear waves, respectively.
A solution in terms of $\omega(k)$ could also be explicitly computed but has a more complex expression and will not be shown here, while a plot of these relations is reported in Fig.~\ref{fig:disp_curves_freq_depe_negative_Cau}.

We can see that the relation between $k$ and $\omega$ is no longer linear as for the dispersion relations (\ref{eq:disp_relations_Cau_express})  of the classical (non-frequency-dependent) Cauchy model, which means that the dispersion curves issued from eq.(\ref{eq:disp_relations_freq_depe}) are then able to account for dispersion and for band-gaps (the argument of the square root can become negative, thus triggering evanescent waves which stop propagation).
\subsection{Time domain models associated to the presented frequency-dependent model}

In this section, we demonstrate how, 
starting from a frequency dependent model in the frequency domain,
we can derive various models in the time domain via suitable changes of variables and the use of the inverse time-Fourier transform. \angcor{It is clear that, since our procedure is based on the introduction of auxiliary variables, this choice cannot be unique, thus implying the possibility of different time-dependent models.} 
\angcor{If one only considers the introduction of auxiliary variables to transform a frequency-dependent model into a frequency-independent micromorphic-type model, one could thus end up with a large number of models in the time domain. 

However, admissible macroscopic models in the time-domain must be mechanically consistent, so that from the many possible models one should select only those satisfying the following minimal requirements:
\begin{enumerate}
    \item existence of an action functional,
    \item positive definiteness of the action functional,
    \item Conservation of the total energy (if the observed system is conservative),
    \item Galilean invariance.
\end{enumerate}
As a matter of fact, points 1) and 2) automatically imply that no creation of energy occurs in the considered mechanical system (thermodynamic consistency), while point 3) is a check that the candidate model does not have internal flows, since the conservation of the total energy must be always satisfied when considering non-dissipative systems. Finally point 4) guarantees that the elastic energy does not change when changing the observer. 

A macroscopic model in the time domain which verifies the aforementioned four properties can thus safely be considered to be well posed from a mechanical point of view. Imposing 1) and 2) drastically restricts the number of possible emerging models in the time domain. Indeed, it becomes quite difficult to identify auxiliary variables that give rise to suitable action functionals that are also positive-definite. However, the imposition of 1) and 2) is not sufficient to isolate a unique consistent macro-model. Imposing 3) ulteriorly restricts the set of possible macro-models, allowing to unveil models' imprecisions that would remain otherwise undetected. Imposing 4) is necessary and further reduces the number of possible models in the time domain.}


\angcor{
We will show  that all these models in the time domain yield the same frequency-dependent model when subjected to the time-Fourier transform. 
Indeed, one could argue that those models in the time domain are equivalent because they yield the same dispersion relations through $\F_{x,t}$ when examined across the entire space-time $\R^3\times\R$, and because the well-posedness of one implies the well-posedness of the others. However, they may also exhibit distinct behaviors, such as differences in infinitesimal Galilean invariance or the conservation of total energy. This section is devoted to the exploration of these issues.}


In particular, for the proposed frequency-dependent model \eqref{eq:poly_rho_freq_depe_negative}, we will derive two different PDE systems in the time domain using \coran{two distinct changes of variables and subsequently applying} $\F_{x,t}^{-1}$. The first model corresponds to a fourth-order system \coran{with only the displacement as unknown field}, while the second one corresponds to a second-order system with extra degrees of freedom with respect to the displacement alone.

We will investigate Galilean invariance and the conservation of total energy for both of them, demonstrating that the first one fails to meet both requirements, whereas the second one addresses the issue of total energy conservation but still does not satisfy the infinitesimal Galilean principle\footnote{This is a well-established fact that, in general, geometrically linear models do not satisfy the Galilean invariance principle (as formulated in the general non-linear framework). Nevertheless, an invariance condition can be derived ``linearizing" the Galilean principle as demonstrated in the Appendix \ref{sec:appGalilean}.}. To address this issue related to Galilean invariance, we will subsequently propose other enriched models so as to finally achieve an enriched time-dependent model that addresses all needed physical requirements (see Section \ref{sec:CC_freq}).
%
%
%
\subsubsection{First attempt: a higher order model}

The first attempt is to derive a time-dependent model from the frequency-dependent model \eqref{eq:poly_rho_freq_depe_negative} directly applying an inverse Fourier transform (as done in \ref{sec:new_ang} for classical linear elasticity in the time domain).
Starting from \eqref{eq:poly_rho_freq_depe_negative} we obtain\footnote{
\angcor{\label{footnote omega}
%
The equivalence stated in equation \eqref{eq:fourth_order_model_fourier} remains valid even when considering $\omega\in\R$. This is due to the nature of the Fourier transform as an integral transform, where its value at a specific point is defined  up to a subset of measure zero. This means that, when starting from the time-domain model \eqref{eq:fourth_order_model_fourier}, one can choose if eliminating eventual roots $\omega\in\left\{\pm\sqrt{\frac{a}{b}}\right\}$ in the associated frequency-domain model or not. 
}
}
\begin{align}
&
&
-
\rho \, \.\omega^2\.\widehat u
\,
-
\,
\rho \frac{c^2\,\omega^2}{a-b\,\omega^2}\.\omega^2\.\widehat u
&=
\text{Div} \left[ \mathbb{C} \, \text{sym}\nabla \widehat u\right]
\, ,
&
&
\hspace{-7mm}
\forall\,\omega\in\R\,\setminus\left\{\pm\sqrt{\frac{a}{b}}\right\}
\label{eq:Fourier_freq_main}
\\*[5pt]
&\Longleftrightarrow
&\quad
-
(a-b\,\omega^2)\rho \, \.\omega^2\.\widehat u
\,
-
\,
\rho \, c^2\,\.\omega^4\.\widehat u
&=
(a-b\,\omega^2)\text{Div} \left[ \mathbb{C} \, \text{sym}\nabla \widehat u\right]
\, ,
&
&
\hspace{-7mm}
\forall\,\omega\in\R\,\setminus\left\{\pm\sqrt{\frac{a}{b}}\right\}
\label{eq:Fourier_freq_second}
\\*[5pt]
&\Longleftrightarrow
&\quad
-
a\,\rho\,\omega^2\.\widehat u + b \, \rho\, \omega^4 \, \widehat u
\,
-
\,
\rho \, c^2\,\omega^4\.\widehat u
&
=
(a-b\,\omega^2)\text{Div} \left[ \mathbb{C} \, \text{sym}\nabla\widehat u\right]
\, ,
&
&
\hspace{-7mm}
\forall\,\omega\in\R\,\setminus\left\{\pm\sqrt{\frac{a}{b}}\right\}
\nonumber
\\
&\xra{}
&
\mathcal{F}_t^{-1}\big[-
a\,\rho\,\omega^2\.\widehat u + b \, \rho\, \omega^4 \, \widehat u
\,
-
\,
\rho \, c^2\,\omega^4\.\widehat u\big]
&
=
\mathcal{F}_t^{-1}\big[
(a-b\,\omega^2)\text{Div} \left[ \mathbb{C} \, \text{sym}\nabla \widehat u\right]
\big]
\, ,
&
&
\hspace{-7mm}
\forall\,\omega\in\R\,\setminus\left\{\pm\sqrt{\frac{a}{b}}\right\}
\nonumber
\\*[5pt]
&\Longleftrightarrow
&
a \,\rho \, \ddot{u}
+b \,\rho \, \ddot{\ddot{u}}
-c^2 \rho \, \ddot{\ddot{u}}
&
=
a \, \text{Div} \left[ \mathbb{C} \, \text{sym}\nabla u\right]
+ b \, \text{Div} \left[ \mathbb{C} \, \text{sym}\nabla \ddot{u}\right]
\, .
\label{eq:fourth_order_model_fourier}
\end{align}
Later in this discussion, we will show explicitly that higher-order models (higher than 2) involving highest derivatives with respect to time can exhibit undesirable behaviors, such as failing to conserve the total energy of the system (even if they involve only even-order derivatives). 
One potential solution to this issue involves the introduction of supplementary kinematical fields as we will show in the paragraph \ref{enriched_model_first}. 
%
%
%
%
\paragraph{Existence of an action functional and positive-definiteness}
The action functional associated with the equilibrium equation (\ref{eq:fourth_order_model_fourier}) is
\begin{equation}
\mathcal{A}
=
\iint\displaylimits_{\Omega\times [0,T]}
\underbrace{
\frac{1}{2}
\left( a\,\rho \, \langle \dot{u} , \dot{u} \rangle
+
(c^2-b) \, \rho \, \langle \ddot{u} , \ddot{u} \rangle
+
b 
\, \langle \mathbb{C} \, \text{sym}\nabla \dot{u}, \text{sym}\nabla \dot{u} \rangle
\right)
}_{\text{K - kinetic energy density}}
-
\underbrace{
\frac{1}{2}
\,
a 
\, \langle \mathbb{C} \, \text{sym}\nabla u, \text{sym}\nabla u \rangle
}_{\text{W - strain energy density}}
\mathrm{d}x\,,
\label{eq:energy_negative_rho_freq_depe_direct}
\end{equation}
where for positive definiteness it is required that
\begin{align}
a>0 \, ,
\qquad
\text{eig}(\mathbb{C})>0 \, ,
\qquad
\rho >0 \, ,
\qquad
b \geq0 \, ,
\qquad
c^2 > b
\, ,
\end{align}
where eig$(\mathbb{C})>0$ means that the eigenvalues of $\mathbb{C}$ are required to be greater than zero.
We underline that if the positive-definiteness condition $c^2 > b$ is respected, \coranR{the band-gap loses its upper bound, preventing the possibilities of having optic branches,} so the model written in the time domain retains the same limits in terms of positive-definiteness as it was the case for the frequency-dependent model.
We also emphasize that the action functional \eqref{eq:energy_negative_rho_freq_depe_direct} allows for a true time-dependent variable $u(x,t)$ that abandons the frequency domain where the original frequency-dependent equilibrium equation \eqref{eq:poly_rho_freq_depe_negative} is defined.
The associated (Neumann) boundary conditions are
\begin{align}
(a\,\mathbb{C} \, \text{sym}\nabla u
+
b\,\mathbb{C} \, \text{sym}\nabla \ddot{u})n=0 \, .
\label{eq:bound_cond_rho_freq_depe_final_negative_direct}
\end{align}
%
%
%
%
\paragraph{Energy conservation}
To ensure that the enriched model is conservative, we have to guarantee that 
\begin{equation}
\frac{\mathrm d}{\mathrm dt}
\int\displaylimits_{\Omega}
E(\dot u,\ddot u,\nabla\dot u,\nabla u) \, \mathrm{d}x
=
\int\displaylimits_{\Omega
}
\frac{\mathrm d}{\mathrm dt}\left[K(\dot u,\ddot u,\nabla\dot u)+W(\nabla u)\right] \mathrm{d}x 
=
0
\,,
\label{eq:conser_ene_rho_freq_depe_final_negative_direct}
\end{equation}
where $\Omega$ is the considered domain.
Substituting the expressions of $K$ and $W$ from eq.(\ref{eq:energy_negative_rho_freq_depe_direct}) into eq.(\ref{eq:conser_ene_rho_freq_depe_final_negative_direct}) we compute
\begin{align}
\int\displaylimits_{\Omega}
\frac{\mathrm dE}{\mathrm dt}\, \mathrm{d}x
=&
\int\displaylimits_{\Omega}
a\,\rho \,\langle \ddot{u} , \dot{u} \rangle
+
(c^2-b) \rho \,\langle \dot{\ddot{u}} , \ddot{u} \rangle
+
b\,\langle \mathbb{C} \, \text{sym}\nabla \ddot{u}, \text{sym}\nabla \dot{u} \rangle
+a\,\langle \mathbb{C} \, \text{sym}\nabla u, \text{sym}\nabla \dot{u} \rangle
\, \mathrm{d}x
\notag
\\
=&
\int\displaylimits_{\Omega}
(a\,\rho\, \langle \ddot{u} , \dot{u} \rangle
+
\frac{\mathrm{d}}{\mathrm{d}t}
((c^2-b) \rho \,\langle \dot{\ddot{u}} , \dot{u} \rangle )
-
(c^2-b) \rho\, \langle \ddot{\ddot{u}} , \dot{u} \rangle
+
\text{div} \, [ ( b\, \mathbb{C} \, \text{sym}\nabla \ddot{u} )^{\text{T}} \dot{u} ]
\label{eq:cons_energy_rho_freq_depe_negative_direct}
\\*
&
\phantom{\int\displaylimits_{\Omega}}
-
\langle \text{Div} \, [b \, \mathbb{C} \, \text{sym}\nabla \ddot{u} ], \dot{u} \rangle
+
\text{div} \, [ ( a\, \mathbb{C} \, \text{sym}\nabla u )^{\text{T}} \dot{u} ]
-
\langle \text{Div} \, [a\,\mathbb{C} \, \text{sym}\nabla u ], \dot{u} \rangle
\, \mathrm{d}x.
\notag
\end{align}
Finally, using the divergence theorem, we can write
\begin{align}
\int\displaylimits_{\Omega}
\frac{\mathrm dE}{\mathrm dt}\, \mathrm{d}x
&
=
\int\displaylimits_{\Omega}
\langle
a
\,\rho \, \ddot{u}
- a \, \text{Div} \,  [ \mathbb{C} \, \text{sym}\nabla u ]
- (c^2-b) \rho \, \ddot{\ddot{u}}
- b \, \text{Div} \, [ \mathbb{C} \, \text{sym}\nabla \ddot{u} ]
,
\dot{u}
\rangle\, \mathrm{d}x
\label{eq:cons_energy_rho_freq_depe_negative_direct_2}
\\*
&
\quad
+
\int\displaylimits_{\partial \Omega}
\langle
(a \, \mathbb{C} \, \text{sym}\nabla u
+ b \, \mathbb{C} \, \text{sym}\nabla \ddot{u}) n
,
\dot{u}
\rangle\, \mathrm{d}s
+
(c^2-b) \, \rho \, 
\frac{d}{dt}
\int\displaylimits_{\Omega}
\langle
\dot{\ddot{u}}
,
\dot{u}
\rangle
\, \mathrm{d}x
\, .
\notag
\end{align}
The first term in eq.(\ref{eq:cons_energy_rho_freq_depe_negative_direct_2}) vanishes because of the equilibrium equations (\ref{eq:fourth_order_model_fourier}), the second term vanishes because of the boundary conditions (\ref{eq:bound_cond_rho_freq_depe_final_negative_direct}),
while the last term $(c^2-b) \, \rho \, 
\frac{d}{dt}\langle\dot{\ddot{u}}
,
\dot{u}\rangle_{L^2(\Omega)}$
will in general be non-zero, causing that the model does not conserve energy.

%
%
%
\paragraph{Infinitesimal Galilean invariance}
As a last check, it is necessary to assess whether the model respects infinitesimal Galilean invariance, which requires the invariance of the equilibrium equations eq.(\ref{eq:fourth_order_model_fourier}) with respect to the following extended infinitesimal Galilean transformation
\begin{equation}
    \corrections{
    u\to\overline u=u+ A(t)\.x+ r(t), 
    \qquad \ddot A(t)= 0
    \quad
    \textrm{and}
    \quad
    \ddot r(t)=0\,,
    \quad\text{for all}
    \quad
    A\in C^2(\R,\mathfrak{so}(3))\,,\; r\in C^2(\R,\R^3)\,.
    }
    \label{eq:ap_Galilean_2}
\end{equation}
We 
now substitute $u$ with $\overline{u}$ from eq.(\ref{eq:ap_Galilean_2}) in eq.(\ref{eq:fourth_order_model_fourier})
\begin{align}
&
a \,\rho \, \ddot{\overline u}
- (c^2-b) \rho \, \ddot{\ddot{\overline u}}
- b \, \text{Div} \left[ \mathbb{C} \, \text{sym}\nabla \ddot{\overline u}\right]
=
a \, \text{Div} \left[ \mathbb{C} \, \text{sym}\nabla \overline u\right]
\, ,
\label{eq:equi_equa_gali_inva_2a_2}
\\*
\Rightarrow
\qquad
&
a \,\rho \, \frac{\text{d}^2}{\text{d}t^2}(u+ A(t)\.x+ r(t))
- (c^2-b) \rho \, \frac{\text{d}^4}{\text{d}t^4}(u+ A(t)\.x+ r(t))
\label{eq:equi_equa_gali_inva_2b_2}
\\*
&
\qquad\qquad\qquad
- b \, \text{Div} \left[ \mathbb{C} \, \text{sym}\nabla \frac{\text{d}^2}{\text{d}t^2}(u+A(t)\.x+ r(t))\right]
=
a \, \text{Div} \left[ \mathbb{C} \, \text{sym}\nabla (u+ A(t)\.x+ r(t))\right]
\notag
\\*
\Rightarrow
\qquad
&
a \,\rho \, \ddot{u}
- (c^2-b) \rho \, \ddot{\ddot{u}}
- b \, \text{Div} \left[ \mathbb{C} \, \text{sym}\nabla \ddot{u}\right]
=
a \, \text{Div} \left[ \mathbb{C} \, \text{sym}\nabla u\right]
\, ,
\label{eq:equi_equa_gali_inva_2c_2}
\end{align}
where we also observe that $\text{sym}\nabla (A(t)\.x)=\text{sym}\,A(t)=0$.
As can be seen by comparing  eq.(\ref{eq:fourth_order_model_fourier}) and eq.(\ref{eq:equi_equa_gali_inva_2c_2}), it is possible to see that \cor{they} exactly match, making \cor{them} invariant with respect to extended infinitesimal Galilean transformations (for further details see Appendix \ref{sec:appGalilean_inf}).

\begin{tcolorbox}[colback=BlueGreen!15!white,colframe=BlueGreen!50!white,coltitle=black,title={\centering \textbf{Summary:} \coran{Direct} time domain counterpart of the $\rho(\omega)$ model}]
\textbf{Original frequency-dependent model (frequency domain):}
\\[1em]
$
-\,\widetilde{\rho}(\omega)\,\omega^2 \, \widehat u
=
\text{Div} \left[ \mathbb{C} \, \text{sym}\nabla\widehat u\right]
$
\qquad
with
\qquad
$
\widetilde{\rho}(\omega)
=
\rho\left( 1+\dfrac{c^2 \, \omega^2}{a - b \, \omega^2
}
\right),
\qquad
(a>0,b\geq0,c^2>b)
$
\\[1em]
\textbf{Time domain counterpart of the $\rho(\omega)$ model:}
\\[1em]
The action functional associated to the time domain model obtained from the original frequency-dependent model through the inverse time-Fourier transform is
\\[1em]
$
\mathcal{A}
=
\displaystyle\iint\displaylimits_{\Omega\times [0,T]}
\underbrace{
\frac{1}{2}
\left( a\,\rho \, \langle \dot{u} , \dot{u} \rangle
+
(c^2-b) \, \rho \,  \langle \ddot{u} , \ddot{u} \rangle
+
b \, \langle \mathbb{C} \, \text{sym}\nabla \dot{u}, \text{sym}\nabla \dot{u} \rangle
\right)
}_{\text{K - kinetic energy density}}
-
\underbrace{
\frac{1}{2}
a \, \langle \mathbb{C} \, \text{sym}\nabla u, \text{sym}\nabla u \rangle
}_{\text{W - strain energy density}}
\mathrm{d}x\.\mathrm{d}t\,.
$
\\[1em]
The equilibrium equations in $\Omega$ are:
$
\quad
a \,\rho \, \ddot{u}
- a \, \text{Div} \left[ \mathbb{C} \, \text{sym}\nabla u\right]
- (c^2-b) \rho \, \ddot{\ddot{u}}
- b \, \text{Div} \left[ \mathbb{C} \, \text{sym}\nabla \ddot{u}\right]
=0
\, ,
$
\\[1em]
and the Neumann boundary conditions on $\partial\Omega\times[0,T]$ are:
$
\quad
(a\,\mathbb{C} \, \text{sym}\nabla u
+
b\,\mathbb{C} \, \text{sym}\nabla \ddot{u}) \, n=0 \, .
$
\\[1em]
\textbf{Consistency checks of the model in the time domain:}
\\[1em]
\text{positive-definiteness} \!\!\!\quad\textcolor{YellowOrange}{\faWarning}\quad (yes, but does not allow for band-gaps) \hspace{10mm}
\text{energy conservation}\quad\textcolor{Red}{\xmark}
%
\text{infinitesimal Galilean invariance \eqref{eq:ap_Galilean}}\quad\textcolor{Green}{\cmark} \hspace{5mm}
\text{extended infinitesimal Galilean invariance \eqref{eq:ap_GalileanExtended}}\quad\textcolor{Green}{\cmark}
\\[1em]
(Note that considering $b=0$ is permitted but does not restore energy conservation.)
\end{tcolorbox}
\begin{remark}
    The problem with energy conservation is related to the appearance of $\ddot u$ in the action functional. The last two authors vividly remember a comment of the late Gérard Maugin at a conference in \cort{Cisterna di Latina (Italy)} in 2014 pointing into the same direction.
\end{remark}
%
%
%
\subsubsection{Second attempt: an associated enriched model, its formulation and positive-definiteness conditions}
\label{enriched_model_first}

\corrections{Let us consider again the frequency-dependent Cauchy model from eq.(\ref{eq:poly_rho_freq_depe_negative}):}
\begin{equation}
-
\rho\left(
1 + \frac{c^2 \, \omega^2}{a - b \, \omega^2}
\right) \.\omega^2\.\widehat u
=
\text{Div} \left[ \mathbb{C} \, \text{sym}\nabla \widehat u\right]
\qquad\quad
\text{with}
\qquad\quad
\omega\neq \pm\sqrt{\frac{a}{b}}
\, .
\label{eq:equi_equa_rho_freq_depe_negative_1}
\end{equation}
Our objective is to present a procedure that enables the construction of a frequency-independent enriched model. This model yields the same $\omega-$parameterized family of differential equations in the frequency domain as in eq.\eqref{eq:equi_equa_rho_freq_depe_negative_1},  simultaneously resolving concerns pertaining to energy conservation. To illustrate, in the context of the aforementioned problem, we will incorporate an extra kinematical field
\cort{$\widehat v:\R^3_x\times(\R_\omega\setminus\{\pm\,\sqrt{\nicefrac{a}{b}}\})\subset\R^3_x\times\R_\omega\rightarrow\R^{3}$} in the frequency domain as
\begin{equation}\label{Fourier_change}
\widehat v(x,\omega)
\coloneqq
\displaystyle
\frac{c \, \omega^2}{a - b \, \omega^2} \, \widehat{u}(x,\omega). 
\end{equation}
By substituting eq.\eqref{Fourier_change} into eq.\eqref{eq:Fourier_freq_main} we arrive at the family of systems parameterized by $\omega$
\cort{
\begin{align}
\left\{ \begin{aligned}
-
\rho \left(
\.\omega^2\.\widehat u + c \; \omega^2 \,\widehat v \right)
&
=
\text{Div} \left[ \mathbb{C} \, \text{sym}\nabla \widehat u\right]
\, ,
\\[5pt]
\widehat v
&
\coloneqq
\displaystyle
\frac{c \, \omega^2}{a - b \, \omega^2} \, \widehat{u} \, ,
\end{aligned}
\right.
\hspace{2cm}
\forall\omega\in\R\,\setminus\left\{\pm\sqrt{\frac{a}{b}}\right\}
\label{Fourier_change2}
\end{align}
}
giving 
\begin{align}
\left\{ \begin{aligned}
-
\rho \left(
\.\omega^2\.\widehat u + c \; \omega^2 \,\widehat v \right)
&
=
\text{Div} \left[ \mathbb{C} \, \text{sym}\nabla \widehat u\right]
\, ,
\\[5pt]
a \, \widehat v - b \, \omega^2 \, \widehat v
&
=
c \, \omega^2 \, \widehat{u} \, ,
\end{aligned}
\right.
\hspace{2cm}
\forall\omega\in\R\,\setminus\left\{\pm\sqrt{\frac{a}{b}}\right\}
\label{Fourier_change3}
\end{align}
and applying the inverse time-Fourier transform $\F_t^{-1}$ we finally obtain the coupled system (same remark as in footnote \ref{footnote omega})
\begin{equation}
\left\{ 
\begin{aligned}
\rho \left(\ddot{u} + c \, \ddot{v}\right) \, 
&
=
\text{Div} \left[ \mathbb{C} \, \text{sym}\nabla u\right]
\, ,
\\
c \, \ddot{u}
+ b \, \ddot{v}
+ a \, v 
&
=
0,
\,
\end{aligned}
\right.
\label{Fourier_change4}
\end{equation}
where $v$ has the dimension of a displacement. This is a second order system of PDEs.  
We have thus replaced the frequency-dependent Cauchy problem in the frequency domain (eq.(\ref{eq:equi_equa_rho_freq_depe_negative_1})) with an extended continuum model in the time domain (eq.(\ref{Fourier_change4})), in which all material parameters are constants that do not depend on frequency. 
In particular, just the even powers of $\omega$ are allowed in order  to avoid imaginary contributions applying the inverse Fourier transform.
Other constraints on the admissible expressions for \angcor{$\widetilde \rho(\omega)$} are given by energy conservation arguments that will be discussed later (see Section~\ref{sec:freq_depe_rho_cons_ene}).
%
%
%
\paragraph{Existence of an action functional and positive-definiteness}
The action functional associated with the equilibrium equations (\ref{Fourier_change4}) is
\begin{equation}
\mathcal{A}
=
\iint\displaylimits_{\Omega\times [0,T]}
\underbrace{
\frac{1}{2}
\rho
\left( \langle \dot{u} , \dot{u} \rangle
+
2 \, c \, \langle \dot{u} , \dot{v} \rangle
+
b\, \langle \dot{v},\dot{v} \rangle
\right)
}_{\text{K - kinetic energy density}}
-
\underbrace{
\frac{1}{2}
\left(
\langle \mathbb{C} \, \text{sym}\nabla u, \text{sym}\nabla u \rangle
+
\rho \, a \, \langle v , v \rangle
\right)
}_{\text{W - strain energy density}}
\mathrm{d}x\.\mathrm{d}t\,,
\label{eq:energy_negative_rho_freq_depe}
\end{equation}
where for positive definiteness it is required that
\begin{align}
a>0 \, ,
\qquad
\text{eig}(\mathbb{C})>0 \, ,
\qquad
\rho >0 \, ,
\qquad
b >0 \, ,
\qquad
c^2 < b
\, ,
\end{align}
where eig$(\mathbb{C})>0$ means that the eigenvalues of $\mathbb{C}$ are required to be greater than zero.
Positive definiteness conditions are of primary importance, and their validity should always be guaranteed when choosing numerical values for the material parameters.
\angcor{Indeed, when considering elastic problems, the existence of an action functional behind the observed phenomenon together with the requirement of its positive-definiteness guarantees that no creation of energy can occur, thus ensuring the respect of the second principle of thermodynamics. This implies, in other words, that so-called « passivity » in the parlance of  \cite{srivastava2015causality} is automatically satisfied.}

However, these crucial conditions of positive-definiteness are often disregarded when dealing with frequency-dependent models.

The associated homogeneous Neumann boundary conditions on $\partial \Omega$ are now
\begin{align}
\sigma \, n = 0
\, ,
\qquad\qquad
\text{with}
\qquad\qquad
\sigma=\mathbb{C} \, \text{sym}\nabla u
\, ,
\label{eq:bound_cond_rho_freq_depe_final_negative}
\end{align}
where $n$ is the normal to the boundary $\partial \Omega$.
We highlight that these boundary conditions together with the PDEs (\ref{Fourier_change4}) can be systematically derived by requiring the minimization of the action $\mathcal{A}$ in eq.(\ref{eq:energy_negative_rho_freq_depe}).

Since the frequency-dependent model contains the frequency $\omega$ as a parameter in the PDEs (\ref{eq:equi_equa_rho_freq_depe_negative_1}), positive-definiteness must be checked for each frequency value.
In particular, we can say that the Cauchy frequency-dependent model is \cors{positive-definite} for $\omega=\omega_0$ if 
\begin{equation}
\widetilde{\rho}(\omega_0)>0
\qquad
\text{and}
\qquad
\text{eig}(\mathbb{C})>0
\, .
\label{eq:well_pose_rho_freq_depe_negative_1}
\end{equation}
We remark that the condition (\ref{eq:well_pose_rho_freq_depe_negative_1})$_1$ is violated by the originary frequency-dependent model when a bang-gap region occurs, starting from a local resonance frequency.
%
%
%
%
\paragraph{Energy conservation}
\label{sec:freq_depe_rho_cons_ene}
Once the enriched continuum (\ref{Fourier_change4}) in time domain corresponding to the
given frequency-dependent model (\ref{eq:equi_equa_rho_freq_depe_negative_1}) is established, energy conservation must be checked to finally validate the choice of the expression of $\widetilde{\rho}(\omega)$.
Indeed, if a given expression of $\widetilde{\rho}(\omega)$ gives rise to an enriched model whose energy is not conserved, this implies that the chosen $\widetilde{\rho}(\omega)$ is not physically acceptable.
To ensure that the enriched model is conservative, we have to guarantee that 
\begin{equation}
\frac{\mathrm d}{\mathrm dt}\int\displaylimits_{\Omega}E(\dot u,\dot v,\nabla u, v) \, \mathrm{d}x=
\int\displaylimits_{\Omega}
\frac{\mathrm d}{\mathrm dt}\left[K(\dot u,\dot v)+W(\nabla u,v)\right] \mathrm{d}x =
0
\,,
\label{eq:conser_ene_rho_freq_depe_final_negative}
\end{equation}
where $\Omega$ is the considered domain.
Substituting the expressions of $K$ and $W$ from eq.(\ref{eq:energy_negative_rho_freq_depe}) into eq.(\ref{eq:conser_ene_rho_freq_depe_final_negative}) we compute
\begin{align}
\int\displaylimits_{\Omega}
\frac{\mathrm dE}{\mathrm dt}\, \mathrm{d}x
=&
\int\displaylimits_{\Omega}
\rho \, 
\langle
\ddot{u},\dot{u}
\rangle
+
c \,
\rho \,
\langle
\ddot{u},\dot{v}
\rangle
+
c \,
\rho \,
\langle
\dot{u},\ddot{v}
\rangle
+
\rho \, b \, 
\langle \ddot{v} , \dot{v} \rangle
+
\langle
\sigma,
\text{sym} \, \nabla \dot{u}
\rangle
+
\rho \, a \,
\langle
v,
\dot{v}
\rangle\, \mathrm{d}x
\notag
\\
=&
\int\displaylimits_{\Omega}
\rho \,
\langle
\ddot{u} + c \,\ddot{v},\dot{u}
\rangle
+
\rho \,
\langle
c \,\ddot{u} + b \, \dot{v} + a \, v,\dot{v}
\rangle
+
\text{div}(\sigma^{\rm T}\.\dot u)
-
\langle
\text{Div}\. \sigma,
\dot{u}
\rangle\, \mathrm{d}x
\label{eq:cons_energy_rho_freq_depe_negative}
\\*
=&
\int\displaylimits_{\Omega}
\langle
\rho \left( \ddot{u} + c \, \ddot{v}\right) -\text{Div} \, \sigma 
,
\dot{u}
\rangle
+
\text{div}\left( \sigma^{\rm T} \, \dot{u}\right)
+
\langle
\rho \left( a \, v + c \, \ddot{u}  + b \, \ddot{v} \right)
,
\dot{v}
\rangle\, \mathrm{d}x
=
0
\, .
\notag
\end{align}
Thanks to the equilibrium equations (\ref{Fourier_change4}), the energy rate (\ref{eq:cons_energy_rho_freq_depe_negative}) becomes
\begin{align}
\frac{\mathrm d}{\mathrm dt}\int\displaylimits_{\Omega}
E\, \mathrm{d}x
=
\int\displaylimits_{\Omega}
\text{div}\left( \sigma^{\rm T} \, \dot{u}\right)\, \mathrm{d}x
=
\int\displaylimits_{\partial \Omega}
\langle
\left(\sigma \, n \right)
,
\dot{u} \rangle \, \mathrm{d}s
=
0 \, ,
\label{eq:cons_energy_rho_freq_depe_final_negative}
\end{align}
which is automatically always satisfied thanks to the homogeneous boundary conditions required in eq.(\ref{eq:bound_cond_rho_freq_depe_final_negative}).

It is clear that if a different expression for $\rho(\omega)$ was chosen in eq.(\ref{eq:equi_equa_rho_freq_depe_negative_1}), that would give rise to a different enriched model (\ref{eq:energy_negative_rho_freq_depe}) which could exhibit a non-trivial condition in order to satisfy the energy conservation requirement.
This implies that the chosen form of $\rho(\omega)$ must be 
selected carefully.

%
%
%
\paragraph{Infinitesimal Galilean invariance}
\label{sec:galilean_invariance}
As a last check, it is necessary to assess whether the model respects infinitesimal Galilean invariance, which requires the invariance of the equilibrium equations eq.(\ref{Fourier_change4}) with respect to the following extended infinitesimal Galilean transformation (cf.\ Appendix \ref{sec:appGalilean})
\begin{align}
u\to\overline u=u+A(t)\.x+r(t)\,,
\qquad 
v\to\overline v&=v+A(t)\.x+r(t)\,,
\qquad 
\ddot{A}(t)=0 \,,
\qquad 
\ddot{r}(t)=0\,,\label{eq:transformation_u_v}
\end{align}
where $A(t)\in\mathfrak{so}(3)$ is a skew-symmetric matrix while $r(t)\in\R^3$ is a vector.
For the sake of clarity, we report below again the equilibrium equations (\ref{Fourier_change4})
\begin{align}
\begin{cases}
\rho \left(\ddot{u} + c \, \ddot{v}\right) \, 
-
\text{Div} \left[ \mathbb{C} \, \text{sym}\nabla u\right]
=0
\, ,
\\[5pt]
c \, \ddot{u}
+ b \, \ddot{v}
+ a \, v 
=
0
\,.
\end{cases}
\label{eq:equi_equa_gali_inva}
\end{align}
We now substitute $u$ and $v$ with $\overline{u}$ and $\overline{v}$ from eq.(\ref{eq:transformation_u_v}), respectively, in eq.(\ref{eq:equi_equa_gali_inva})
\begin{align}
&\begin{cases}
\rho \left(\ddot{\overline{u}} + c \, \ddot{\overline{v}}\right) \, 
-
\text{Div} \left[ \mathbb{C} \, \text{sym}\nabla \overline{u}\right]
=
0
\, ,
\\[5pt]
c \, \ddot{\overline{u}}
+ b \, \ddot{\overline{v}}
+ a \, \overline{v}
=
0
\,,
\end{cases}
\nonumber
\\*
\Rightarrow
&
\begin{cases}
\rho \left(
\ddot{u} 
+ \frac{\text{d}^2}{\text{d}t^2}\left(A(t)\.x\right)
+ \frac{\text{d}^2}{\text{d}t^2}\.r(t)
+ c 
\left(
\ddot{v}
+ \frac{\text{d}^2}{\text{d}t^2}\left(A(t)\.x\right)
+ \frac{\text{d}^2}{\text{d}t^2}\.r(t)
\right)
\right)
\\[5pt]
\hspace{8.25cm}
-\text{Div} \left[ \mathbb{C} \, \text{sym}\nabla 
\left(
u
+ A(t)\.x
+ r(t)
\right)
\right]
=
0
\, ,
\\[5pt]
c \,
\left(
\ddot{u} 
+ \frac{\text{d}^2}{\text{d}t^2}\left(A(t)\.x\right)
+ \frac{\text{d}^2}{\text{d}t^2}\,r(t)
\right)
+ b \,
\left(
\ddot{v}
+ \frac{\text{d}^2}{\text{d}t^2}\left(A(t)\.x\right)
+ \frac{\text{d}^2}{\text{d}t^2}\.\ddot{r}(t)
\right)
+ a \,
\left(
v
+ A(t)\.x
+ \ddot{r}(t)
\right)
=
0
\,.
\end{cases}
\nonumber
\\*[5pt]
\Rightarrow
&
\begin{cases}
\rho \left(\ddot{u} + c \, \ddot{v}\right) \, 
-
\text{Div} \left[ \mathbb{C} \, \text{sym}\nabla u\right]
=0
\, ,
\\[5pt]
c \, \ddot{u}
+ b \, \ddot{v}
+ a \,
\left(
v
+ A(t)\.x
+ r(t)
\right)
=
0
\,.
\end{cases}
\label{eq:equi_equa_gali_inva_2c}
\end{align}
Comparing eq.(\ref{eq:equi_equa_gali_inva}) and eq.(\ref{eq:equi_equa_gali_inva_2c}), it is possible to see that the second equation has an extra term $A(t)\.x+r(t)$, such that the enriched model \textbf{is not invariant} with respect to extended infinitesimal Galilean transformations. Also the simpler infinitesimal Galilean invariance \eqref{eq:ap_Galilean} is not satisfied (see Appendix \ref{sec:appGalilean}).

\begin{figure}[!ht]
\centering
\includegraphics[width=0.49\textwidth]{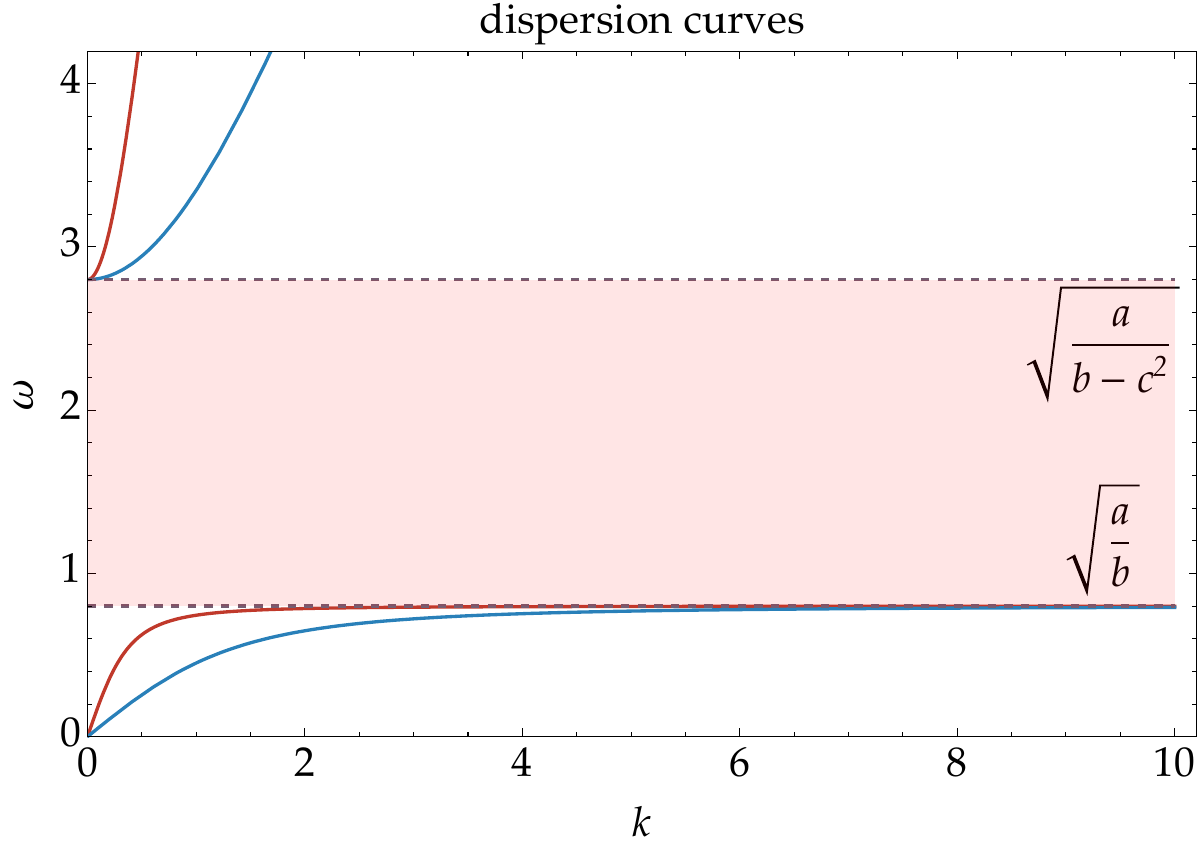}
\caption{
Dispersion curves for the isotropic class of symmetry in which the following values for the parameters have been used: $\rho=900$ kg/m$^3$, $\lambda=2898$ Pa, $\mu=262$ Pa, $a=0.697\;\si{1\per s\squared}$, $b=1.089$, and $c=1$.
The curves for the frequency-dependent model (Fig.\ref{fig:disp_curves_freq_depe_negative_Cau}) and the corresponding enriched model coincide.
However, while the enriched model remains \cors{positive-definite} in the band-gap region, the frequency-dependent one does not since $\widetilde \rho(\omega)<0$.
}
\label{fig:disp_curves_freq_depe_negative}
\end{figure}

To avoid this problem, in Section \ref{sec:CC_freq} we will show how the frequency-dependent density Cauchy model~(\ref{eq:equi_equa_rho_freq_depe_negative_1}) can be modified simply by moving the function of the frequency from the left to the right side of the equation as
\begin{align}\label{orig}
-
\,
\rho \left( 1+\frac{c^2 \, \omega^2}{a - b \, \omega^2}\right)  \, \omega^2 \, \widehat{u}
=
\text{Div} \left[ \mathbb{C}\, \text{sym}\nabla \widehat u\right]
\Longleftrightarrow
-
\,
\rho \, \omega^2 \, \widehat{u}
=
\text{Div} \left[ \left( \widetilde{f}-\frac{\widetilde{c}^2}{\widetilde{a} - \widetilde{b} \, \omega^2}\right) \mathbb{C}
\, \text{sym}\nabla \widehat u\right]
\, ,
\end{align}
where $\widetilde{f}=1+\frac{\widetilde{c}^2}{\widetilde{a}}$.
The correspondence between the two formulations is then given by
\begin{align}
\left(1+\frac{\widetilde{c}^2}{\widetilde{a}}-\frac{\widetilde{c}^2}{\widetilde{a} - \widetilde{b} \, \omega^2}\right)^{-1}
=
\frac{\widetilde{a}^2-\widetilde{a}\,\widetilde{b}\,\omega^2}{\widetilde{a}^2-\widetilde{a}\,\widetilde{b}\,\omega^2-\widetilde{b}\, \widetilde{c}^2\,\omega^2}
=
1+\frac{\widetilde{b} \, \widetilde{c}^2\,\omega^2}{\widetilde{a}^2-\widetilde{b}(\widetilde{a}+\widetilde{c}^2)\,\omega^2}
=
1+\frac{c^2\,\omega^2}{a-b\,\omega^2}
\, .
\label{eq:CC_vs_rho_parameters}
\end{align}
%
%
%
%
\paragraph{Procedure to obtain the dispersion relations for an enriched model}
\label{sec:disp_relations_enriched}
Here, we briefly show the procedure to obtain the dispersion relations for a 2D enriched model with more degrees of freedom than just the displacement field.
As done previously, \coran{one way to proceed is} to apply the space-Fourier transform $\F_x$ to both equations
\begin{equation}
		-
		\rho \left(
		\.\omega^2 \, \widehat u \, (x,\omega)+ c \; \omega^2 \, \widehat v \, (x,\omega) \right)
		=
		\text{Div} \left[ \mathbb{C} \, \text{sym}\nabla \widehat u \, (x,\omega) \right]
		\, ,
		\hspace{2cm}
		\widehat v \, (x,\omega)
		=
		\dfrac{c \, \omega^2}{a - b \, \omega^2} \, \widehat{u} \, (x,\omega) \, ,
\end{equation}
This gives
\begin{equation}
		-
		\rho \left(
		\.\omega^2 \, \widehat{u} \, ( q , \omega ) + c \; \omega^2 \, \widehat v \, ( q , \omega ) \right)
		=
		- \left[ \mathbb{C} \, \text{sym}(\widehat{u} \, ( q , \omega ) \otimes q) \right]\,q
		\, ,
		\hspace{2cm}
		\widehat v \, ( q , \omega )
		=
		\displaystyle
		\frac{c \, \omega^2}{a - b \, \omega^2} \, \widehat{u} \, ( q , \omega ) \, ,
\end{equation}
and hence
\begin{equation}
		-
		\rho \, \omega^2 \, \left(
		1 + c \, \frac{c \, \omega^2}{a - b \, \omega^2} \,  \right) \widehat{u}
		=
		- \left[ \mathbb{C} \, \text{sym}(\widehat{u} \otimes q)\right]\,q
		\, ,
		\hspace{2cm}
		\widehat v
		=
		\dfrac{c \, \omega^2}{a - b \, \omega^2} \, \widehat{u} \, .
	\label{eq.dispersion_curves_mic_new}
\end{equation}
Now, the first equation of \eqref{eq.dispersion_curves_mic_new}$_1$ can be rewritten with the help of a linear map $\mathbb A(\omega,q,\rho,\mathbb{C},a,b,c):\R^3\to\R^3$
\begin{equation}
	 \mathbb A(\omega,q,\rho,\mathbb{C},a,b,c)
     \,
     \widehat{u}
	 \coloneqq
	 - \, \rho \, \omega^2 \,  \left(
	 1 + c \, \frac{c \, \omega^2}{a - b \, \omega^2} \,  \right) \,  \mathds{1} \, 
	 +
	 \left[ \mathbb{C} \, \text{sym}(\widehat{u} \otimes q)\right]\,q
\end{equation}
and the corresponding algebraic problem $\mathbb A(\omega,q,\rho,\mathbb{C},a,b,c)
     \,
     \widehat{u}=0$  admits non trivial solutions if and only if 
\begin{align}
	\text{det}\left[
	\mathbb A(\omega,q,\rho,\mathbb{C},a,b,c)
	\right]=0 \, .
	\label{eq:disp_relations_enriched}
\end{align}
The solutions of equations (\ref{eq:disp_relations_enriched}) can be evaluated in terms of $k(\omega)$ and they read (considering only the positive roots)
\begin{align}
	k_{\rm p}=\sqrt{\frac{\rho \, \omega ^2 \left(a+\omega ^2 \left(c^2-b\right)\right)}{(\lambda + 2 \mu)\left(a-b \, \omega ^2\right)}}\,,
	\qquad
	\qquad
	k_{\rm s}=\sqrt{\frac{\rho \, \omega ^2 \left(a+\omega ^2 \left(c^2-b\right)\right)}{\mu\left(a-b \, \omega ^2\right)}}\,.\label{eq:disp_relations_enriched_express}
\end{align}
The solution $k_{\rm p}$ is associated with the propagation of pressure waves, while $k_{\rm s}$ is associated with the propagation of shear waves.
By direct comparison of eq.(\ref{eq:disp_relations_enriched_express}) with eq.(\ref{eq:disp_relations_freq_depe}), it is possible to see that the dispersion curves of the enriched model coincide with those stemming from the original frequency-dependent model.
In Fig.~\ref{fig:disp_curves_freq_depe_negative} it is possible to see the plot for specific values of parameters.
Also in this case, the dispersion relations can be obtained formally from the frequency-dependent model introducing the space-plane-wave ansatz
\begin{equation}
	\widehat u (x,\omega)
	=
	\xi(\omega) \, e^{i\,\langle x,q \rangle}
	\qquad
	\qquad
	\qquad
	\text{and}
	\qquad
	\qquad
	\qquad
	\widehat v (x,\omega)
	=
	\zeta(\omega) \, e^{i\,\langle x,q \rangle}
\end{equation}
or directly from \eqref{Fourier_change4} by introducing the space-time-plane-wave ansatz
\begin{align}
	u(x,t)=\psi \, e^{i(\langle q,x\rangle-\omega \, t)} \, 
	\qquad
	\qquad
	\qquad
	\text{and}
	\qquad
	\qquad
	\qquad
	v(x,t)=\eta \, e^{i(\langle q,x\rangle-\omega \, t)}.
	\label{eq:time_harmo_ansatz_enriched}
\end{align}
%
%
%
%
\paragraph{Relations between the frequency-dependent model and the enriched equivalent model}
While a classical Cauchy model with frequency-independent parameters gives rise to two linear dispersion relations (see Fig.\ref{fig:classic_Cauchy_disp_curve}), enriched continuum models result in additional dispersion modes (see eq.(\ref{eq:disp_relations_enriched_express}) and Fig.~\ref{fig:disp_curves_freq_depe_negative}) \coran{while having all their parameters to be frequency independent}.
When letting the parameters be frequency-dependent, also a Cauchy model can exhibit dispersion and band-gaps (see eq.(\ref{eq:disp_relations_freq_depe}) and Fig.~\ref{fig:disp_curves_freq_depe_negative}).
We comment here about the fact that the dispersion curves obtained with the frequency-dependent model in eq.(\ref{eq:poly_rho_freq_depe_negative}) and the ones obtained from the enriched model in eq.(\ref{Fourier_change4}) coincide and that the enriched model is always positive-definite, while the frequency-dependent model loses positive-definiteness in the band-gap region.
From eq.(\ref{eq:poly_rho_freq_depe_negative}), it is possible to calculate the frequencies such that 
\begin{align}
\widetilde{\rho} \to \infty
\quad
\Longleftrightarrow
\quad
\omega=\pm\sqrt{\frac{a}{b}}
\, ,
\qquad\qquad\qquad
\widetilde{\rho}=0
\quad
\Longleftrightarrow
\quad
\omega=\pm\sqrt{\frac{a}{b-c^2}}
\, ,
\end{align}
When compared to the associated enriched model, these frequencies correspond to the cut-off frequencies of the optic curves and the asymptotes of the acoustic curves, respectively (see Fig.~\ref{fig:disp_curves_freq_depe_negative}).
In particular, the cut-off frequencies can be obtained from eq.(\ref{eq:rho_freq_depe_disp_rela_expre}) by setting $k_{\rm p},k_{\rm s}=0$, while the asymptotes ($k_{\rm p},k_{\rm s}\to \infty$) can be computed by setting the denominator of (\ref{eq:disp_relations_enriched_express}) to zero.

It can also be checked that, in the frequency-dependent model's band-gap range, the effective density $\widetilde{\rho}(\omega)$ is negative and this makes the speed of propagation (which is also frequency-dependent) imaginary
\begin{align}
k_{\rm p}
&
=
\omega\sqrt{\frac{\widetilde{\rho}(\omega)}{\lambda + 2 \mu}}
=
\omega\sqrt{\frac{-\lvert \widetilde{\rho}(\omega) \rvert}{\lambda + 2 \mu}}
=
i \, \omega \,\sqrt{\frac{ \lvert \widetilde{\rho}(\omega) \rvert}{\lambda + 2 \mu}}\, ,
&&
\omega \in \left(\sqrt{\frac{a}{b}},\sqrt{\frac{a}{b-c^2}} \, \right),
\\
k_{\rm s}&=\omega\sqrt{\frac{\widetilde{\rho}(\omega)}{\mu}}
=
\omega\sqrt{\frac{-\lvert \widetilde{\rho}(\omega) \rvert}{\mu}}
=
i \, \omega \,\sqrt{\frac{ \lvert \widetilde{\rho}(\omega) \rvert}{\mu}}\, ,
&&
\notag
\end{align}
where 
$k_{\rm p}$ and $k_{\rm s}$ represent the wavenumber for pressure and shear waves, respectively.
Given the negative value of $\widetilde{\rho}\,(\omega)$, the frequency-dependent model is not \cors{positive-definite} in the band-gap region.

In the same frequency interval, the wavenumber for the enriched model is also imaginary, but this time, because of the interpretation of the parameters $a,b$, and $c$ as material parameters, it retains the positive-definiteness
\begin{align}
k_{\rm p}
&= \,\sqrt{\frac{\rho \, \omega ^2 \left(a+\omega ^2 \left(c^2-b\right)\right)}{(\lambda + 2 \mu) \left(a-b \, \omega ^2\right)}}
=i \, \omega\sqrt{\rho\left\lvert\left( 1+\frac{c^2 \, \omega^2}{a - b \, \omega^2} \right)\right\rvert\frac{1}{\lambda + 2 \mu}}\, ,&&\omega \in \left(\sqrt{\frac{a}{b}},\sqrt{\frac{a}{b-c^2}} \, \right)\,,
\label{eq:rho_freq_depe_disp_rela_expre}
\\
k_{\rm s}
&
=
 \,\sqrt{\frac{\rho \, \omega ^2 \left(a+\omega ^2 \left(c^2-b\right)\right)}{\mu \left(a-b \, \omega ^2\right)}}
=
i \, \omega\sqrt{\rho\left\lvert\left( 1+\frac{c^2 \, \omega^2}{a - b \, \omega^2} \right)\right\rvert\frac{1}{\mu}}\, ,
\notag
\end{align}
and the imaginary wavenumber can be directly associated with the triggering of evanescent waves.
\begin{tcolorbox}[colback=Salmon!15!white,colframe=Salmon!50!white,coltitle=black,title={\centering \textbf{Summary:} micromorphic enriched model stemming from the $\rho(\omega)$ frequency-dependent model}]
\textbf{Original frequency-dependent model (frequency domain):}
\\[1em]
$
-
\,
\widetilde{\rho}(\omega)\,\omega^2\.\widehat u
=
\text{Div} \left[ \mathbb{C} \, \text{sym}\nabla\widehat u\right]
$
\qquad
with
\qquad
$
\widetilde{\rho}(\omega)
=
\rho\left( 1+\dfrac{c^2 \, \omega^2}{a - b \, \omega^2
}
\right),
\qquad
(a>0,b>0,c^2<b)
$
\textbf{Introduction of the new variable:}
$
\widehat v(x,\omega)
\coloneqq
\displaystyle
\frac{c \, \omega^2}{a - b \, \omega^2} \, \widehat{u}(x,\omega).
$
\\[1em]
\textbf{Enriched model (time domain):}
\\[1em]
The action functional associated to the time domain model obtained from the original frequency-dependent model through the inverse time-Fourier transform and the introduction of the additional kinematical field $v$ is
\\[1em]
$
\mathcal{A}
=
\displaystyle\iint\displaylimits_{\Omega\times [0,T]}
\underbrace{
\frac{1}{2}
\rho
\left( \langle \dot{u} , \dot{u} \rangle
+
2 \, c \, \langle \dot{u} , \dot{v} \rangle
+
b\, \langle \dot{v},\dot{v} \rangle
\right)
}_{\text{K - kinetic energy density}}
-
\underbrace{
\frac{1}{2}
\left(
\langle \mathbb{C} \, \text{sym}\nabla u, \text{sym}\nabla u \rangle
+
\rho \, a \, \langle v , v \rangle
\right)
}_{\text{W - strain energy density}}
\mathrm{d}x\.\mathrm{d}t\,.
$
\\[1em]
The equilibrium equations are:
$
\quad
\rho \left(\ddot{u} + c \, \ddot{v}\right) \, 
=
\text{Div}\,\sigma
\, ,
\qquad
c \, \ddot{u}
+ b \, \ddot{v}
+ a \, v 
=
0
\, ,
$
\\[1em]
and the Neumann boundary conditions on $\partial\Omega\times[0,T]$ are:
$
\quad
\sigma \, n = 0
\text{ with }
\,
\sigma=\mathbb{C} \, \text{sym}\nabla u
$.
\\[1em]
\textbf{Consistency checks of the enriched model:}
\\[1em]
\text{positive-definiteness}\quad\textcolor{Green}{\cmark}
\text{energy conservation}\quad\textcolor{Green}{\cmark}
\text{infinitesimal Galilean invariance \eqref{eq:ap_Galilean}}\quad\textcolor{Red}{\xmark}
\text{extended infinitesimal Galilean invariance \eqref{eq:ap_GalileanExtended}}\quad\textcolor{Red}{\xmark}
\end{tcolorbox}
%
%
%
%
\section{A Cauchy model with frequency-dependent stiffness tensor and associated enriched continuum}
\label{sec:CC_freq}
What has been done in Section~\ref{sec:rho_freq} with a frequency-dependent density model, can be repeated by considering a frequency-dependent elasticity tensor as a starting point.
Let us start considering the equilibrium equations \cor{in eq.(\ref{orig})
for a Cauchy model in which 
the elasticity tensor depends on the frequency $\omega$ as}
\begin{equation}
-
\rho\;\omega^2\.\widehat u
=
\text{Div} \left[ \widetilde{\mathbb{C}}(\omega) \, \text{sym}\nabla\widehat u\right]
\qquad
\text{where}
\qquad
\widetilde{\mathbb{C}}(\omega)
=
\left( \widetilde{f} - \frac{\widetilde{c}^2}{\widetilde{a} - \widetilde{b} \, \omega^2}\right) \mathbb{C}
\qquad
\text{and}
\qquad
\widetilde{f}=1+\frac{\widetilde{c}^2}{\widetilde{a}}
\, ,
\label{eq:poly_CC_freq_depe_negative}
\end{equation}
where $\widetilde{c}$ and $\widetilde{a}$ are dimensionless coefficients, $\widetilde{b}$ has the dimension of [$\si{s\squared}$].
Note that $\displaystyle\lim_{\omega \to 0}\widetilde{\mathbb{C}}(\omega)=\mathbb{C}$, which means that the stiffness tensor approaches the classical value in the long-wave limit.

We explicitly remark that this frequency-dependent stiffness model is equivalent to the frequency-dependent density model of Section \ref{sec:rho_freq} 
\corf{in the frequency domain} (see eqs.(\ref{eq:CC_vs_rho_parameters})).
However, we will show in this section that the enriched models stemming from the frequency-dependent elasticity tensor are Galilean invariant, while those stemming from the frequency-dependent mass density are not.

To make the expressions easier to read, we removed the $\sim$ from the coefficients $a,b$ and $c$ in the remainder of this section.
%
%
%
\subsection{Formulation of the enriched model and positive-definiteness conditions: form I}

Introducing an additional tensor field $\widehat Q:\cors{\R^3_x\times \big(\R_\omega\,\setminus\big\{\pm\sqrt{\nicefrac{a}{b}}\big\}\big)\subset\R^3_x\times\R_\omega}\rightarrow\R^{3\times3}$ in the frequency domain satisfying\footnote{We only need to define the symmetric part of $\widehat Q$.}
\begin{equation}
\mathbb{C} \, \text{sym}\, \widehat Q
=
-
\frac{c}{a - b \, \omega^2} \, \mathbb{C} \, \text{sym}\nabla \widehat u \, , 
\end{equation}
equation~(\ref{eq:poly_CC_freq_depe_negative}) can be rewritten as
\begin{align}
&
\hspace{-5mm}
\begin{cases}
\displaystyle 
-
\rho\;\omega^2\.\widehat u
=
\text{Div} \left[ \left( f-\frac{c^2}{a - b \, \omega^2}\right) \mathbb{C}
\, \text{sym}\nabla \widehat u\right]
\, ,
\\[20pt]
\displaystyle 
\mathbb{C} \, \text{sym}\, \widehat Q
=
-
\frac{c}{a - b \, \omega^2} \, \mathbb{C} \, \text{sym}\nabla \widehat u \, ,
\end{cases}
\hspace{-5mm}
\Longleftrightarrow
\begin{cases}
\displaystyle 
-
\rho\;\omega^2\.\widehat u
=
\text{Div} \left[ \left( f-\frac{c^2}{a - b \, \omega^2}\right) \mathbb{C}
\, \text{sym}\nabla \widehat u\right]
\, ,
\\[20pt]
\displaystyle 
(a - b \, \omega^2) \; \mathbb{C} \, \text{sym}\, \widehat Q
=
-
c \, \mathbb{C} \, \text{sym}\nabla \widehat u \, ,
\end{cases}
\\[3mm]
&
\hspace{4cm}
\xLeftrightarrow[\F^{-1}_t]{\F_t}
\begin{cases}
\rho \, \ddot{u} 
=
f \, \text{Div} \left[ \mathbb{C} \, \text{sym}\nabla u\right]
+ c \, \text{Div} \left[ \mathbb{C} \,\text{sym}\, Q\right]
\, ,
\\[15pt]
c \, \mathbb{C} \, \text{sym} \nabla u 
+ a \, \mathbb{C} \, \text{sym}\, Q
+ b \, \mathbb{C} \, \text{sym}\, \ddot{Q}
=
0 \, ,
\end{cases}
\label{eq:equi_equa_CC_freq_depe_negative_2}
\end{align}
where $Q$ has the dimension of $\nabla u$, i.e. is dimensionless.

\bigskip

We have thus replaced the frequency-dependent Cauchy problem in eq.(\ref{eq:poly_CC_freq_depe_negative}) with an extended continuum model in eq.(\ref{eq:equi_equa_CC_freq_depe_negative_2}), in which all the material parameters are constants that do not depend on frequency.
It is underlined again that just even powers of $\omega$ are allowed in the choice of the expression for $\mathbb{C}(\omega)$ in order not to have imaginary contributions in the energy.

%
%
%
\subsubsection{Existence of an action functional and positive-definiteness}
The action functional associated with the PDEs system (\ref{eq:equi_equa_CC_freq_depe_negative_2}) is
\begin{align}
\mathcal{A}
=
&
\iint\displaylimits_{\Omega\times [0,T]}
\underbrace{
\frac{1}{2}
\left(
\rho
\,
\langle \dot{u} , \dot{u} \rangle
+
b 
\, 
\langle \mathbb{C} \, \text{sym} \, \dot{Q} , \text{sym} \, \dot{Q} \rangle
\right)
}_{\text{K - kinetic energy density}}
\label{eq:energy_negative_CC_freq_depe}
\\*
&
\phantom{\iint\displaylimits_{\Omega\times [0,T]}}
-
\underbrace{
\frac{1}{2}
\left(
f \, \langle \mathbb{C} \, \text{sym}\nabla u, \text{sym}\nabla u \rangle
+
2 \, c \, \langle \mathbb{C} \, \text{sym}\nabla u, \text{sym} \, Q \rangle
+
a \, \langle \mathbb{C} \, \text{sym} \, Q, \text{sym} \, Q \rangle
\right)
}_{\text{W - strain energy density}}
\mathrm{d}x\.\mathrm{d}t\,,
\notag
\end{align}
where for positive definiteness it is required that (we remind that $f=1+\frac{c^2}{a}$)
\begin{align}
a>0 \, ,
\qquad
\text{eig}(\mathbb{C})>0 \, ,
\qquad
\rho >0 \, ,
\qquad
b >0
\, .
\end{align}
The associated homogeneous Neumann boundary conditions are
\begin{align}
\left(
f \, \sigma 
+ c \, \mathbb{C} \, \text{sym} \, Q
\right) \, n = 0 \, ,
\label{eq:bound_cond_CC_freq_depe_final_negative}
\end{align}
where $\sigma=\mathbb{C} \, \text{sym}\nabla u$ and $n$ is the normal to the boundary.

Given the fact that the frequency-dependent model (\ref{eq:poly_CC_freq_depe_negative}) contains the frequency $\omega$ as a parameter, positive-definiteness must be checked for all values of $\omega$.
In particular, we can say that the Cauchy frequency-dependent model is \cors{positive-definite} at $\omega_0$ if 
\begin{equation}
\rho>0
\qquad
\text{and}
\qquad
\text{eig}(\widetilde{\mathbb{C}}(\omega_0))>0
\, .
\end{equation}
%
%
%
\subsubsection{Energy conservation}
To ensure that the resulting model is conservative, we have to guarantee that 
\begin{equation}
\frac{\mathrm d}{\mathrm dt}\int\displaylimits_{\Omega}
E(\dot u,\dot Q,\nabla u,Q) \, \mathrm{d}x=
\int\displaylimits_{\Omega}
\frac{\mathrm d}{\mathrm dt}\left[K(\dot u,\dot v)+W(\nabla u,Q)\right] \mathrm{d}x=
0
\,,
\end{equation}
where $\Omega$ is the considered domain.
With $\sigma=\mathbb{C} \, \text{sym} \, \nabla u$, we compute
\begin{align}
\int\displaylimits_{\Omega}
\frac{\mathrm d E}{\mathrm dt}\, \mathrm{d}x
=&
\int\displaylimits_{\Omega}
\rho\,
\langle \ddot{u} , \dot{u} \rangle
+
b \, \langle \mathbb{C} \, \text{sym} \, \ddot{Q} , \text{sym} \, \dot{Q} \rangle
+
f \, \langle \sigma, \text{sym}\nabla \dot{u} \rangle
\notag
\\*[-10pt]
&
\phantom{\int\displaylimits_{\Omega}}
+
c \, \langle \mathbb{C} \, \text{sym}\nabla \dot{u}, \text{sym} \, Q \rangle
+
c \, \langle \mathbb{C} \, \text{sym} \, \nabla u, \text{sym} \, \dot{Q} \rangle
+
a \, \langle \mathbb{C} \, \text{sym} \, Q, \text{sym} \, \dot{Q} \rangle\, \mathrm{d}x
\notag
\\*[-10pt]
=&
\int\displaylimits_{\Omega}
\rho\,
\langle \ddot{u} , \dot{u} \rangle
+
\langle
b \, \mathbb{C} \, \text{sym}\, \ddot{Q}
+ c \, \mathbb{C} \, \text{sym} \nabla u 
+ a \, \mathbb{C} \, \text{sym}\, Q
,
\dot{Q}
\rangle
\label{eq:cons_energy_CC_freq_depe_negative}
\\
&
\phantom{\int\displaylimits_{\Omega}}
+
f \, \text{div}(\sigma^{\rm T}\.\dot u)
-
f \, \langle \text{Div}\. \sigma, \dot{u} \rangle
+
c \, \text{div}([\mathbb{C}\,\text{sym}Q]^{\rm T}\.\dot u)
-
c \, \langle \text{Div}[\mathbb{C}\,\text{sym}Q], \dot{u} \rangle
\, \mathrm{d}x\notag
\\*[-10pt]
=&
\int\displaylimits_{\Omega}
\langle
\rho \, \ddot{u} 
- f \, \text{Div} \, \sigma 
- c \, \text{Div} \left[ \mathbb{C} \, \text{sym} \, Q \right]
,
\dot{u}
\rangle
+
\text{div}\left( \sigma^{\rm T} \, \dot{u}\right)
+
\text{div}\left( c \, [\mathbb{C} \, \text{sym} \, Q]^{\rm T} \, \dot{u}\right)
\notag
\\*[-10pt]
&
\phantom{\int\displaylimits_{\Omega}}
+
\langle
c \, \mathbb{C} \, \text{sym} \nabla u 
+ a \, \mathbb{C} \, \text{sym}\, Q
+ b \, \mathbb{C} \, \text{sym}\, \ddot{Q}
,
\dot{Q}
\rangle\, \mathrm{d}x
=
0
\, .
\notag
\end{align}
Thanks to the equilibrium equations (\ref{eq:equi_equa_CC_freq_depe_negative_2}), the condition (\ref{eq:cons_energy_CC_freq_depe_negative}) becomes
\begin{align}
\frac{\mathrm d}{\mathrm dt}\int\displaylimits_{\Omega}
E\, \mathrm{d}x
&=
\int\displaylimits_{\Omega}
f \, \text{div}\left( \sigma^{\rm T} \, \dot{u}\right)
+
\text{div}\left( c \, [\mathbb{C} \, \text{sym} \, Q]^{\rm T} \, \dot{u}\right)\, \mathrm{d}x
=
\int\displaylimits_{\partial \Omega}
\langle
\left(
f \, \sigma 
+ c \, \mathbb{C} \, \text{sym} \, Q
\right)n,
\dot{u}
\rangle\, \mathrm{d}s
=
0 \, 
\label{eq:cons_energy_CC_freq_depe_final_negative}
\end{align}
which is automatically always satisfied thanks to the homogeneous boundary conditions required in eq.(\ref{eq:bound_cond_CC_freq_depe_final_negative}).

%
%
%
\subsubsection{Infinitesimal Galilean invariance}
\label{sec:galilean_invariance_1b}
As a last check, it is necessary to assess whether the model respects Galilean invariance, which requires the invariance of the equilibrium equations eq.(\ref{eq:equi_equa_CC_freq_depe_negative_2}) with respect to the following extended infinitesimal Galilean transformation (cf.\ Appendix \ref{sec:appGalilean})
\begin{align}
u\to\overline u=u+A(t)\.x+r(t)\,,\qquad Q\to\overline Q=Q+A(t)\,,\qquad \ddot{A}(t)=0,\qquad\ddot{r}(t)=0\,,\label{eq:transformation_u_v_1b}
\end{align}
where $A(t)\in\mathfrak{so}(3)$ is a skew-symmetric matrix while $r(t)\in\R^3$ is a vector.
For the sake of clarity, we report below the equilibrium equations (\ref{eq:equi_equa_CC_freq_depe_negative_2})
\begin{equation}
\begin{cases}
\rho \, \ddot{u} \, 
=
f \, \text{Div} \left[ \mathbb{C} \, \text{sym}\nabla u\right]
+ c \, \text{Div} \left[ \mathbb{C} \, \text{sym} \, Q\right]
\, ,
\\[5pt]
c \, \mathbb{C} \, \text{sym} \nabla u 
+ a \, \mathbb{C} \, \text{sym}\, Q
+ b \, \mathbb{C} \, \text{sym}\, \ddot{Q}
=
0
\,.
\end{cases}
\label{eq:equi_equa_CC_wave_depe_3_gali_inva}
\end{equation}
We now substitute $u$ and $Q$ with $\overline{u}$ and $\overline{Q}$ from eq.(\ref{eq:transformation_u_v_1b}), respectively, in eq.(\ref{eq:equi_equa_CC_wave_depe_3_gali_inva})
\begin{align}
&
\begin{cases}
\rho \, \ddot{\overline{u}} \, 
=
f \, \text{Div} \left[ \mathbb{C} \, \text{sym}\nabla \overline{u}\right]
+ c \, \text{Div} \left[ \mathbb{C} \, \text{sym} \, \overline{Q}\right]
\, ,
\\[5pt]
c \, \mathbb{C} \, \text{sym} \nabla \overline{u} 
+ a \, \mathbb{C} \, \text{sym}\, \overline{Q}
+ b \, \mathbb{C} \, \text{sym}\, \ddot{\overline{Q}}
=
0
\,.
\end{cases}
\nonumber
\\*
\Rightarrow
&
\begin{cases}
\rho \, \left(
\ddot{u} 
+ \frac{\text{d}^2}{\text{d}t^2}\left(A(t)\.x\right)
+ \frac{\text{d}^2}{\text{d}t^2}\.r(t)
\right) \, 
=
\\
\hspace{2cm}
f \, \text{Div} \left[ \mathbb{C} \, \text{sym}\nabla \left(
u 
+ A(t)\.x + r(t)
\right)\right]
+ c \, \text{Div} \left[ \mathbb{C} \, \text{sym} \, \left(Q+A(t)\right)\right]
\, ,
\\[5pt]
c \, \mathbb{C} \, \text{sym} \nabla \left(
u 
+ A(t)\.x + r(t)
\right) 
+ a \, \mathbb{C} \, \text{sym}\, \left(Q+A(t)\right)
+ b \, \mathbb{C} \, \text{sym}\, \left(\ddot{Q}+\frac{\text{d}}{dt}\.A(t)\right)
=
0
\,.
\end{cases}
\nonumber
\\[5pt]
\Rightarrow
&
\begin{cases}
\rho \, \ddot{u} \, 
=
f \, \text{Div} \left[ \mathbb{C} \, \text{sym}\nabla u\right]
+ c \, \text{Div} \left[ \mathbb{C} \, \text{sym} \, Q\right]
\, ,
\\[5pt]
c \, \mathbb{C} \, \text{sym} \nabla u 
+ a \, \mathbb{C} \, \text{sym}\, Q
+ b \, \mathbb{C} \, \text{sym}\, \ddot{Q}
=
0
\,.
\end{cases}
\label{eq:equi_equa_gali_inva_2c_4}
\end{align}
Comparing eq.(\ref{eq:equi_equa_CC_wave_depe_3_gali_inva}) and eq.(\ref{eq:equi_equa_gali_inva_2c_4}), it is possible to see that they exactly match, making them invariant with respect to extended infinitesimal Galilean transformations.
%
%
%
\subsection{Relations between the frequency-dependent model and the equivalent enriched model: form I}
The dispersion curves obtained with the frequency-dependent model in eq.(\ref{eq:poly_CC_freq_depe_negative}) are also obtained from the enriched model in eq.(\ref{eq:equi_equa_CC_freq_depe_negative_2}) which also posses the extra root $\omega=\sqrt{\nicefrac{a}{b}}$ (where as always we are only considering the positive roots).
However, the non trivial solution in terms of the kinematic fields $u$ and $Q$ associated with this extra root is
\begin{align}
u_1 = 0 \, ,
\qquad
u_2 = 0 \, ,
\qquad
Q_{11} = Q_{22} \left(\frac{2 \mu }{\lambda + 2 \mu}-1\right) \, ,
\qquad
Q_{12} = 0 \, ,
\qquad
\text{for}
\quad
\omega=\sqrt{\frac{a}{b}} \, .
\end{align}
Since $u_1=u_2=0$, this solution corresponds to a trivial one in the frequency-dependent model (\ref{eq:poly_CC_freq_depe_negative}), and cannot be associated with a dispersion curve.
Moreover, as long as a finite domain is taken into account and it is guaranteed that $u\neq 0$ in some subset of the domain, this extra solution vanishes.

We explicitly remark that the enriched models obtained with the procedure presented in this paper may sometimes show additional constant roots in the dispersion diagrams compared to the frequency-dependent models
for those frequencies at which the original frequency-dependent model is not well-defined (here $\omega=\sqrt{\nicefrac{a}{b}}$).
These extra roots may account for special behaviours such as local resonances that could not be caught in the frequency-dependent model.
Those singularity values of the frequency-dependent model, e.g. frequencies that make the density (or stiffness) vanishing or infinite, correspond to additional \angcor{constant} roots $\omega(k)$ that can appear in the associated enriched model\footnote{\angcor{We remarked that these extra constant roots, when present, only involve a contribution to the solution for the micro-distortion and not for the displacement. In other words, the solution for the macro-displacement is never affected by the presence of such extra constant roots. This points to the fact that such constant roots might be related to some micro-scale resonances that do not affect the overall macroscopic displacement $u$ (for example resonance of the internal mass $m_2$ in Fig.\ref{fig:2} which does not provoke a movement of the external mass $m_1$). Given that $u$ is the only kinematical field of the frequency-dependent model, such micro-resonances, although possible, can be only caught by the enriched model. }}.

\begin{figure}[!ht]
\centering
\includegraphics[width=0.49\textwidth]{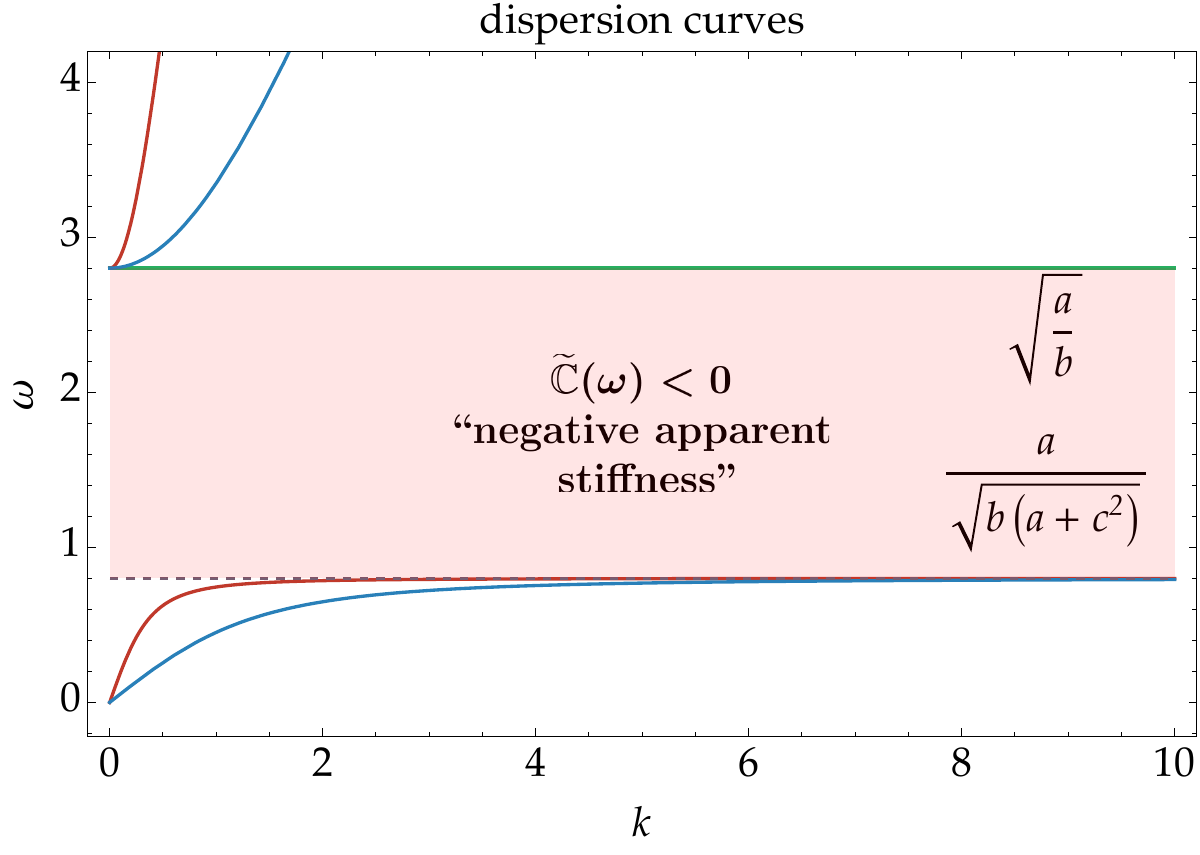}
\caption{
Dispersion curves for an isotropic class of symmetry in which the following values for the parameters have been used: $\rho=900$ kg/m$^3$, $\lambda=2898$ Pa, $\mu=262$ Pa, $a=0.089$, $b=0.011\.\si{s\squared}$, and $c=1$.
The curves for the frequency-dependent model are also reproduced by the corresponding enriched model, although the enriched model has the additional solution $\omega=\sqrt{\nicefrac{a}{b}}$.
While the enriched model remains \cors{positive-definite} in the band-gap region, the frequency-dependent one does not.
}
\label{fig:disp_curves_CC_freq_depe_negative}
\end{figure}

From eq.(\ref{eq:poly_CC_freq_depe_negative}), it is possible to calculate the frequencies such that 
\begin{align}
\omega=\pm\sqrt{\frac{a}{b}}
\quad
\Longleftrightarrow
\quad
\widetilde{\mathbb{C}} \to \infty
\, ,
\qquad\qquad\qquad
\omega=\pm\frac{a}{\sqrt{b \left(a+c^2\right)}}
\quad
\Longleftrightarrow
\quad
\widetilde{\mathbb{C}}=0 \, ,
\end{align}
where these frequencies correspond to the cut-off frequency of the optic curves and the asymptote of the acoustic curves, respectively (see Fig.~\ref{fig:disp_curves_CC_freq_depe_negative}).

\begin{tcolorbox}[colback=Salmon!15!white,colframe=Salmon!50!white,coltitle=black,title={\centering \textbf{Summary:} micromorphic enriched model stemming from the $\mathbb{C}(\omega)$ frequency-dependent model. \textbf{Form I}}]
\textbf{Original frequency-dependent model (frequency domain):}
\\[1em]
$
-\rho\.\omega^2\.\widehat u
=
\text{Div} \left[ \widetilde{\mathbb{C}}(\omega) \, \text{sym}\nabla\widehat u\right]
$
\qquad
with
\qquad
$
\widetilde{\mathbb{C}}(\omega)
=
\left( f - \dfrac{c^2}{a - b \, \omega^2}\right) \mathbb{C}
\qquad
(a>0,b>0)
$
\\[1em]
\textbf{Introduction of the new variable:}
$
\mathbb{C} \, \text{sym}\, \widehat Q
=
-
\dfrac{c}{a - b \, \omega^2} \, \mathbb{C} \, \text{sym}\nabla \widehat u \,.
$
\\[1em]
\textbf{Enriched model (time domain):}
\\[1em]
the action functional associated to the time domain model obtained from the original frequency-dependent model through the inverse time-Fourier transform and the introduction of the additional kinematical field $Q$ is
\\[1em]
$
\mathcal{A}
=
\displaystyle\iint\displaylimits_{\Omega\times [0,T]}
\underbrace{
\frac{1}{2}
\left(
\rho
\,
\langle \dot{u} , \dot{u} \rangle
+
b \, \langle \mathbb{C} \, \text{sym} \, \dot{Q} , \text{sym} \, \dot{Q} \rangle
\right)
}_{\text{K - kinetic energy density}}
$
\\*
\hspace*{3cm}
$
-
\underbrace{
\frac{1}{2}
\left(
f \, \langle \mathbb{C} \, \text{sym}\nabla u, \text{sym}\nabla u \rangle
+
2 \, c \, \langle \mathbb{C} \, \text{sym}\nabla u, \text{sym} \, Q \rangle
+
a \, \langle \mathbb{C} \, \text{sym} \, Q, \text{sym} \, Q \rangle
\right)
}_{\text{W - strain energy density}}
\mathrm{d}x\.\mathrm{d}t\,.
$
\\[1em]
The equilibrium equations are:
$
\quad
\rho \, \ddot{u} \, 
=
f \, \text{Div}\,\sigma
+ c \, \text{Div}\,\tau
 ,
c \, \sigma
+ a \, \tau
+ b \, \ddot{\tau}
=
0
\, ,
$
\\[1em]
and the Neumann boundary conditions on $\partial\Omega\times[0,T]$ are:
$
\quad
\left(
f \, \sigma 
+ c \, \tau
\right) \, n = 0 \, ,
\,
$

with
$
\;
\sigma=\mathbb{C} \, \text{sym}\nabla u
\,,
\,
\tau=\mathbb{C} \, \text{sym}\.Q
$.
\\[1em]
\textbf{Consistency checks of the model in the time domain:}
\\[1em]
\text{positive-definiteness}\quad\textcolor{Green}{\cmark}
\text{energy conservation}\quad\textcolor{Green}{\cmark}
\text{infinitesimal Galilean invariance \eqref{eq:ap_Galilean}}\quad\textcolor{Green}{\cmark}
\text{extended infinitesimal Galilean invariance \eqref{eq:ap_GalileanExtended}}\quad\textcolor{Green}{\cmark}
\end{tcolorbox}
%
%
%
%
\subsection{Formulation and positive-definiteness conditions: form II}
We can introduce the additional kinematic field $\widehat w:\cors{\R^3_x\times \big(\R_\omega\,\setminus\big\{\pm\sqrt{\nicefrac{a}{b}}\big\}\big)\subset\R^3_x\times\R_\omega}\to\R^{3}$ in the frequency domain such that
\begin{equation}
\text{Div} \left[
\mathbb{C} \, \text{sym}\nabla \widehat w
\right]
\coloneqq
-\frac{c}{a - b \, \omega^2}
\;
\text{Div} \left[ 
\mathbb{C} \, \text{sym}\nabla \widehat u
\right] \, .
\end{equation}
In this way, we obtain
\begin{align}
&
\begin{cases}
-
\,
\rho \; \omega^2 \widehat{u}
=
\text{Div} \left[ \left( f - \dfrac{c^2}{a - b \, \omega^2} \right) \mathbb{C} \, \text{sym}\nabla \widehat{u}\right]
\, ,
\nonumber
\\[20pt]
\text{Div} \left[
\mathbb{C} \, \text{sym}\nabla \widehat w
\right]
\coloneqq
-\dfrac{c}{a - b \, \omega^2}
\;
\text{Div} \left[ 
\mathbb{C} \, \text{sym}\nabla \widehat u
\right] \, ,
\end{cases}
\hspace{-0.55cm}
\nonumber
\\*[15pt]
\xLeftrightarrow[\hphantom{\F^{-1}_t}]{}
&
\begin{cases}
-
\,
\rho \; \omega^2 \widehat{u}
=
\text{Div} \left[ 
f \, \mathbb{C} \, \text{sym}\nabla \widehat u
+
c \, \mathbb{C} \, \text{sym}\nabla \widehat w
\right]
\, ,
\\[15pt]
(a - b \, \omega^2)
\,
\text{Div} \left[
\mathbb{C} \, \text{sym}\nabla \widehat  w
\right]
+
c
\,
\text{Div} \left[
\mathbb{C} \, \text{sym}\nabla \widehat u
\right]
\, 
=0\, ,
\end{cases}
\label{eq:equi_equa_CC_freq_depe_negative_2b}
\\*[15pt]
\xLeftrightarrow[\F^{-1}_t]{\F_t}
&
\begin{cases}
\rho\,\ddot{u}
-
\text{Div} \left[ 
\mathbb{C} \, 
\text{sym}
\left(
f \, \nabla u
+ c \, \nabla w
\right)
\right]
=
0
\, ,
\\[15pt]
\text{Div} \left[
\mathbb{C} \,
\text{sym}
\left(
a \, \nabla w
+ b \, \nabla \ddot{w}
+ c \, \nabla u
\right)
\right]
=0
\, ,
\end{cases}
\qquad
\xLeftrightarrow[\hphantom{\F^{-1}_t}]{}
\qquad
\begin{cases}
\rho\,\ddot{u}
-
\text{Div} \left[ 
f \, \sigma
+c \, \tau
\right]
=
0
\, ,
\\[15pt]
\text{Div} \left[
a \, \tau
+b \, \ddot{\tau}
+c \, \sigma
\right]
=0
\, ,
\end{cases}
\label{eq:equi_equa_C_freq_depe}
\end{align}
where $\sigma=\mathbb{C} \, \text{sym} \, \nabla u$ and $\tau=\mathbb{C} \, \text{sym} \, \nabla w$.
%
%
%
\subsubsection{Existence of an action functional and positive-definiteness}
The associated resulting functional is
\begin{align}
\mathcal{A}
=
\iint\displaylimits_{\Omega\times [0,T]}
&
\underbrace{
\frac{1}{2}
\left(
\rho
\,
\langle \dot{u} , \dot{u} \rangle
+
b
\,
\langle \mathbb{C} \, \text{sym} \nabla\dot{w} , \text{sym}\nabla\dot{w} \rangle
\right)
}_{\text{K - kinetic energy density}}
\\*
-
&
\underbrace{
\frac{1}{2}
\left(
\langle f \, \mathbb{C} \, \text{sym}\nabla u, \text{sym}\nabla u \rangle
+
2
\,
\langle c \, \mathbb{C} \, \text{sym}\nabla u, \text{sym}\nabla w \rangle
+
\langle a \, \mathbb{C} \, \text{sym}\nabla w, \text{sym}\nabla w \rangle
\right)
}_{\text{W - strain energy density}}
\mathrm{d}x\.\mathrm{d}t\,,
\label{eq:energy_negative_CC_freq_depe_z}
\end{align}
where for positive definiteness it is required that (we remind that $f=1+\frac{c^2}{a}$)
\begin{align}
a>0 \, ,
\qquad
\text{eig}(\mathbb{C})>0 \, ,
\qquad
\rho >0 \, ,
\qquad
b > 0 \, .
\end{align}
The associated homogeneous Neumann boundary conditions are
\begin{align}
\left(
f \, \sigma + c \, \tau
\right) \, n = 0 \, ,
\qquad
\qquad
\left(
c \, \sigma + a \, \tau + b \, \Ddot{\tau}
\right) \, n = 0 \, .
\label{eq:bound_cond_CC_rho_freq_depe_final_negative_2}
\end{align}
where $n$ is the normal on the boundary.
%
%
%
\subsubsection{Energy conservation}

To ensure that the resulting model is conservative, we have to guarantee that 
\begin{equation}
\frac{\mathrm d}{\mathrm dt}\int\displaylimits_{\Omega}
E(\dot{u},\nabla\dot{w},\nabla u,\nabla w) \, \mathrm{d}x=
\int\displaylimits_{\Omega}
\frac{\mathrm d}{\mathrm dt}\left[K(\dot{u},\nabla\dot{w})+W(\nabla u,\nabla w)\right] \, \mathrm{d}x=
0
\,,
\end{equation}
where $\Omega$ is the domain.
\begin{align}
\int\displaylimits_{\Omega}
\frac{\mathrm dE}{\mathrm dt}\, \mathrm{d}x
=&
\int\displaylimits_{\Omega}
\rho
\,
\langle \ddot{u} , \dot{u} \rangle
+
b
\,
\langle \mathbb{C} \, \text{sym} \nabla\ddot{w} , \text{sym}\nabla\dot{w} \rangle
+
f
\,
\langle \mathbb{C} \, \text{sym}\nabla u, \text{sym}\nabla \dot{u} \rangle
\notag
\\*[-10pt]
&
\phantom{\int\displaylimits_{\Omega}}
+
c
\,
\langle \mathbb{C} \, \text{sym}\nabla u, \text{sym}\nabla \dot{w} \rangle
+
c
\,
\langle \mathbb{C} \, \text{sym}\nabla \dot{u}, \text{sym}\nabla w \rangle
+
a
\,
\langle \mathbb{C} \, \text{sym}\nabla v, \text{sym}\nabla \dot{w} \rangle\, \mathrm{d}x
\notag
\\*[-10pt]
=&
\int\displaylimits_{\Omega}
\rho
\,
\langle \ddot{u} , \dot{u} \rangle
+
b
\,
\langle \ddot{\tau} , \text{sym}\nabla\dot{w} \rangle
+
f
\,
\langle \sigma, \text{sym}\nabla \dot{u} \rangle
\notag
\\*[-10pt]
&
\phantom{\int\displaylimits_{\Omega}}
+
c
\,
\langle \sigma, \text{sym}\nabla \dot{w} \rangle
+
c
\,
\langle \text{sym}\nabla \dot{u}, \tau \rangle
+
a
\,
\langle \tau, \text{sym}\nabla \dot{w} \rangle
\, \mathrm{d}x
\label{eq:cons_energy_C_freq_depe}
\\*[-10pt]
=&
\int\displaylimits_{\Omega}
\langle
\rho \, \ddot{u}
-  \, \text{Div} \left[f\, \sigma + c\, \tau\right]
,
\dot{u}
\rangle
+
\text{div}\left[\left( f \, \sigma^{\rm T} + c \, \tau^{\rm T}\right)\dot{u}\right]
\notag
\\*[-10pt]
&
\phantom{\int\displaylimits_{\Omega}}
-
\langle
\text{Div} \left[c\, \sigma + a\, \tau + b\, \ddot{\tau}\right]
,
\dot{w}
\rangle
+\text{div}\left[\left(c\, \sigma^{\rm T} + a\, \tau^{\rm T} + b\, \ddot{\tau}^{\rm T}\right)\dot{w}\right]\mathrm{d}x
=
0
\,,
\notag
\end{align}
where again, $\sigma=\mathbb{C} \, \text{sym} \, \nabla u$ and $\tau=\mathbb{C} \, \text{sym} \, \nabla w$.
Thanks to the equilibrium equations (\ref{eq:equi_equa_C_freq_depe}), the condition (\ref{eq:cons_energy_C_freq_depe}) becomes
\begin{align}
\frac{\mathrm d}{\mathrm dt}\int\displaylimits_{\Omega}
E\, \mathrm{d}x
=&
\int\displaylimits_{\Omega}
\text{div}\left[
\left( f \, \sigma^{\rm T} + c \, \tau^{\rm T}\right)\dot{u}
+\left(c\, \sigma^{\rm T} + a\, \tau^{\rm T} + b\, \ddot{\tau}^{\rm T}\right)\dot{w}
\right]
\mathrm{d}x
\\*[5pt]
=&
\int\displaylimits_{\partial\Omega}
\langle \left( f \, \sigma + c \, \tau \right)n,\dot{u} \rangle
+\langle \left(c\, \sigma + a\, \tau + b\, \ddot{\tau}\right)n,\dot{w} \rangle\, \mathrm{ds}
=
0
\, ,
\notag
\end{align}
which is automatically always satisfied thanks to the homogeneous boundary conditions reported in eq.(\ref{eq:bound_cond_CC_rho_freq_depe_final_negative_2}).

%
%
%
\subsubsection{Infinitesimal Galilean invariance}
\label{sec:galilean_invariance_1d}
As a last check, it is necessary to assess whether the model respects Galilean invariance, which requires the invariance of the equilibrium equations eq.(\ref{eq:equi_equa_C_freq_depe}) with respect to the following extended infinitesimal Galilean transformation (cf.\ Appendix \ref{sec:appGalilean})
\corrections{
\begin{align}
u\to\overline u=u+A(t)\.x+r(t)\,,\qquad w\to\overline w&=w+A(t)\.x+r(t)\,,\qquad \ddot{A}(t)=0,\qquad \ddot{r}(t)=0\,,\label{eq:transformation_u_v_1d}
\end{align}}
where $A(t)\in\mathfrak{so}(3)$ is a skew-symmetric matrix while $r(t)\in\R^3$ is a vector.
For the sake of clarity, we report below the equilibrium equations (\ref{eq:equi_equa_C_freq_depe})
\begin{align}
\begin{cases}
\rho\,\ddot{u}
-
\text{Div} \left[ 
\mathbb{C} \, 
\text{sym}
\left(
f \, \nabla u
+ c \, \nabla w
\right)
\right]
=
0
\, ,
\\[5pt]
\text{Div} \left[
\mathbb{C} \,
\text{sym}
\left(
a \, \nabla w
+ b \, \nabla \ddot{w}
+ c \, \nabla u
\right)
\right]
=0
\, .
\end{cases}
\label{eq:equi_equa_gali_inva_1d}
\end{align}
We now substitute $u$ and $w$ with $\overline{u}$ and $\overline{w}$ from eq.(\ref{eq:transformation_u_v_1d}), respectively, in eq.(\ref{eq:equi_equa_gali_inva_1d})
\begin{align}
&
\begin{cases}
\rho\,\ddot{\overline{u}}
-
\text{Div} \bigg[ 
\mathbb{C} \, 
\text{sym}
\left(
f \, \nabla \overline{u}
+ c \, \nabla \overline{w}
\right)
\bigg]
=
0
\, ,
\\[5pt]
\text{Div} \bigg[
\mathbb{C} \,
\text{sym}
\left(
a \, \nabla \overline{w}
+ b \, \nabla \ddot{\overline{w}}
+ c \, \nabla \overline{u}
\right)
\bigg]
=0
\, ,
\end{cases}
\nonumber
\\*
\Rightarrow
&
\begin{cases}
\rho\,
\left(
\ddot{u} 
+ \frac{\text{d}^2}{\text{d}t^2}\left(A(t)\.x\right)
+ \frac{\text{d}^2}{\text{d}t^2}\.r(t)
\right)
-
\text{Div} \bigg[
\mathbb{C} \, 
\text{sym}
\bigg(
f \, \nabla 
\bigg(
u
+ A(t)\.x
+ r(t)
\bigg)
\\*
\hspace{8cm}
+ c \, \nabla 
\bigg(
v
+ A(t)\.x
+ r(t)
\bigg)
\bigg)
\bigg]
=
0
\, ,
\\[5pt]
\text{Div} \bigg[
\mathbb{C} \,
\text{sym}
\bigg(
a \, \nabla 
\bigg(
v
+ A(t)\.x
+ r(t)
\bigg)
+ b \, \nabla
\left(
\ddot{w} 
+ \frac{\text{d}^2}{\text{d}t^2}\left(A(t)\.x\right)
+ \frac{\text{d}^2}{\text{d}t^2}\left(r(t)\right)
\right)
\\*
\hspace{8cm}
+ c \, \nabla 
\bigg(
u
+ A(t)\.x
+ r(t)
\bigg)
\bigg)
\bigg]
=0
\, .
\end{cases}
\nonumber
\\*[5pt]
\Rightarrow
&
\begin{cases}
\rho\,\ddot{u}
-
\text{Div} \left[ 
\mathbb{C} \, 
\text{sym}
\left(
f \, \nabla u
+ c \, \nabla w
\right)
\right]
=
0
\, ,
\\[5pt]
\text{Div} \left[
\mathbb{C} \,
\text{sym}
\left(
a \, \nabla v
+ b \, \nabla \ddot{w}
+ c \, \nabla u
\right)
\right]
=0
\, .
\end{cases}
\label{eq:equi_equa_gali_inva_2d_3}
\end{align}
Comparing eq.(\ref{eq:equi_equa_gali_inva_1d}) and eq.(\ref{eq:equi_equa_gali_inva_2d_3}), it is possible to see that the two sets of equations coincide, making the enriched model invariant with respect to extended infinitesimal Galilean transformations.
%
%
%
\subsection{Relations between the frequency-dependent model and the equivalent enriched model: form II}
The curves for the frequency-dependent model eq.(\ref{eq:poly_CC_freq_depe_negative}) are also reproduced by the corresponding enriched model eq.(\ref{eq:equi_equa_C_freq_depe}), although the enriched model has the additional solution $k=0$.
The presence of this extra root does not affect the overall metamaterial response and is associated with the fact that the auxiliary variable in eq.(\ref{eq:equi_equa_CC_freq_depe_negative_2b}) is introduced trough its divergence.
It is possible to combine the model presented in Section \ref{sec:rho_freq} and Section \ref{sec:CC_freq}, and the calculations are shown in Appendix \ref{sec:appC} for the sake of brevity.
\begin{figure}[!ht]
\centering
\includegraphics[width=0.49\textwidth]{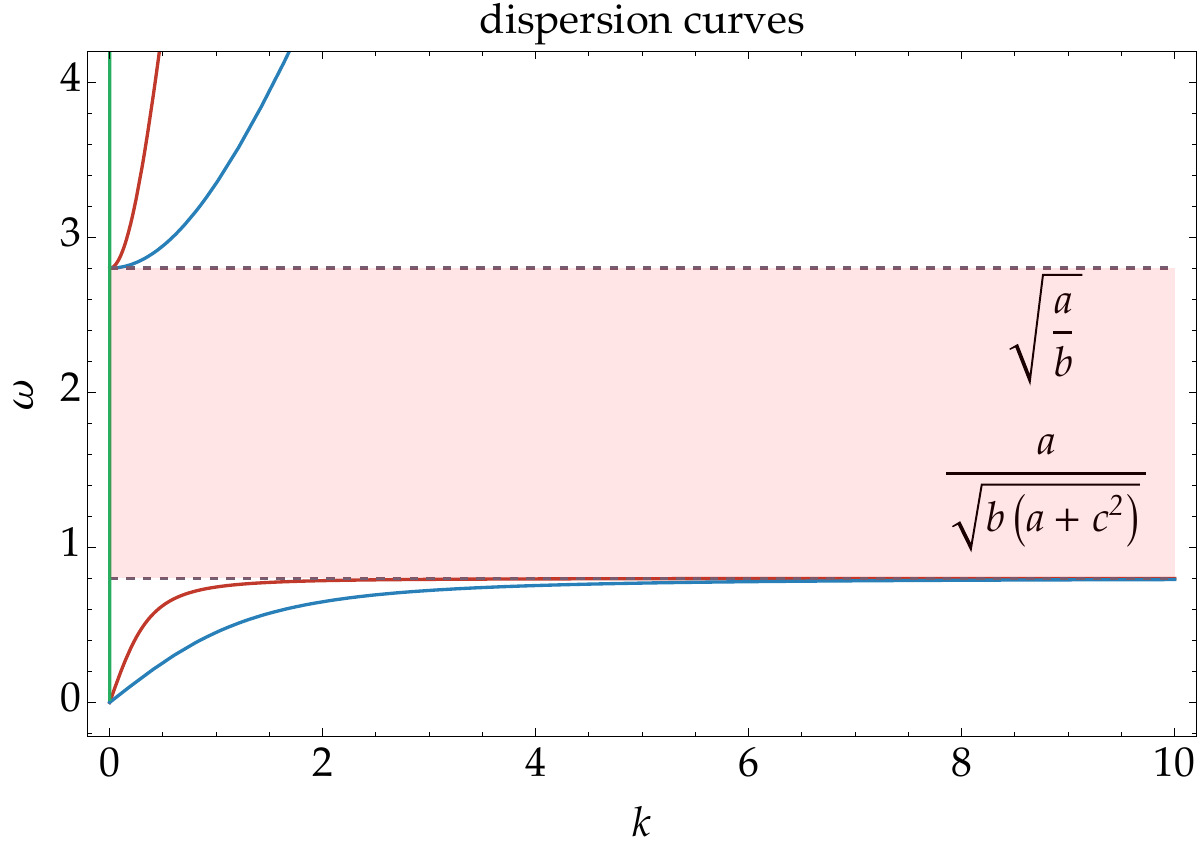}
\caption{
Dispersion curves for an isotropic class of symmetry in which the following values for the parameters have been used: $\rho=900$ kg/m$^3$, $\lambda=2898$ Pa, $\mu=262$ Pa, $a=0.089$, $b=0.011\.\si{s\squared}$, and $c=1$.
The curves for the frequency-dependent model are also reproduced by the corresponding enriched model coincide, although the enriched model has the additional solution $k=0$.
While the enriched model remains \cors{positive-definite} in the band-gap region, the frequency-dependent one does not.}
\label{fig:disp_curves_CC_freq_depe_negative_2}
\end{figure}

\begin{tcolorbox}[colback=Salmon!15!white,colframe=Salmon!50!white,coltitle=black,title={\centering \textbf{Summary:} micromorphic enriched model stemming from the $\mathbb{C}(\omega)$ frequency-dependent model. \textbf{Form II}}]
\textbf{Original frequency-dependent model (frequency domain):}
\\[1em]
$
-\rho \, \omega^2 \. \widehat u
=
\text{Div} \left[ \widetilde{\mathbb{C}}(\omega) \, \text{sym}\nabla\widehat u\right]
$
\qquad
with
\qquad
$
\widetilde{\mathbb{C}}(\omega)
=
\left( f - \dfrac{c^2}{a - b \, \omega^2}\right) \mathbb{C}
\qquad
(a>0,b>0)
$
\\[1em]
\textbf{Introduction of the new variable:}
$
\text{Div} \left[
\mathbb{C} \, \text{sym}\nabla \widehat w
\right]
\coloneqq
-\dfrac{c}{a - b \, \omega^2}
\;
\text{Div} \left[ 
\mathbb{C} \, \text{sym}\nabla \widehat u
\right] \, .
$
\\[1em]
\textbf{Enriched model (time domain):}
\\[1em]
The action functional associated to the time domain model obtained from the original frequency-dependent model through the inverse time-Fourier transform and the introduction of the additional kinematical field $v$ is
\\[1em]
$
\mathcal{A}
=
\displaystyle\iint\displaylimits_{\Omega\times [0,T]}
\underbrace{
\frac{1}{2}
\left(
\rho
\,
\langle \dot{u} , \dot{u} \rangle
+
b
\,
\langle \mathbb{C} \, \text{sym} \nabla\dot{w} , \text{sym}\nabla\dot{w} \rangle
\right)
}_{\text{K - kinetic energy density}}
$
\\*
\hspace*{2.5cm}
$
-
\underbrace{
\frac{1}{2}
\left(
\langle f \, \mathbb{C} \, \text{sym}\nabla u, \text{sym}\nabla u \rangle
+
2
\,
\langle c \, \mathbb{C} \, \text{sym}\nabla u, \text{sym}\nabla w \rangle
+
\langle a \, \mathbb{C} \, \text{sym}\nabla w, \text{sym}\nabla w \rangle
\right)
}_{\text{W - strain energy density}}
\mathrm{d}x\.\mathrm{d}t\,.
$
\\[1em]
The equilibrium equations are:
$
\quad
\rho\,\ddot{u}
-
\text{Div} \left[ 
f \, \sigma
+c \, \tau
\right]
=
0
\, ,
\qquad
\text{Div} \left[
a \, \tau
+b \, \ddot{\tau}
+c \, \sigma
\right]
=0
\, ,
$
\\[1em]
and the Neumann boundary conditions on $\partial\Omega\times[0,T]$ are:
\\[1em]
$
\left(
f \, \sigma + c \, \tau
\right) \, n = 0 \, ,\qquad
\left(
c \, \sigma + a \, \tau + b \, \Ddot{\tau}
\right) \, n = 0
\qquad\text{with}\qquad
\sigma=\mathbb{C} \, \text{sym}\nabla u\,,\qquad
\tau=\mathbb{C} \, \text{sym}\nabla w.
$
\\[1em]
\textbf{Consistency checks of the model in the time domain:}
\\[1em]
\text{positive-definiteness}\quad\textcolor{Green}{\cmark}
\text{energy conservation}\quad\textcolor{Green}{\cmark}
\text{infinitesimal Galilean invariance \eqref{eq:ap_Galilean}}\quad\textcolor{Green}{\cmark}
\text{extended infinitesimal Galilean invariance \eqref{eq:ap_GalileanExtended}}\quad\textcolor{Green}{\cmark}
\end{tcolorbox}
%
%
%
%
%
%
\section{An example from the literature}
We consider the following 1D example (see \cite{shen2018analysis}) with both a frequency-dependent effective density\footnote{\corrections{The identical effective mass $M_{\rm eff}(\omega)=m_1+m_2+\frac{m_2^2\,\omega^2}{k_2-m_2\,\omega^2}$ can also be found in \cite{fedele2023effective} eq.(2.21) describing the same spring-interconnected mass-in-mass cell \cor{lattices}.}} $\rho (\omega)$ and Young modulus $E (\omega)$
\begin{align}
\begin{cases}
\displaystyle 
\; \overline{\rho}(\omega) =
\frac{m_1+m_2}{A \, L}
\left(
1
+
\frac{m_2^2 \, \omega ^2}{\left(m_1+m_2\right) \left(k_2-m_2 \omega ^2\right)}
\right)
\, ,
\\*[20pt]
\displaystyle 
\overline{E}(\omega) =
E_{0} \left(1-
\frac{\omega ^2}{4}
\frac{m_{1}+m_{2}}{k_{1}}
\left( 1+ \frac{m_{2}^2 \, \omega ^2}{(m_{1}+m_{2}) \left(k_{2}-m_{2} \, \omega ^2\right)} \right)
\right)
\, ,
\end{cases}
\label{eq:densy_young_freq_depe_litt_example}
\end{align}
with additional material constants $A,L,m_1,m_2,k_1,k_2>0$.
The accounted bulk equation is
\begin{equation}
-
\,
\overline{\rho}(\omega)\,\omega^2\.\widehat u_{1}
=
\overline{E}(\omega) \, \widehat u_{1,11}
\, ,
\label{eq:equi_equi_litt_example}
\end{equation}
which after substituting $\overline{\rho}(\omega)$ and $\overline{E}(\omega)$ becomes
\begin{align}
-
\,
\frac{m_1+m_2}{A \, L}
\left(
1
+
\frac{m_2^2 \, \omega ^2}{\left(m_1+m_2\right) \left(k_2-m_2 \, \omega ^2\right)}
\right)\,\omega^2  \, \widehat{u}_{1}
\hspace{5cm}
\label{eq:equi_equa_litt_example}
\\*
\vphantom{\int\displaylimits^A}
=
E_{0} \left(1-
\frac{\omega ^2}{4}
\frac{m_{1}+m_{2}}{k_{1}}
\left( 1+ \frac{m_{2}^2 \, \omega ^2}{(m_{1}+m_{2}) \left(k_{2}-m_{2} \, \omega ^2\right)} \right)
\right) \, \widehat u_{1,11}
\, .
\notag
\end{align}
This model is \cors{positive-definite} if both the frequency-dependent Young modulus $\overline{E}$ and the density $\overline{\rho}$ are positive:
\begin{equation}
\begin{array}{ccl}
\overline{\rho}>0
&
:
&
\begin{cases}
\omega^2
<
\frac{k_2}{m_2}
\, ,
\\[5pt]
\text{or}
\\[5pt]
\omega^2
>
k_2\,\frac{m_1+m_2}{m_1 \, m_2}
\, .
\end{cases}
\, ,
\\[12mm]
\overline{E}>0
&
:
&
\frac{k_{2}}{m_{2}}<
\omega^2
<\frac{4 k_{1} m_{2}+k_{2} (m_{1}+m_{2})+\sqrt{2 k_{2} m_{1} m_{2} (k_{2}-4 k_{1})+m_{2}^2 (4 k_{1}+k_{2})^2+k_{2}^2 m_{1}^2}}{2 m_{1} m_{2}}
\, ,
\\[5mm]
\begin{cases}
\overline{\rho}>0
\\[5pt]
\text{and}
\\[5pt]
\overline{E}>0 
\end{cases}
&
:
&
\begin{cases}
\omega^2
<
\frac{4 k_{1} m_{2}+k_{2} (m_{1}+m_{2})-\sqrt{2 k_{2} m_{1} m_{2} (k_{2}-4 k_{1})+m_{2}^2 (4 k_{1}+k_{2})^2+k_{2}^2 m_{1}^2}}{2 m_{1} m_{2}}
\, ,
\\[5pt]
\text{or}
\\[5pt]
k_2\,\frac{m_1+m_2}{m_1 \, m_2}
<
\omega^2
<
\frac{4 k_{1} m_{2}+k_{2} (m_{1}+m_{2})+\sqrt{2 k_{2} m_{1} m_{2} (k_{2}-4 k_{1})+m_{2}^2 (4 k_{1}+k_{2})^2+k_{2}^2 m_{1}^2}}{2 m_{1} m_{2}}
\, ,
\end{cases}
\, ,
\end{array}
\label{eq:well_posed_freq_depe_litt_example}
\end{equation}
where particular emphasis is put on the fact that $\omega\neq\sqrt{\nicefrac{k_2}{m_2}}$ in order to have a finite density and Young modulus.
The domain of positive-definiteness in eq.(\ref{eq:well_posed_freq_depe_litt_example}) is represented in Fig.~\ref{fig:disp_curves_freq_depe_negative_litt_example}.
\corrections{
Developing we obtain
\begin{align}
-\,\rho_0 \, \omega^2 \, \widehat{u}_{1}
=
E_0
\left(
t-\frac{q^2}{r-s\, \omega^2}
-h \, \omega^2
\right) \, \widehat u_{1,11}
\, ,
\label{eq:equi_equa_litt_example_2}
\end{align}
}
where
\begin{align}
\rho_0 = \frac{m_{1}+m_{2}}{A \, L}\, ,
\qquad
r = \frac{m_{1} \, q^2}{m_{2}}\, ,
\qquad
s = \frac{m_{1}^2 \, q^2}{k_{2} (m_{1}+ m_{2})}\, ,
\qquad
t = \frac{m_{1}+m_{2}}{m_{1}}\, ,
\qquad
h = \frac{m_{1}+m_{2}}{4 k_{1}}\,.
\label{eq:para_relation_litt_example}
\end{align}
For the mass-in-mass lattice model \eqref{eq:densy_young_freq_depe_litt_example} the dimensionless parameter $q$ is not necessary and can be chosen arbitrarily.
However, we keep the parameter nevertheless in order to build a more complete associated enriched model.
%
%
%
\subsection{Formulation and positive-definiteness conditions}
Introducing the additional kinematic field
$\widehat v_1:\cors{\R_x\times \big(\R_\omega\,\setminus\big\{\pm\sqrt{\nicefrac{r}{s}}\big\}\big)\subset\R_x\times\R_\omega}\to\R$ in the frequency domain such that
\begin{equation}
\widehat v_{1,11}
=
-
\,
\dfrac{q}{r - s \, \omega^2} \, \widehat u_{1,11}
\, ,    
\end{equation}
we establish the following:
\begin{align}
&
\begin{cases}
-\,\rho_0 \, \omega^2 \, \widehat{u}_{1}
=
E_0
\left(
t-\dfrac{q^2}{r-s\, \omega^2}
-h \, \omega^2
\right) \, \widehat u_{1,11}
\, ,
\\[3mm]
\widehat v_{1,11}
=
-\dfrac{q}{r - s \, \omega^2} \, \widehat u_{1,11}
\, ,
\end{cases}
\\*[3mm]
\Longleftrightarrow
&
\begin{cases}
-\,\rho_0 \, \omega^2 \,  \, \widehat{u}_{1}
=
E_0
\left(
t - h \, \omega^2
\right) \, \widehat u_{1,11}
+E_{0} \, q \, \widehat v_{1,11}
\, ,
\\[3mm]
r \, \widehat v_{1,11} -s \, \omega^2 \, \widehat v_{1,11} + q \, \widehat u_{1,11}
=
0
\, ,
\end{cases}
\label{eq:equi_equa_rho_freq_depe_negative_litt_example}
\end{align}
Hence, utilizing the inverse time-Fourier transform $\F_t^{-1}$,
equation (\ref{eq:equi_equa_rho_freq_depe_negative_litt_example}) entails
\begin{align}
\rho_0
\, \ddot{u}_{1}
=
E_0
\left(
t \, u_{1,11}
+ h \, \ddot{u}_{1,11}
+ q \, v_{1,11}
\right)
\, ,
\hspace{1cm}
r \, v_{1,11}
+
s \, \ddot{v}_{1,11}
+
q \, u_{1,11}
=
0
\,,
\label{eq:equi_equa_rho_freq_depe_negative_litt_example_2}
\end{align}
where $v$ has the dimension of a displacement.
%
%
%
\subsubsection{Existence of an action functional and positive-definiteness}
Setting $\Omega=[0,L]$, the associated action functional is
\begin{align}
\mathcal{A}
=
&
\int\displaylimits_0^T\int\displaylimits_{0}^L
\underbrace{
\frac{1}{2}
\rho_0
\left(
\dot{u}_{1}^2
+
\frac{E_0}{\rho_0}
(h \, \dot{u}_{1,1}^2
+
s \, \dot{v}_{1,1}^2)
\right)
}_{\text{K - kinetic energy density}}
-
\underbrace{
\frac{E_{0}}{2}
\left(
t\,u_{1,1}^2
+
2q\,u_{1,1} \, v_{1,1}
+
r\,v_{1,1}^2
\right)
}_{\text{W - strain energy density}}
\mathrm{d}x\.\mathrm{d}t\,,
\label{eq:energy_negative_litt_example}
\end{align}
where for positive definiteness it is required that
\begin{align}
\rho_0>0 \, ,
\qquad
\quad
E_0>0
\, ,
\qquad
\quad
s>0 \, ,
\qquad
\quad
h>0 \, ,
\qquad
\quad
r>0\, ,
\qquad
\quad
q^2<r \, t
\, .
\end{align}
The associated homogeneous Neumann boundary conditions are
\begin{align}
E_{0} \left( t \, u_{1,1} + h \, \ddot{u}_{1,1} + q \, v_{1,1} \right)=  0 \, ,
\qquad
\qquad
E_{0} \left( r \, v_{1,1} + s \, \ddot{v}_{1,1} + q \, u_{1,1} \right)=  0 \, .
\label{eq:bound_cond_rho_freq_depe_final_negative_litt_example}
\end{align}
%
%
%
\subsubsection{Energy conservation}
To ensure that the resulting model is conservative, we have to guarantee that 
\begin{equation}
\frac{\mathrm d}{\mathrm dt}
\int\displaylimits_{0}^L
E(\dot{u}_1,\dot{u}_{1,1},\dot{v}_{1,1},u_{1,1},v_{1,1}) \, \mathrm{d}x=
\int\displaylimits_{0}^L
\frac{\mathrm d}{\mathrm dt}\left[
K(\dot{u}_1,\dot{u}_{1,1},\dot{v}_{1,1})
+W(u_{1,1},v_{1,1})
\right] \mathrm{d}x=
0
\,.
\end{equation}
\begin{align}
\int\displaylimits_{0}^L
\frac{\mathrm dE}{\mathrm dt}\, \mathrm{d}x
=&
\int\displaylimits_{0}^L
\rho_0
\left(
\ddot{u}_{1} \, \dot{u}_{1}
+
\frac{E_0}{\rho_0}
(h \, \ddot{u}_{1,1} \, \dot{u}_{1,1}
+
s \, \ddot{v}_{1,1} \, \dot{v}_{1,1})
\right)
\notag
\\[-10pt]
&
\phantom{\int\displaylimits_{\Omega}}
+
E_{0}
\left(
t\,u_{1,1}\,\dot{u}_{1,1}
+
q\,u_{1,1} \, \dot{v}_{1,1}
+
q\,\dot{u}_{1,1} \, v_{1,1}
+
r\,v_{1,1}\,\dot{v}_{1,1}
\right) \mathrm{d}x
\label{eq:cons_energy_rho_freq_depe_negative_litt_example}
\\[-10pt]
=
&
\int\displaylimits_{0}^L
\rho_0 \, 
\ddot{u}_{1} \, \dot{u}_{1}
+
E_0 \, h
\left[
\left(
\ddot{u}_{1,1} \, \dot{u}_{1}
\right)_{,1}
-
\ddot{u}_{1,11} \, \dot{u}_{1}
\right]
+
E_0 \, s
\left[
\left(
\ddot{v}_{1,1} \, \dot{v}_{1}
\right)_{,1}
-
\ddot{v}_{1,11} \, \dot{v}_{1}
\right]
\notag
\\[-10pt]
&
\phantom{\int\displaylimits_{\Omega}}
+
E_0 \, t
\left[
\left(
u_{1,1} \, \dot{u}_{1}
\right)_{,1}
-
u_{1,11} \, \dot{u}_{1}
\right]
+
E_0 \, q
\left[
\left(
v_{1,1} \, \dot{u}_{1}
\right)_{,1}
-
v_{1,11} \, \dot{u}_{1}
\right]
\notag
\\[-10pt]
&
\phantom{\int\displaylimits_{\Omega}}
+
E_0 \, q
\left[
\left(
u_{1,1} \, \dot{v}_{1}
\right)_{,1}
-
u_{1,11} \, \dot{v}_{1}
\right]
+
E_0 \, r
\left[
\left(
v_{1,1} \, \dot{v}_{1}
\right)_{,1}
-
v_{1,11} \, \dot{v}_{1}
\right] \mathrm{d}x
\notag
\\[-10pt]
=
&
\int\displaylimits_{0}^L
\left[
\rho_0 \, 
\ddot{u}_{1}
-
E_0
\left(
h \, \ddot{u}_{1,11}
+ t \, u_{1,11}
+ q \, v_{1,11}
\right)
\right]
 \, \dot{u}_{1}
\notag
-
E_0
\left(
s \, \ddot{v}_{1,11}
+ q \, u_{1,11}
+ r \, v_{1,11}
\right)
\notag
\\[-10pt]
&
\phantom{\int\displaylimits_{\Omega}}
+
E_0
\left[
\left(
h \, \ddot{u}_{1,11}
+ t \, u_{1,11}
+ q \, v_{1,11}
\right)
\dot{u}_{1}
+
\left(
s \, \ddot{v}_{1,11}
+ q \, u_{1,11}
+ r \, v_{1,11}
\right)
\dot{v}_{1}
\right]_{,1}\, \mathrm{d}x
=
0
\, .
\notag
\end{align}
Thanks to the equilibrium equations (\ref{eq:equi_equa_rho_freq_depe_negative_litt_example_2}), the condition (\ref{eq:cons_energy_rho_freq_depe_negative_litt_example}) becomes
\begin{align}
\frac{\mathrm d}{\mathrm dt}
\int\displaylimits_{0}^L
E\, \mathrm{d}x
&=
\int\displaylimits_{0}^L
E_0
\left[
\left(
h \, \ddot{u}_{1,11}
+ t \, u_{1,11}
+ q \, v_{1,11}
\right)
\dot{u}_{1}
+
\left(
s \, \ddot{v}_{1,11}
+ q \, u_{1,11}
+ r \, v_{1,11}
\right)
\dot{v}_{1}
\right]_{,1} \mathrm{d}x
\\*
&=
E_{0}
\left(
h \, \ddot{u}_{1,11}
+ t \, u_{1,11}
+ q \, v_{1,11}
\right)\dot{u}_1 
\Big|_{(0,\cdot)}^{(L,\cdot)}
+
E_0
\left(
s \, \ddot{v}_{1,11}
+ q \, u_{1,11}
+ r \, v_{1,11}
\right) \dot{v}_1
\Big|_{(0,\cdot)}^{(L,\cdot)}
=
0 \, ,
\notag
\end{align}
which is automatically always satisfied thanks to the boundary conditions required in eq.(\ref{eq:bound_cond_rho_freq_depe_final_negative_litt_example}) in the case of zero externals surface traction.
%
%
%
\subsubsection{Infinitesimal Galilean invariance}
With arguments similar to that presented in Section \ref{sec:galilean_invariance}, \ref{sec:galilean_invariance_1b}, and \ref{sec:galilean_invariance_1d}, it is easy to check that eqs.(\ref{eq:equi_equa_rho_freq_depe_negative_litt_example_2}) are extended infinitesimal Galilean invariant \eqref{eq:ap_GalileanExtended}.
%
%
%
\subsection{Relations between the frequency-dependent model and the enriched equivalent model}
\begin{figure}[H]
\centering
\includegraphics[width=0.49\textwidth]{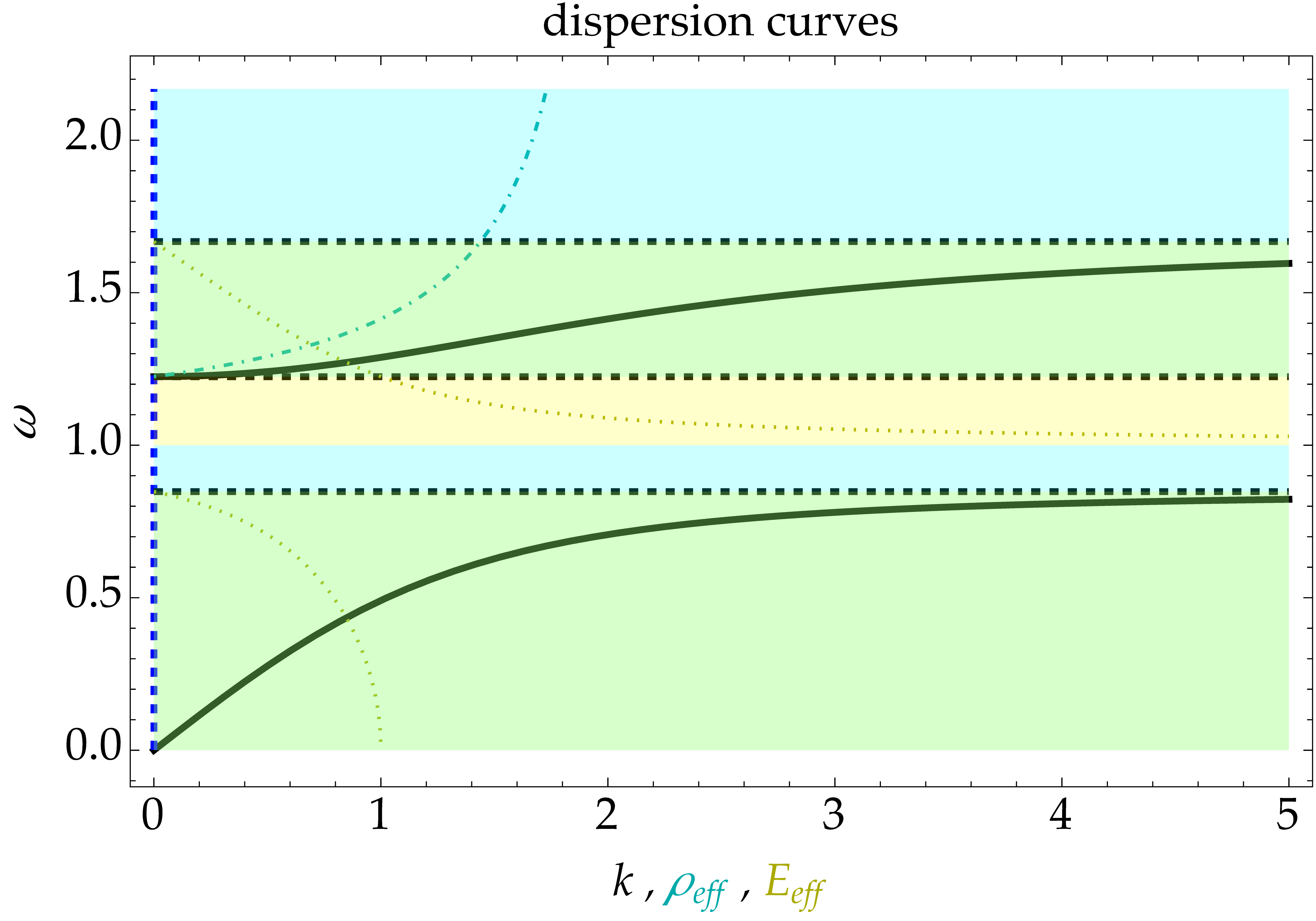}
\caption{
Dispersion curves plot:
the solid black lines are the dispersion curves.
The region where only the frequency-dependent Young modulus $\overline{E}$ is positive is yellow.
The region where only the frequency-dependent density $\overline{\rho}$ is positive is light blue.
The region where the frequency-dependent Young modulus $\overline{E}$ and the density $\overline{\rho}$ are both positive is in green.
The black dashed lines are the asymptotes/cut-off.
The light blue dot-dashed curve represents the values of the density $\overline{\rho}$ while changing the frequency and the brown dotted curve represents the values of the Young modulus $\overline{E}$ while changing the frequency.
The blue dashed line represents the extra imaginary dispersion curve which correspond to $k=0$.
The values used for the parameters are
$k_1 = 1, k_2 = 1, m_1 = 2, m_2 = 1, E_0 = 1, A = 1,\text{ and } L = 1$.
}
\label{fig:disp_curves_freq_depe_negative_litt_example}
\end{figure}

The dispersion curves associated with the equilibrium equations (\ref{eq:equi_equa_litt_example}), or equivalently of the equation~(\ref{eq:equi_equa_rho_freq_depe_negative_litt_example_2}) are reported in Fig.~\ref{fig:disp_curves_freq_depe_negative_litt_example}.
The system (\ref{eq:equi_equa_rho_freq_depe_negative_litt_example_2}) has an extra imaginary dispersion curve which corresponds to $k=0$ (blue dashed line).

\begin{tcolorbox}     [colback=Salmon!15!white,
                       colframe=Salmon!50!white,
                       coltitle=black,
                       breakable,
                       pad at break=1mm,
                       title={\centering \textbf{Summary:} enriched model stemming from a frequency-dependent example from the literature}]
\textbf{Original frequency-dependent model (frequency domain):}
\\[1em]
$
-\overline{\rho}(\omega)\,\omega^2\.\widehat u_{1}
=
\overline{E}(\omega) \, \widehat u_{1,11}
$
\quad
with
\quad
$
\begin{cases}
\overline{\rho}(\omega) =
\dfrac{m_1+m_2}{A \, L}
\left(
1
+
\dfrac{m_2^2 \, \omega ^2}{\left(m_1+m_2\right) \left(k_2-m_2 \omega ^2\right)}
\right)
\\*[15pt]
\overline{E}(\omega) =
E_{0} \left(1-
\dfrac{\omega ^2}{4}
\dfrac{m_{1}+m_{2}}{k_{1}}
\left( 1+ \dfrac{m_{2}^2 \, \omega ^2}{(m_{1}+m_{2}) \left(k_{2}-m_{2} \, \omega ^2\right)} \right)
\right)
\end{cases}
$
\\[1em]
\textbf{Introduction of the new variable:}
$
\widehat v_{1,11}(x,\omega)
=
-
\,
\dfrac{q}{r - s \, \omega^2} \, \widehat u_{1,11}(x,\omega)
\, .
$
\\[1em]
\textbf{Enriched model (time domain):} the action functional associated to the time domain model obtained from the original frequency-dependent model through the inverse time-Fourier transform and the introduction of the additional kinematical field $v$ is
\[
\mathcal{A}
=
\int\displaylimits_0^T\int\displaylimits_{0}^L
\underbrace{
\frac{1}{2}
\rho_0
\left(
\dot{u}_{1}^2
+
\frac{E_0}{\rho_0}
(h \, \dot{u}_{1,1}^2
+
s \, \dot{v}_{1,1}^2)
\right)
}_{\text{K - kinetic energy density}}
-
\underbrace{
\frac{E_{0}}{2}
\left(
t\,u_{1,1}^2
+
2q\,u_{1,1} \, v_{1,1}
+
r\,v_{1,1}^2
\right)
}_{\text{W - strain energy density}}
\mathrm{d}x\.\mathrm{d}t\,.
\]
The equilibrium equations are:
$
\quad
\rho_0
\, \ddot{u}_{1}
=
E_0
\left(
t \, u_{1,11}
+ h \, \ddot{u}_{1,11}
+ q \, v_{1,11}
\right)
,
\;
r \, v_{1,11}
+
s \, \ddot{v}_{1,11}
+
q \, u_{1,11}
=
0
\, ,
$
\\[1em]
and the Neumann boundary conditions on $\{0,L\}\times[0,T]$ are:
\\[1em]
$
E_{0} \left( t \, u_{1,1} + h \, \ddot{u}_{1,1} + q \, v_{1,1} \right)=  0 \, ,
\qquad
E_{0} \left( r \, v_{1,1} + s \, \ddot{v}_{1,1} + q \, u_{1,1} \right)=  0.
$
\\[1em]
\textbf{Consistency checks of the model in the time domain:}
\\[1em]
\text{positive-definiteness}\quad\textcolor{Green}{\cmark}
\text{energy conservation}\quad\textcolor{Green}{\cmark}
\text{infinitesimal Galilean invariance \eqref{eq:ap_Galilean}}\quad\textcolor{Green}{\cmark}
\text{extended infinitesimal Galilean invariance \eqref{eq:ap_GalileanExtended}}\quad\textcolor{Green}{\cmark}
\end{tcolorbox}
%
%
%
%
%
%
\section{Conclusions}
In the present paper, we have shown an explicit procedure allowing to transform specific frequency-dependent Cauchy continuum models into their frequency-independent micromorphic counterparts.
While frequency-dependent models fail to respect positive definiteness in those frequency ranges which are close to local-resonance frequencies of the internal masses, their micromorphic counterpart remains positive-definite in the whole range of the considered frequencies.
Moreover, consistency checks of the obtained micromorphic models on (i) existence of an action functional, (ii) total energy conservation, and  (iii) Galilean invariance are performed so as to guarantee their physical grounds.
The proposed procedure will be extended to wavenumber-dependent models in forthcoming papers to show how more and more complex enriched continua can be generated to describe larger classes of metamaterials by keeping a reasonably low number of constitutive material parameters.

{
\begin{spacing}{0.5}
\scriptsize
\subparagraph{{\scriptsize Acknowledgements.}}
Angela Madeo and Gianluca Rizzi acknowledge support from the European Commission through the funding of the ERC Consolidator Grant META-LEGO, N$^\circ$ 101001759.
Patrizio Neff acknowledges support in the framework of the DFG-Priority Programme 2256 ``Variational Methods for Predicting Complex Phenomena in Engineering Structures and Materials'', Neff 902/10-1, Project-No. 440935806.
\end{spacing}
}

\footnotesize{
\begingroup
\setstretch{0.35}
\setlength\bibitemsep{1.55pt}

\printbibliography

\endgroup
}

\section*{Appendix}
\footnotesize{
\appendix
%
%
%
\section{Dispersion analysis of the \texorpdfstring{$\widetilde\rho(\omega)$}{rho} model}
\label{disp_analysis_app}
%
%
%
\subsection{Dispersion relations obtained through the space-Fourier transform}
In order to derive the dispersion relations of the considered model, let us apply the space-Fourier transform $\F_x$ to both sides of the equation \eqref{eq:poly_rho_freq_depe_negative}.
This gives\footnote{Indeed, 
\begin{equation}
     \F_x\big[\nabla_x(\widehat u(x,\omega)\big]
     =
     i\, q \otimes \, \widehat u(q,\omega)
     \qquad
     \text{and}
     \qquad
     \F_x\Big[\text{Div}_x \big[ \C\,\text{sym}\nabla_x \widehat u(x,\omega) \big]\Big]
     =
     -\Big(\C\,\text{sym}\big( q \otimes \widehat u(q,\omega)\big)\Big) \, q 
     \, ,
\end{equation}}

\begin{equation}
-
\,
\rho \, \left( 1+\frac{c^2 \, \omega^2}{a - b \, \omega^2 } \right) \. \omega^2 \, \widehat u(x,\omega)
=
-\big[\C\,\text{sym} \, \big(\widehat u(q,\omega) \otimes q\big)\big] \, q, 
\qquad
\qquad
\omega\in\R\,\setminus\left\{\pm\sqrt{\frac{a}{b}}\right\} 
\, ,
\label{eq.dispersion_rel_higher}
\end{equation}
which can be rewritten as a family (a perturbation problem) of non-linear eigenvalue problems with respect to $q\in\R^3$ i.e. for every $q\in\R^3$ we look for the $\omega$ (as then functions of $q$) such that there exist non-trivial solutions $\widehat u(q,\omega)\in\R^3$ satisfying
\begin{equation}
\underline{\mathbb A}(\omega,q,\rho,\C)
\,
\widehat u(q,\omega)
=
\bigg(
-
\,
\rho \, \left( 1+\frac{c^2 \, \omega^2}{a - b \, \omega^2 } \right) \. \omega^2 \, \widehat u(q,\omega)
+
\big[\C\,\text{sym}(\widehat u(q,\omega) \otimes q)\big]  \, q 
=
0.
\end{equation}
The stated algebraic problem admits non-trivial solutions if and only if 
\begin{equation}
\det
\underline{\mathbb A}(\omega,q,\rho,\C)
=
0.
\label{eq.det}
\end{equation}
Accounting for an isotropic medium, 
equation \eqref{eq.det} gives
\begin{equation}
\det
\bigg(
-
\,
\rho \, \left( 1+\frac{c^2 \, \omega^2}{a - b \, \omega^2 } \right) \. \omega^2 \, \mathds{1}
+
\mu \, k^2 \, \mathds{1} + (\mu+\lambda) \, q\otimes q \bigg)
=
0,
\qquad
\qquad
\omega\in\R\setminus\left\{\pm\sqrt{\frac{a}{b}}\right\} 
\, ,
\end{equation}
Considering now the 2D case, we obtain
\begin{align}
    \det
    &
    \begin{pmatrix}
        -\,\rho\,\omega^{2}\,\left(1+\dfrac{c^{2}\,\omega^{2}}{a-b\,\omega^{2}}\right)+\mu\,k^{2}+(\mu+\lambda)\,k_{1}^{2} & (\mu+\lambda)\,k_{1}\,k_{2}
        \\
        (\mu+\lambda)\,k_{1}\,k_{2} & -\,\rho\,\omega^{2}\,\left(1+\dfrac{c^{2}\,\omega^{2}}{a-b\,\omega^{2}}\right)+\mu\,k^{2}+(\mu+\lambda)\,k_{2}^{2}
    \end{pmatrix}
    \\[3mm]
    \\
    &
    =
    \rho^{2}\,\omega^{4} \left(1+\frac{c^{2}\,\omega^{2}}{a-b\,\omega^{2}}\right)^{2}
    -
    \rho\,\omega^{2} \left(1+\frac{c^{2}\,\omega^{2}}{a-b\,\omega^{2}}\right)\,\left(3\,\mu+\lambda\right) \, k^{2}+\mu\,(2\,\mu+\lambda)\,k^{4}.
    \nonumber
\end{align}
Solving with respect to $k^2$ we obtain
\begin{equation}
    k^2
    =
    \rho\,\omega^{2}\,\left(1+\frac{c^{2}\,\omega^{2}}{a-b\,\omega^{2}}\right)\,\frac{\left(3\,\mu+\lambda\right)\pm\sqrt{\left(3\,\mu+\lambda\right)^{2}-4\,\mu\,(2\,\mu+\lambda)}}{2\,\mu\,(2\,\mu+\lambda)}
\end{equation}
giving the two roots
\begin{equation}
    k^2_{\rm p}
    =
    \left(1+\frac{c^{2}\,\omega^{2}}{a-b\,\omega^{2}}\right)\,\frac{\rho\,\omega^{2}}{2\,\mu+\lambda}
    \qquad
    \text{and}
    \qquad
    k^2_{\rm s}
    =
    \left(1+\frac{c^{2}\,\omega^{2}}{a-b\,\omega^{2}}\right)\,\frac{\rho\,\omega^{2}}{\mu}.
\end{equation}
\corc{By taking the positive square root from this, we obtain $(k_{\rm p},k_{\rm s})$ as in eq.\eqref{eq:disp_relations_freq_depe}.}
%
%
%
%
\subsection{Dispersion relations obtained through the space-plane wave ansatz}
As we have seen previously, we can obtain the dispersion relations starting from the model in the frequency domain by setting
$
    \angcor{\widehat u(x,\omega)
     =
      \psi(\omega)}
      \,e^{i\,\langle x, q\rangle}
$
and inserting it into equation  eq.\eqref{eq.dispersion_rel_higher} 
\begin{equation}
-
\,
\rho \, \left( 1+\frac{c^2 \, \omega^2}{a - b \, \omega^2 } \right) \. \omega^2 \, \psi \, e^{i\,\langle x, q\rangle}
=
-\big[\C\,\text{sym}(\psi\otimes q)\big] \, q \, e^{i\,\langle x, q\rangle},
\qquad
\qquad
\omega\in\R\,\setminus\left\{\pm\sqrt{\frac{a}{b}}\right\} 
\, .
\end{equation}
Simplifying the $e^{i\,\langle x, q\rangle}$ factor we finally obtain 
$\mathbb A(\omega,q,\rho,\C)\,\psi=0$, giving the same algebraic problems 
$
    \det
    \,
     \mathbb A(\omega,q,\rho,\C)
    =
    0
$
as eq.\eqref{eq.det}.
%
%
%
%
\section{Galilean invariance}\label{sec:appGalilean}
The \emph{Galilean transformation} formulates the transformation of coordinates between two reference frames which only differ by a steady motion. In the setting of nonlinear elasticity where the deformation mapping $\varphi\colon\Omega\to\R^3$ describes the material in its current state, the corresponding Galilean transformation reads as
\begin{equation}
    \varphi\to\overline\varphi=\overline R\,\varphi+ r(t)\,,\qquad \ddot{r}(t)=0\,,\qquad\text{for all}\quad\overline R\in\mathrm{SO}(3)\,,\; r\in C^2(\R;\R^3)\,.\label{eq:ap_GalileanNonlinear}
\end{equation}
In the hyperelastic framework with an energy density function $W\colon\mathrm{GL}^{\!+}(3)\to\R$, the equilibrium equation for nonlinear elasticity is
\begin{equation}
    \rho\,\ddot\varphi=\mathrm{Div}\,\mathrm S_1(\nabla\varphi)=\mathrm{Div}\.\mathrm DW(\nabla\varphi)\,,
    \label{eq:ap_nonlinEl}
\end{equation}
where $S_1(\nabla\varphi)=\mathrm DW(\nabla\varphi)$ is the first Piola-Kirchhoff stress tensor.
All objective\footnote{An energy function is called objective (or frame-indifferent) if $W(\overline R\.F)=W(F)$ for all $F\in\mathrm{GL}^{\!+}(3)$ and $\overline R\in\mathrm{SO}(3)$.} hyperelastic energy functions are Galilean invariant, i.e. the corresponding equilibrium equation transform as follows
\begin{align}
    &\rho\,\ddot{\overline\varphi}=\mathrm{Div}\.\mathrm DW(\nabla\overline\varphi)
    \;
    \iff
    \;
    \rho\,\frac{\mathrm d^2}{\mathrm d t^2}\left[\overline R\.\varphi+ r(t)\right]=\mathrm{Div}\.\mathrm DW(\nabla[\overline R\.\varphi+ r(t)])
    \;
    \iff
    \;
    \rho\,(\overline R\,\ddot\varphi+\ddot{r}(t))=\mathrm{Div}\.\mathrm DW(\overline R\nabla\varphi)
    \\
    \iff& \rho\,\overline R\,\ddot\varphi=\mathrm{Div}\left[\overline R\.\mathrm DW(\nabla\varphi)\right]
    \;
    \iff
    \;
    \overline R\,(\rho\,\ddot\varphi)=\overline R\,\mathrm{Div}\.\mathrm DW(\nabla\varphi)
    \;
    \iff
    \; 
    \rho\,\ddot\varphi=\mathrm{Div}\.\mathrm DW(\nabla\varphi)\,\notag
\end{align}
such that the form of the equations remains the same (form-invariance).
%
%
%
\newcommand{\id}{{\boldsymbol{\mathbbm{1}}}}
\subsection{Infinitesimal Galilean invariance}\label{sec:appGalilean_inf}
Although it is not customary, linear elasticity can be written as well in terms of the deformation $\varphi$. For this, we define the quadratic energy density
\begin{align}
    W_{\rm lin}(\nabla\varphi)&=\frac12\langle\mathbb C\,\mathrm{sym}(\nabla\varphi-\id)\,,\mathrm{sym}(\nabla\varphi-\id)\rangle\,,
    \qquad
    \qquad
    \quad
    \C\colon\mathrm{Sym}(3)\to\mathrm{Sym}(3),\notag
    \\
    \rho\,\ddot \varphi&=\mathrm{Div}\,\mathrm DW_{\rm lin}(\nabla\varphi)=\mathrm{Div}\left[\mathbb C\,\mathrm{sym}(\nabla\varphi-\id)\right].\label{eq:ap_linEl}
\end{align}
It is then clear that the linearized equation of motion \eqref{eq:ap_linEl} does not remain invariant under the transformation presented in \eqref{eq:ap_GalileanNonlinear}, since $W_{\rm lin}(\overline R\.\nabla\varphi)\neq W_{\rm lin}(\nabla\varphi)$ \cite{munch2018rotational}.
For example,
\begin{equation}
    \mathrm{sym}(\overline R\.\nabla\varphi-\id)\neq\overline R\,\mathrm{sym}(\nabla\varphi-\id)\,.
\end{equation}
Therefore, in the process of linearizing the equation (\ref{eq:ap_nonlinEl}), one cannot expect invariance of the response under the transformation \eqref{eq:ap_GalileanNonlinear}.
Let us therefore turn to the equilibrium equation for linear elasticity of motion in the traditional displacement form
\begin{equation}
    \rho\,\ddot u=\mathrm{Div}\,\sigma\,,\qquad\sigma=\C\.\mathrm{sym}\nabla u\,,\qquad\mathbb C\colon\mathrm{Sym}(3)\to\mathrm{Sym}(3)
\end{equation}
with the displacement $u(x,t)=\varphi(x,t)-x(t)$ where $\sigma\in\mathrm{Sym}(3)$ is the symmetric Cauchy force stress tensor.  
Now, we need to infer the corresponding invariances by due linearization.
Since any orthogonal matrix $\overline R\in\mathrm{SO}(3)$ can be written as 
\begin{equation}
    \overline R=\exp(\overline A)=\id+\overline A+\hdots\qquad\text{with}\quad\overline A\in\mathfrak{so}(3)\,,
\end{equation}
it is possible to transform the Galilean invariance in nonlinear elasticity \eqref{eq:ap_GalileanNonlinear} into a corresponding statement for small strains by dropping higher-order terms
\begin{align}
    &&\overline\varphi(x,t)&=\overline R\.\varphi(x,t)+r(t)\notag\\
    &\iff& x(t)+\overline u(x,t)&=(\id+\overline A+\hdots)(x(t)+u(x,t))+r(t)\\
    &\iff& x(t)+\overline u(x,t)&=x(t)+u(x,t)+\overline A\.x(t)+\overline A\.u(x,t)+r(t)+\hdots\notag\\
    &\iff& \overline u(x,t)&=u(x,t)+\overline A\.x(t)+r(t)+\hdots\,,\notag
\end{align}
with some constant skew-symmetric matrix $\overline A\in\mathfrak{so}(3)$. Thus we arrive at, what we call, \emph{infinitesimal Galilean transformations}
\begin{equation}
    u\to\overline u=u+\overline A\.x+ r(t), \qquad \ddot{r}(t)=0\,,\qquad\text{for all}\quad\overline A\in\mathfrak{so}(3)\,,\; r\in C^2(\R,\R^3)\,.\tag{IGI}\label{eq:ap_Galilean}
\end{equation}
Indeed, linear elasticity \eqref{eq:ap_linEl} is \emph{infinitesimal Galilean invariant} \eqref{eq:ap_Galilean} because of the following identifications
\begin{align}
    \nabla \overline u&=\nabla\left(u+\overline A\.x+ r(t)\right)=\nabla u+\overline A\,,\notag\\
    \mathrm{sym}\nabla \overline u&=\mathrm{sym}\nabla\left(u+\overline A\.x+ r(t)\right)=\mathrm{sym}(\nabla u+\overline A)=\mathrm{sym}\nabla u\,,\label{eq:ap_gradUIdentification}\\
    \ddot{\overline u}&=\frac{\mathrm d^2}{\mathrm d t^2}\left[u+\overline A\.x+ r(t)\right]=\ddot u+ \ddot{r}(t)=\ddot u\,.\notag
\end{align}
For an enriched kinematic variable $P:\Omega\times\R\subset\R^3\times\R\to\R^{3\times3}$ without a unit, e.g. appearing as microdistortion in micromorphic models or microrotation in Cosserat models, we assume the \textbf{same} transformation behavior as for the displacement gradient \eqref{eq:ap_gradUIdentification}$_1$, we must therefore consider
\begin{equation}
     P\to\overline P=P+\overline A\,,\qquad\text{for all}\quad\overline A\in\mathfrak{so}(3)\,.\label{eq:ap_Galilean2}
\end{equation}
Then it holds for the expressions used in these models, e.g.
\begin{align}
    \mathrm{sym} \, \overline P=\mathrm{sym}(P+\overline A)=\mathrm{sym}P\,,\qquad\ddot{\overline P}=\frac{\mathrm d^2}{\mathrm d t^2}\left[P+\overline A\right]=\ddot P\,.
\end{align}
On the other hand, for an enriched kinematic variable $v:\Omega\times\R\subset\R^3\times\R\to\R^3$ whose unit is meter (as the displacement $u(x,t)$ itself) which is used in this work, we also require the same transformation as for the displacement, i.e.
\begin{equation}
    v\to\overline v=v+\overline A\.x+ r(t), \qquad \ddot{r}(t)=0\,,\qquad\text{for all}\quad\overline A\in\mathfrak{so}(3)\,,\; r\in C^2(\R;\R^3)\,.\label{eq:ap_Galilean3}
\end{equation}
Indeed, linear classical generalized continuum models (Cosserat, micromorphic, second gradient, etc.) satisfy infinitesimal Galilean invariance in this sense.
%
%
%
\subsection{Extended infinitesimal Galilean invariance}
We note that in linear elasticity, it is possible to generalize the infinitesimal Galilean invariance by extending the constant matrix $\overline A$ to a function $A\in C^2(\R;\mathfrak{so}(3))$ with $\ddot{A}(t)=0$. Thus we introduce the novel concept of \emph{extended infinitesimal Galilean transformations}
\begin{equation}
    \begin{array}{l}
         u\to\overline u=u+A(t)\.x+\overline r(t)\,,
         \\[2mm]
         \hspace{-0.13cm}P\to\overline P=P+A(t)\,,
    \end{array} 
    \quad
    \hfill \ddot{A}(t)=0,\quad \ddot{r}(t)=0\,,\qquad\text{for all}\quad A\in C^2(\R;\mathfrak{so}(3))\,,\; r\in C^2(\R;\R^3)\,.\tag{EIGI}\label{eq:ap_GalileanExtended}
\end{equation}
Again, linear elasticity is \emph{extended infinitesimal Galilean invariant} \eqref{eq:ap_GalileanExtended} because of the following identifications
\begin{align}
    \nabla\overline u&=\nabla\left(u+A(t)\.x+ r(t)\right)=\nabla u+A(t)\,,\notag\\
    \mathrm{sym}\nabla\overline u&=\mathrm{sym}\nabla\left(u+A(t)\.x+ r(t)\right)=\mathrm{sym}(\nabla u+A(t))=\mathrm{sym}\nabla u\,,\label{eq:ap_identificationExtended}\\
    \ddot{\overline u}&=\frac{\mathrm d^2}{\mathrm d t^2}\left[u+ A(t)\.x+ r(t)\right]=\ddot u+\ddot{A}(t)\.x+ \ddot{r}(t)=\ddot u\,.\notag
\end{align}
We note that all linear classical generalized continuum models also satisfy this \emph{extended infinitesimal Galilean invariance}.
However, there is no equivalent geometrically condition using $Q(t)\in\mathrm{SO}(3)$ in the nonlinear case.
Nevertheless, it seems reasonable to us to ask for all linear enriched continuum models to also ensure extended infinitesimal Galilean invariance if used as a homogenized surrogate model because the underlying microstructured linear Cauchy model always satisfies this new invariance condition.
It is therefore this condition \eqref{eq:ap_GalileanExtended} that we check in the main body of this paper.
%
%
%
%
\section{A Cauchy model with uncoupled frequency-dependent stiffness tensor and density and associated enriched continuum}
\label{sec:appC}
We present here the derivation of an enriched model stemming from a Cauchy model in which both the density and the elastic tensor are frequency-dependent.
Let us start considering the equilibrium equations
in the frequency domain
for a Cauchy model in which the density and the stiffness tensor depend on the frequency $\omega$ as
\begin{equation}
-\,\overline{\rho}(\omega)\.\omega^2\.\widehat{u}
=
\text{Div} \left[ \overline{\mathbb{C}}(\omega) \, \text{sym}\nabla\widehat u\right] \, ,
\qquad
\begin{cases}
\overline{\rho}(\omega)
=
\left(
\dfrac{f_5(a_1,b_1,a_2,b_2)-f_6(a_1,b_1,a_2,b_2) \, \omega^2}{f_7(a_1,b_1,a_2,b_2)-f_8(a_1,b_1,a_2,b_2) \, \omega^2}
\right)
\rho
\, ,
\\[4mm]
\overline{\mathbb{C}}(\omega)
=
\left(
\dfrac{f_1(a_1,b_1,a_2,b_2)+f_2(a_1,b_1,a_2,b_2) \, \omega^2}{f_3(a_1,b_1,a_2,b_2)+f_4(a_1,b_1,a_2,b_2) \, \omega^2}
\right)
\mathbb{C}
\, ,    
\end{cases}
\label{eq:poly_CC_rho_freq_depe_negative}
\end{equation}
where $\{f_i(a_1,b_1,a_2,b_2)\}_{i=1}^8$ are suitable functions and we must guarantee that $\displaystyle\lim_{\omega \to 0}\overline{\mathbb{C}}(\omega)=\mathbb{C}$ and $\displaystyle\lim_{\omega \to 0}\overline{\rho}(\omega)=\rho$.
This model can be seen as a combination of the two models presented in Section~\ref{sec:rho_freq} and \ref{sec:CC_freq}.
%
%
%
\subsection{Formulation of the enriched model and positive-definiteness conditions: form I}
Starting from eq.(\ref{eq:poly_CC_rho_freq_depe_negative}) and by moving the dependency on $\omega$ to the right side of the equation,
we can equivalently write
\begin{equation}
-
\. \rho \, \omega^2\.\widehat{u}
=
\text{Div} \left[ \left( f - \frac{(a_1-b_1\,\omega^2-1)+(a_2-b_2\,\omega^2-1)}{(a_1 - b_1 \, \omega^2)(a_2 - b_2 \, \omega^2)-1}\right) \mathbb{C}
\, \text{sym}\nabla \widehat u\right]
\, ,
\label{eq:equi_equa_CC_rho_freq_depe_negative_1}
\end{equation}
where $f=1+\frac{a_1+a_2-2}{a_1\,a_2-1}$.
If we now introduce two additional tensor fields 
$\widehat P:\text{Dom}\,\widehat P\subset\R^3_x\times\R_\omega\to\R^{3\times3}$ 
and 
$ \widehat Q:\text{Dom}\,\widehat Q\subset\R^3_x\times\R_\omega\to\R^{3\times3}$, equation~(\ref{eq:equi_equa_CC_rho_freq_depe_negative_1}) can be rewritten as\footnote{Also, in this case, we only need to define the symmetric part of $P$ and $Q$.}
\begin{align}
&
\begin{cases}
-
\. \rho \, \omega^2\.\widehat{u}
=
\text{Div} \left[ \left( f - \dfrac{(a_1-b_1\,\omega^2-1)+(a_2-b_2\,\omega^2-1)}{(a_1 - b_1 \, \omega^2)(a_2 - b_2 \, \omega^2)-1}\right) \mathbb{C}
\, \text{sym}\nabla u\right]
\, ,
\\[3mm]
\mathbb{C}\,\text{sym}\,\widehat P
=
-\dfrac{a_2-b_2\,\omega^2-1}{(a_1 - b_1 \, \omega^2)(a_2 - b_2 \, \omega^2)-1} \, \mathbb{C}\,\text{sym} \nabla \widehat u
\, ,
\\[3mm]
\mathbb{C}\,\text{sym}\,\widehat Q
=
-\dfrac{a_1-b_1\,\omega^2-1}{(a_1 - b_1 \, \omega^2)(a_2 - b_2 \, \omega^2)-1} \, \mathbb{C}\,\text{sym} \nabla \widehat u
\, ,
\end{cases}
\nonumber
\\[3mm]
&
\begin{cases}
-
\. \rho \, \omega^2\.\widehat{u}
=
\text{Div} \left[ f\,\mathbb{C}\,\text{sym}\nabla \widehat u + \mathbb{C}\,\text{sym}\,\widehat P + \mathbb{C}\,\text{sym}\,\widehat Q \right]
\, ,
\\[2mm]
\mathbb{C}\,\text{sym}\,\widehat P
=
-\dfrac{1}{a_1 - b_1 \, \omega^2} \, \mathbb{C}\left(\text{sym} \nabla \widehat u+\text{sym}\,\widehat Q\right)
\, ,
\\[2mm]
\mathbb{C}\,\text{sym}\,\widehat Q
=
-\dfrac{1}{a_2 - b_2 \, \omega^2} \, \mathbb{C}\left(\text{sym} \nabla \widehat u+\text{sym}\,\widehat P\right)
\, ,
\end{cases}
\nonumber
\\[3mm]
\xLeftrightarrow[\F^{-1}_t]{\F_t}
&
\begin{cases}
\rho \, \ddot{u}
=
\text{Div} \left[ f\,\mathbb{C}\,\text{sym}\nabla u + \mathbb{C}\,\text{sym}\,P + \mathbb{C}\,\text{sym}\,Q \right]
\, ,
\\[5pt]
a_1 \, \mathbb{C}\,\text{sym}\,P
+ b_1 \, \mathbb{C}\,\text{sym}\,\ddot{P}
+ \mathbb{C}\,\text{sym}\nabla u
+ \mathbb{C}\,\text{sym}\,Q
=
0 \, ,
\\[5pt]
a_2 \, \mathbb{C}\,\text{sym}\,Q
+ b_2 \, \mathbb{C}\,\text{sym}\,\ddot{Q}
+ \mathbb{C}\,\text{sym}\nabla u
+ \mathbb{C}\,\text{sym}\,P
=
0
\,.
\end{cases}
\label{eq:equi_equa_CC_rho_freq_depe_negative}
\end{align}
We have thus obtained from the frequency-dependent Cauchy problem in eq.(\ref{eq:equi_equa_CC_rho_freq_depe_negative_1})  an extended continuum model in eq.(\ref{eq:equi_equa_CC_rho_freq_depe_negative}), in which all elastic parameters are material constants that do not depend on frequency.
%
%
%
\subsubsection{Existence of an action functional and positive-definiteness}
The associated resulting action functional is
\begin{align}
\mathcal{A}
=
&
\iint\displaylimits_{\Omega\times [0,T]}
\underbrace{
\frac{1}{2}
\bigg(
\rho 
\, \langle \dot{u} , \dot{u} \rangle
+
b_1 \, \langle \mathbb{C}\,\text{sym}\,\dot{P},\text{sym}\,\dot{P} \rangle
+
b_2 \, \langle \mathbb{C}\,\text{sym}\,\dot{Q},\text{sym}\,\dot{Q} \rangle
}_{\text{K - kinetic energy density}}
\notag
\\*
&
\phantom{\iint\displaylimits_{\Omega\times [0,T]}}
-
\underbrace{
\color{black}{
f \, \langle \mathbb{C}\,\text{sym}\nabla u,\text{sym}\nabla u \rangle
+
a_1 \, \langle \mathbb{C}\,\text{sym}\, P,\text{sym}\, P \rangle
+
a_2 \, \langle \mathbb{C}\,\text{sym}\, Q,\text{sym}\, Q \rangle
}
}_{\text{W - strain energy density}}
\label{eq:energy_negative_CC_rho_freq_depe}
\\*
&
\phantom{\iint\displaylimits_{\Omega\times [0,T]}}
\underbrace{
+
\,
2 \, \langle \mathbb{C}\,\text{sym}\nabla u,\text{sym}\, P \rangle
+
2 \, \langle \mathbb{C}\,\text{sym}\nabla u,\text{sym}\, Q \rangle
+
2 \, \langle \mathbb{C}\,\text{sym}\, P,\text{sym}\, Q \rangle
\bigg)
}_{\text{W - strain energy density}}
\mathrm{d}x\.\mathrm{d}t\,,
\notag
\end{align}
where for positive definiteness it is required that
\begin{align}
a_1>0 \, ,
\qquad
b_1 >0 \, ,
\qquad
a_2>0 \, ,
\qquad
b_2>0 \, ,
\qquad
a_1\,a_2 >0 \, ,
\qquad
\text{eig}(\mathbb{C})>0 \, ,
\qquad
\rho >0 \, .
\end{align}
The associated homogeneous Neumann boundary conditions are
\begin{align}
\left(
f \, \mathbb{C} \, \text{sym} \, \nabla u
+ \mathbb{C} \, \text{sym} \, P
+ \mathbb{C} \, \text{sym} \, Q
\right) \, n = 0 \, .
\label{eq:bound_cond_CC_rho_freq_depe_final_negative}
\end{align}
%
%
%
\subsubsection{Energy conservation}
To ensure that the resulting model is conservative, we have to guarantee that 
\begin{equation}
\frac{\mathrm d}{\mathrm dt}\int\displaylimits_{\Omega}
E(\dot u, \dot{P},\dot{Q},\nabla u,P,Q) \, \mathrm{d}x=
\int\displaylimits_{\Omega}
\frac{\mathrm d}{\mathrm dt}\left[K(\dot u, \dot{P},\dot{Q})+W(\nabla u,P,Q)\right] \mathrm{d}x=
0\,,
\end{equation}
where $\Omega$ is the considered domain.
With $\sigma=\mathbb{C} \, \text{sym} \, \nabla u$, $\tau=\mathbb{C} \, \text{sym} \, P$, $\eta=\mathbb{C} \, \text{sym} \, Q$ and similar to eq.(\ref{eq:cons_energy_rho_freq_depe_negative}) and eq.(\ref{eq:cons_energy_CC_freq_depe_negative}), we compute
\begin{align}
\int\displaylimits_{\Omega}
\frac{\mathrm dE}{\mathrm dt}\, \mathrm{d}x
=&
\int\displaylimits_{\Omega}
\rho\, \langle \ddot{u} , \dot{u} \rangle
+ b_1 \langle \mathbb{C}\,\text{sym}\,\ddot{P},\text{sym}\,\dot{P} \rangle
+ b_2 \langle \mathbb{C}\,\text{sym}\,\ddot{Q},\text{sym}\,\dot{Q} \rangle
\notag
\\*[-10pt]
&
\phantom{\int\displaylimits_{\Omega}}
+ f \langle \mathbb{C}\,\text{sym}\nabla u,\text{sym}\nabla \dot{u} \rangle
+ a_1 \langle \mathbb{C}\,\text{sym}\, P,\text{sym}\, \dot{P} \rangle
+ a_2 \langle \mathbb{C}\,\text{sym}\, Q,\text{sym}\, \dot{Q} \rangle
\notag
\\*[-10pt]
&
\phantom{\int\displaylimits_{\Omega}}
+ \, \langle \mathbb{C}\,\text{sym}\nabla u,\text{sym}\, \dot{P} \rangle
+ \langle \mathbb{C}\,\text{sym}\nabla u,\text{sym}\, \dot{Q} \rangle
+ \langle \mathbb{C}\,\text{sym}\, P,\text{sym}\, \dot{Q} \rangle
\notag
\\*[-10pt]
&
\phantom{\int\displaylimits_{\Omega}}
+ \, \langle \mathbb{C}\,\text{sym}\nabla \dot{u},\text{sym}\, P \rangle
+ \langle \mathbb{C}\,\text{sym}\nabla \dot{u},\text{sym}\, Q \rangle
+ \langle \mathbb{C}\,\text{sym}\, \dot{P},\text{sym}\, Q \rangle\, \mathrm{d}x
\label{eq:cons_energy_CC_rho_freq_depe_negative}
\\[-10pt]
=&
\int\displaylimits_{\Omega}
\rho \,\langle \ddot{u} , \dot{u} \rangle
+ b_1 \langle \ddot{\tau},\text{sym}\,\dot{P} \rangle
+ b_2 \langle \ddot{\eta},\text{sym}\,\dot{Q} \rangle
+ f \langle \sigma,\text{sym}\nabla \dot{u} \rangle
+ a_1 \langle \tau,\text{sym}\, \dot{P} \rangle
+ a_2 \langle \eta,\text{sym}\, \dot{Q} \rangle
\notag
\\*[-10pt]
&
\phantom{\int\displaylimits_{\Omega}}
+ \, \langle \sigma,\text{sym}\, \dot{P} \rangle
+ \langle \sigma,\text{sym}\, \dot{Q} \rangle
+ \langle \tau,\text{sym}\, \dot{Q} \rangle
+ \, \langle \tau,\text{sym}\nabla \dot{u} \rangle
+ \langle \eta,\text{sym}\nabla \dot{u} \rangle
+ \langle \eta,\text{sym}\, \dot{P} \rangle\, \mathrm{d}x
\notag
\\[-10pt]
=&
\int\displaylimits_{\Omega}
\langle
\rho \,\ddot{u}
- f \, \text{Div} \, \sigma
- \text{Div} \, \tau
- \text{Div} \, \eta
,
\dot{u}
\rangle
+
\text{div} \left[ (f \, \sigma+\tau+\eta)^{\rm T}\dot{u} \right]
\notag
\\*[-10pt]
&
\phantom{\int\displaylimits_{\Omega}}
+
\langle
a_1\,\tau + b_1\,\ddot{\tau} + \sigma + \eta
,
\dot{P}
\rangle
+
\langle
a_2\,\eta + b_2\,\ddot{\eta} + \sigma + \tau
,
\dot{Q}
\rangle\, \mathrm{d}x
=
0
\, .
\notag
\end{align}
Thanks to the equilibrium equations (\ref{eq:equi_equa_CC_rho_freq_depe_negative}), the condition (\ref{eq:cons_energy_CC_rho_freq_depe_negative}) becomes
\begin{align}
\frac{\mathrm d}{\mathrm dt}\int\displaylimits_{\Omega}
E\, \mathrm{d}x
=
\int\displaylimits_{\Omega}
\text{div} \left[ (\sigma+\tau+\eta)^{\rm T}\dot{u} \right] \mathrm{d}x
=
\int\displaylimits_{\partial \Omega}
\langle
(f\,\sigma+\tau+\eta) n,
\dot{u}
\rangle\, \mathrm{d}s
=
0 \, .
\label{eq:cons_energy_CC_rho_freq_depe_final_negative}
\end{align}
which is automatically always satisfied thanks to the homogeneous boundary conditions reported in eq.(\ref{eq:bound_cond_CC_rho_freq_depe_final_negative}).
%
%
%
\subsubsection{Infinitesimal Galilean invariance}
With arguments similar to that presented in Section \ref{sec:galilean_invariance}, \ref{sec:galilean_invariance_1b}, and \ref{sec:galilean_invariance_1d}, it is easy to check that eqs.(\ref{eq:equi_equa_CC_rho_freq_depe_negative}) are extended infinitesimal Galilean invariant.
%
%
%
\subsection{Relations between the frequency-dependent model and the equivalent enriched model: form I}
\label{sec:relation_mixed_freq_depe_formI}
In this case it is possible to see from the following Figure~\ref{fig:disp_curves_CC_rho_freq_depe} that this model has two cut-offs frequencies and two asymptotes frequencies giving the possibility of creating two separate band-gaps.
Their expressions are
\begin{align}
\text{cut-offs}:
&
&&\omega_1= \sqrt{\frac{\beta_{\rm b}-\sqrt{\beta_{\rm b}^2-4 \beta_{\rm a} \beta_{\rm c}}}{2\beta_{\rm a}}} \, ,
&&\omega_2= \sqrt{\frac{\beta_{\rm b}+\sqrt{\beta_{\rm b}^2-4 \beta_{\rm a} \beta_{\rm c}}}{2\beta_{\rm a}}} \, ,
&&
\label{eq:cut_off_CC_rho_freq_depe}
\\*
\text{asymptotes}:
&
&&\omega_3= \sqrt{\frac{\gamma_{\rm b}-\sqrt{\gamma_{\rm b}^2-4 \gamma_{\rm a} \gamma_{\rm c}}}{2\gamma_{\rm a}}} \, ,
&&\omega_4= \sqrt{\frac{\gamma_{\rm b}+\sqrt{\gamma_{\rm b}^2-4 \gamma_{\rm a} \gamma_{\rm c}}}{2\gamma_{\rm a}}} \, ,
\label{eq:asympto_CC_rho_freq_depe}
\end{align}
where
\begin{equation}
\begin{array}{lll}
&
\beta_{\rm a}= b_1\,b_2 \, ,
\qquad
\beta_{\rm b}= a_2 \, b_1 + a_1 \, b_2 \, ,
\qquad
\beta_{\rm c}= a_1 \, a_2 - 1 \, ,
&
\gamma_{\rm a}= (a_1 + a_2 + a_1 a_2 - 3) b_1 b_2 \, ,
\\*[5pt]
\qquad
&
\gamma_{\rm b}= (a_2 + a_1 a_2 - 3) a_2 b_1 + (a_1 + a_1 a_2 - 3) a_1 b_2 + b_1 + b_2 \, ,
&
\gamma_{\rm c}= (a_1 a_2 - 1)^2\, .
\end{array}
\end{equation}

\begin{figure}[!ht]
\centering
\includegraphics[width=0.49\textwidth]{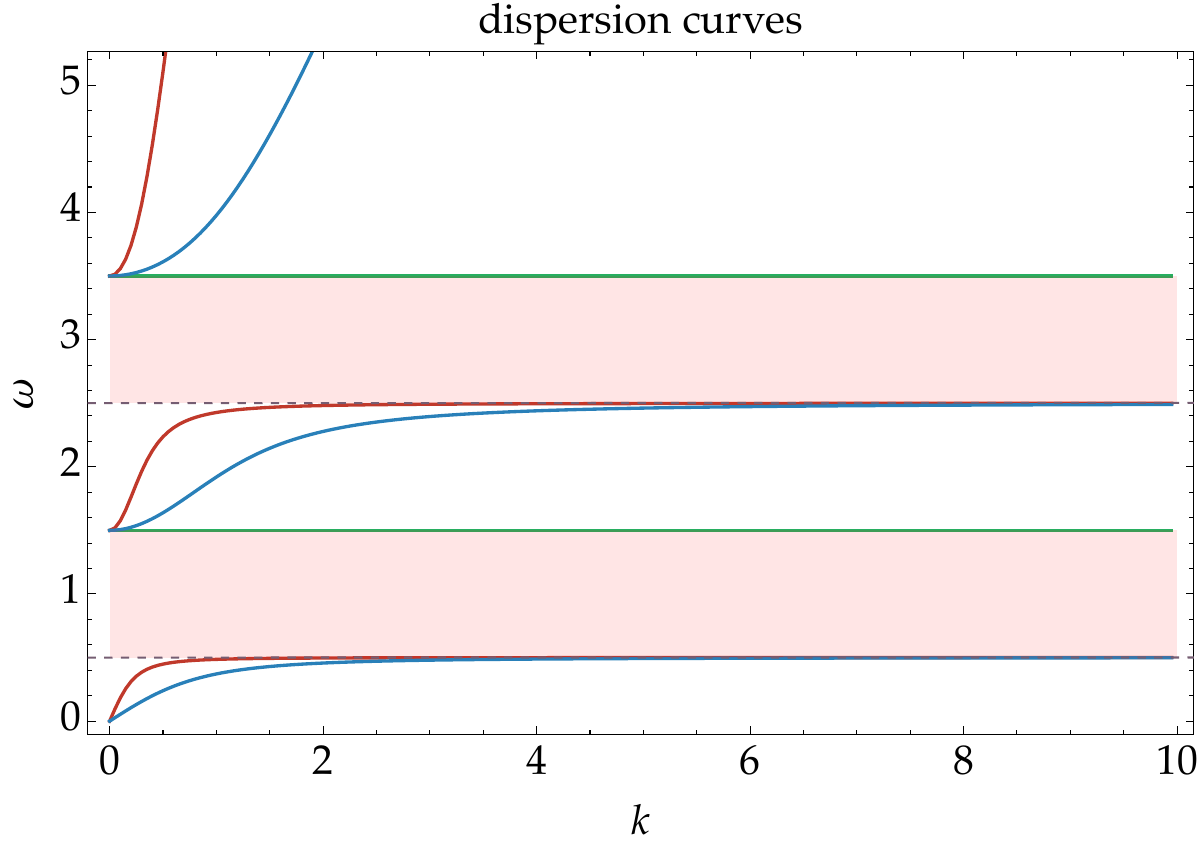}
\caption{\footnotesize{Dispersion curves in which the following values for the parameters have been used: $\rho=900$ kg/m$^3$, $\lambda=2898$ Pa, $\mu=262$ Pa, $a_1=0.079$, $b_1=0.0071\,\text{s}^2$, $a_2=47.36$, $b_2=13.97\,\text{s}^2$.
The curves for the frequency-dependent model are also reproduced by the corresponding enriched model, although the enriched model has two additional constant solutions (green lines).}}
\label{fig:disp_curves_CC_rho_freq_depe}
\end{figure}
The curves for the frequency-dependent model eq.(\ref{eq:poly_CC_rho_freq_depe_negative}) are also reproduced by the corresponding enriched model eq.(\ref{eq:equi_equa_CC_rho_freq_depe_negative}), although the enriched model has the additional solutions which are also singularity values for the original frequency-dependent model.
\vspace{-8mm}
%
%
%
\subsection{Formulation of the enriched model and positive-definiteness conditions: form II}
If we now introduce two additional vector-valued fields 
$\widehat v:\text{Dom}\,\widehat v\subset\R^3_x\times\R_\omega\to\R^{3}$ 
as well as 
$\widehat w:\text{Dom}\,\widehat w\subset\R^3_x\times\R_\omega\to\R^{3}$,
equation~(\ref{eq:equi_equa_CC_rho_freq_depe_negative_1}) can be rewritten as
\begin{align}
&
\begin{cases}
- \,
\rho \, \omega^2 \, \widehat{u}
=
\text{Div} \left[ \left( f - \dfrac{(a_1-b_1\,\omega^2-1)+(a_2-b_2\,\omega^2-1)}{(a_1 - b_1 \, \omega^2)(a_2 - b_2 \, \omega^2)-1}\right) \mathbb{C}
\, \text{sym}\nabla \widehat u\right]
\, ,
\\[4mm]
\text{Div} \left[\mathbb{C}\,\text{sym}\nabla \widehat v\right]
=
-\dfrac{a_2-b_2\,\omega^2-1}{(a_1 - b_1 \, \omega^2)(a_2 - b_2 \, \omega^2)-1} \,
\text{Div} \left[\mathbb{C}\,\text{sym} \nabla \widehat u\right]
\, ,
\\[4mm]
\text{Div} \left[\mathbb{C}\,\text{sym}\nabla \widehat w\right]
=
-\dfrac{a_1-b_1\,\omega^2-1}{(a_1 - b_1 \, \omega^2)(a_2 - b_2 \, \omega^2)-1} \, \text{Div} \left[\mathbb{C}\,\text{sym} \nabla \widehat u\right]
\, ,
\end{cases}
\nonumber
\\[5mm]
&
\begin{cases}
- \,
\rho \, \omega^2 \, \widehat{u}
=
\text{Div} \left[ f\,\mathbb{C}\,\text{sym}\nabla \widehat u + \mathbb{C}\,\text{sym}\,\nabla \widehat v + \mathbb{C}\,\text{sym}\,\nabla \widehat w \right]
\, ,
\\[3mm]
\text{Div} \left[\mathbb{C}\,\text{sym}\nabla \widehat v\right]
=
-\dfrac{1}{a_1 - b_1 \, \omega^2} \, \text{Div} \left[\mathbb{C}\left(\text{sym} \nabla \widehat u+\text{sym}\,\nabla \widehat w\right)\right]
\, ,
\\[3mm]
\text{Div} \left[\mathbb{C}\,\text{sym}\nabla \widehat w\right]
=
-\dfrac{1}{a_2 - b_2 \, \omega^2} \, \text{Div} \left[\mathbb{C}\left(\text{sym} \nabla \widehat u+\text{sym}\,\nabla \widehat v\right)\right]
\, .
\end{cases}
\nonumber
\\[5mm]
\xLeftrightarrow[\F_t]{\F_t^{-1}}
&
\begin{cases}
\rho \, \ddot{u}
=
\text{Div} \left[ f\,\mathbb{C}\,\text{sym}\nabla u + \mathbb{C}\,\text{sym}\,\nabla v + \mathbb{C}\,\text{sym}\,\nabla w \right]
\, ,
\\[5pt]
\text{Div} \left[ 
a_1 \, \mathbb{C}\,\text{sym}\,\nabla v
+ b_1 \, \mathbb{C}\,\text{sym}\,\nabla\ddot{v}
+ \mathbb{C}\,\text{sym}\nabla u
+ \mathbb{C}\,\text{sym}\,\nabla w
\right]
=
0 \, ,
\\[5pt]
\text{Div} \left[
a_2 \, \mathbb{C}\,\text{sym}\,\nabla w
+ b_2 \, \mathbb{C}\,\text{sym}\,\nabla \ddot{w}
+ \mathbb{C}\,\text{sym}\nabla u
+ \mathbb{C}\,\text{sym}\,\nabla v
\right]
=
0
\,,
\end{cases}
\label{eq:equi_equa_CC_rho_freq_depe_negative_z}
\end{align}
\corc{where $v$ and $w$ have the dimension of a displacement.}
We have thus \corc{obtained from the frequency-dependent Cauchy problem in eq.(\ref{eq:equi_equa_CC_rho_freq_depe_negative_1}) an extended} continuum model in eqs.(\ref{eq:equi_equa_CC_rho_freq_depe_negative_z}), in which all elastic parameters are material constants that do not depend on frequency.
%
%
%
\subsubsection{Existence of an action functional and positive-definiteness}
The associated resulting action functional is
\begin{align}
\mathcal{A}
=
&
\iint\displaylimits_{\Omega\times [0,T]}
\underbrace{
\frac{1}{2}
(
\rho \, \langle \dot{u} , \dot{u} \rangle
+
b_1 \langle \mathbb{C}\,\text{sym}\,\nabla \dot{v},\text{sym}\,\nabla \dot{v} \rangle
+
b_2 \langle \mathbb{C}\,\text{sym}\,\nabla \dot{w},\text{sym}\,\nabla \dot{w} \rangle
)
}_{\text{K - kinetic energy density}}
\notag
\\
&
\phantom{\int\displaylimits_{\Omega\times [0,T]}}
-
\underbrace{
\frac{1}{2}
\bigg(
f \langle \mathbb{C}\,\text{sym}\nabla u,\text{sym}\nabla u \rangle
+
a_1 \, \langle \mathbb{C}\,\text{sym}\, \nabla v,\text{sym}\, \nabla v \rangle
+
a_2 \, \langle \mathbb{C}\,\text{sym}\, \nabla w,\text{sym}\, \nabla w \rangle
}_{\text{W - strain energy density}}
\label{eq:energy_negative_CC_rho_freq_depe_z}
\\*
&
\phantom{\int\displaylimits_{\Omega\times [0,T]}}
\underbrace{
+
\,
2 \, \langle \mathbb{C}\,\text{sym}\nabla u,\text{sym}\, \nabla v \rangle
+
2 \, \langle \mathbb{C}\,\text{sym}\nabla u,\text{sym}\, \nabla w \rangle
+
2 \, \langle \mathbb{C}\,\text{sym}\, \nabla v,\text{sym}\, \nabla w \rangle
\bigg)
}_{\text{W - strain energy density}}
\mathrm{d}x\.\mathrm{d}t\,,
\notag
\end{align}
where for positive definiteness it is required that
\begin{align}
a_1>0 \, ,
\qquad
b_1 >0 \, ,
\qquad
a_2>0 \, ,
\qquad
b_2>0 \, ,
\qquad
a_1\,a_2 >0 \, ,
\qquad
\text{eig}(\mathbb{C})>0 \, ,
\qquad
\rho >0 \, .
\end{align}
The associated homogeneous Neumann boundary conditions are
\begin{align}
\left(
f \, \mathbb{C} \, \text{sym} \, \nabla u
+ \mathbb{C} \, \text{sym} \, \nabla v
+ \mathbb{C} \, \text{sym} \, \nabla w
\right) \, n = 0 \, ,
\notag
\\*[5pt]
\left(
a_1 \, \mathbb{C}\,\text{sym}\,\nabla v
+ b_1 \, \mathbb{C}\,\text{sym}\,\nabla\ddot{v}
+ \mathbb{C}\,\text{sym}\nabla u
+ \mathbb{C}\,\text{sym}\,\nabla w
\right) \, n = 0 \, ,
\label{eq:bound_cond_CC_rho_freq_depe_final_negative_z}
\\*[5pt]
\left(
a_2 \, \mathbb{C}\,\text{sym}\,\nabla w
+ b_2 \, \mathbb{C}\,\text{sym}\,\nabla \ddot{w}
+ \mathbb{C}\,\text{sym}\nabla u
+ \mathbb{C}\,\text{sym}\,\nabla v
\right) \, n = 0 \, .
\notag
\end{align}
%
%
%
\subsubsection{Energy conservation}
To ensure that the resulting model is conservative, we have to guarantee that 
\begin{equation}
\frac{\mathrm d}{\mathrm dt}\int\displaylimits_{\Omega}
E(\dot{u},\nabla \dot{v},\nabla \dot{w},\nabla u,\nabla v,\nabla w) \, \mathrm{d}x=
\int\displaylimits_{\Omega}
\frac{\mathrm d}{\mathrm dt}\left[K(\dot{u},\nabla \dot{v},\nabla \dot{w})+W(\nabla u,\nabla v,\nabla w)\right] \mathrm{d}x=
0\,,
\label{eq:conse_energy_equa_freq_dens_stiff_z}
\end{equation}
where $\Omega$ is the considered domain.
With $\sigma=\mathbb{C} \, \text{sym} \, \nabla u$, $\tau=\mathbb{C} \, \text{sym} \, \nabla v$, $\eta=\mathbb{C} \, \text{sym} \, \nabla w$ and similar to eq.(\ref{eq:cons_energy_CC_rho_freq_depe_negative}), we compute
\begin{align}
\int\displaylimits_{\Omega}
\frac{\mathrm dE}{\mathrm dt}\,\mathrm{d}x
=&
\int\displaylimits_{\Omega}
\rho \,\langle \ddot{u} , \dot{u} \rangle
+ b_1 \langle \mathbb{C}\,\text{sym}\,\nabla\ddot{v},\text{sym}\,\nabla\dot{v} \rangle
+ b_2 \langle \mathbb{C}\,\text{sym}\,\nabla\ddot{w},\text{sym}\,\nabla\dot{w} \rangle
\notag
\\*[-10pt]
&
\phantom{\int\displaylimits_{\Omega}}
+ f \langle \mathbb{C}\,\text{sym}\nabla u,\text{sym}\nabla \dot{u} \rangle
+ a_1 \langle \mathbb{C}\,\text{sym}\, \nabla v,\text{sym}\, \nabla \dot{v} \rangle
+ a_2 \langle \mathbb{C}\,\text{sym}\, \nabla w,\text{sym}\, \nabla \dot{w} \rangle
\notag
\\*[-10pt]
&
\phantom{\int\displaylimits_{\Omega}}
+ \, \langle \mathbb{C}\,\text{sym}\nabla u,\text{sym}\, \nabla \dot{v} \rangle
+ \langle \mathbb{C}\,\text{sym}\nabla u,\text{sym}\, \nabla \dot{w} \rangle
+ \langle \mathbb{C}\,\text{sym}\, \nabla v,\text{sym}\, \nabla \dot{w} \rangle
\notag
\\*[-10pt]
&
\phantom{\int\displaylimits_{\Omega}}
+ \, \langle \mathbb{C}\,\text{sym}\nabla \dot{u},\text{sym}\, \nabla v \rangle
+ \langle \mathbb{C}\,\text{sym}\nabla \dot{u},\text{sym}\, \nabla w \rangle
+ \langle \mathbb{C}\,\text{sym}\, \nabla \dot{v},\text{sym}\, \nabla w \rangle\,\mathrm{d}x\.\mathrm{d}t
\notag
\\[-10pt]
=&
\int\displaylimits_{\Omega}
\rho \,\langle \ddot{u} , \dot{u} \rangle
+ b_1 \langle \ddot{\tau},\text{sym}\,\nabla \dot{v} \rangle
+ b_2 \langle \ddot{\eta},\text{sym}\,\nabla \dot{w} \rangle
+ f \langle \sigma,\text{sym}\nabla \dot{u} \rangle
+ a_1 \langle \tau,\text{sym}\, \nabla \dot{v} \rangle
\label{eq:cons_energy_CC_rho_freq_depe_negative_z}
\\*[-10pt]
&
\phantom{\int\displaylimits_{\Omega}}
+ a_2 \langle \eta,\text{sym}\, \nabla \dot{w} \rangle
+ \, \langle \sigma,\text{sym}\, \nabla \dot{v} \rangle
+ \langle \sigma,\text{sym}\, \nabla \dot{w} \rangle
+ \langle \tau,\text{sym}\, \nabla \dot{w} \rangle
+ \, \langle \tau,\text{sym}\nabla \dot{u} \rangle
\notag
\\*[-10pt]
&
\phantom{\int\displaylimits_{\Omega}}
+ \langle \eta,\text{sym}\nabla \dot{u} \rangle
+ \langle \eta,\text{sym}\, \nabla \dot{v} \rangle\,\mathrm{d}x
\notag
\\[-10pt]
=&
\int\displaylimits_{\Omega}
\langle
\rho \,\ddot{u}
- f \, \text{Div} \, \sigma
- \text{Div} \, \tau
- \text{Div} \, \eta
,
\dot{u}
\rangle
+
\text{div} \left[ (f \, \sigma+\tau+\eta)^{\rm T}\dot{u} \right]
\notag
\\*[-10pt]
&
\phantom{\int\displaylimits_{\Omega}}
-
\langle
\text{Div}\left[a_1\,\tau + b_1\,\ddot{\tau} + \sigma + \eta\right]
,
\dot{v}
\rangle
+
\text{div} \left[ (a_1\,\tau + b_1\,\ddot{\tau} + \sigma + \eta)^{\rm T}\dot{v} \right]
\notag
\\*[-10pt]
&
\phantom{\int\displaylimits_{\Omega}}
-
\langle
\text{Div}\left[a_2\,\eta + b_2\,\ddot{\eta} + \sigma + \tau\right]
,
\dot{w}
\rangle
+
\text{div} \left[ (a_2\,\eta + b_2\,\ddot{\eta} + \sigma + \tau)^{\rm T}\dot{w} \right]\mathrm{d}x
=
0
\, .
\notag
\end{align}
Thanks to the equilibrium equations (\ref{eq:equi_equa_CC_rho_freq_depe_negative_z}), the condition (\ref{eq:cons_energy_CC_rho_freq_depe_negative_z}) becomes
\begin{align}
\frac{\mathrm d}{\mathrm dt}\int\displaylimits_{\Omega}
E\,\mathrm{d}x
=
&
\int\displaylimits_{\Omega}
\text{div} \left[ (f \, \sigma+\tau+\eta)^{\rm T}\dot{u} \right]
+
\text{div} \left[ (a_1\,\tau + b_1\,\ddot{\tau} + \sigma + \eta)^{\rm T}\dot{v} \right]
+
\text{div} \left[ (a_2\,\eta + b_2\,\ddot{\eta} + \sigma + \tau)^{\rm T}\dot{w} \right]\mathrm{d}x
\notag
\\*
=
&
\int\displaylimits_{\partial \Omega}
\langle (f \, \sigma+\tau+\eta) \, n,\dot{u} \rangle
+
\langle (a_1\,\tau + b_1\,\ddot{\tau} + \sigma + \eta) \, n,\dot{v} \rangle
+
\langle (a_2\,\eta + b_2\,\ddot{\eta} + \sigma + \tau) \, n,\dot{w} \rangle\,\mathrm{d}s
=
0 \, .
\label{eq:cons_energy_CC_rho_freq_depe_final_negative_Z}
\end{align}
which is automatically always satisfied thanks to the homogeneous boundary conditions reported in eq.(\ref{eq:bound_cond_CC_rho_freq_depe_final_negative_z}).
%
%
%
\subsubsection{Infinitesimal Galilean invariance}
With arguments similar to that presented in Section \ref{sec:galilean_invariance}, \ref{sec:galilean_invariance_1b}, and \ref{sec:galilean_invariance_1d}, it is easy to check that eqs.(\ref{eq:equi_equa_CC_rho_freq_depe_negative_z}) are extended infinitesimal Galilean invariant \eqref{eq:ap_GalileanExtended}.
%
%
%
\subsection{Relations between the frequency-dependent model and the equivalent enriched model: form II}
It can be seen in Figure~\ref{fig:disp_curves_CC_rho_freq_depe_z} that this model (\ref{eq:equi_equa_CC_rho_freq_depe_negative_z}) is also able to reproduce the curves produced by the frequency-dependent one eq.(\ref{eq:poly_CC_rho_freq_depe_negative}) with all their properties, it posses an additional solution, namely $k=0$, but it does not possess the extra constant roots in $\omega$ of the model eq.(\ref{eq:equi_equa_CC_rho_freq_depe_negative}).

\begin{figure}[!ht]
\centering
\includegraphics[width=0.49\textwidth]{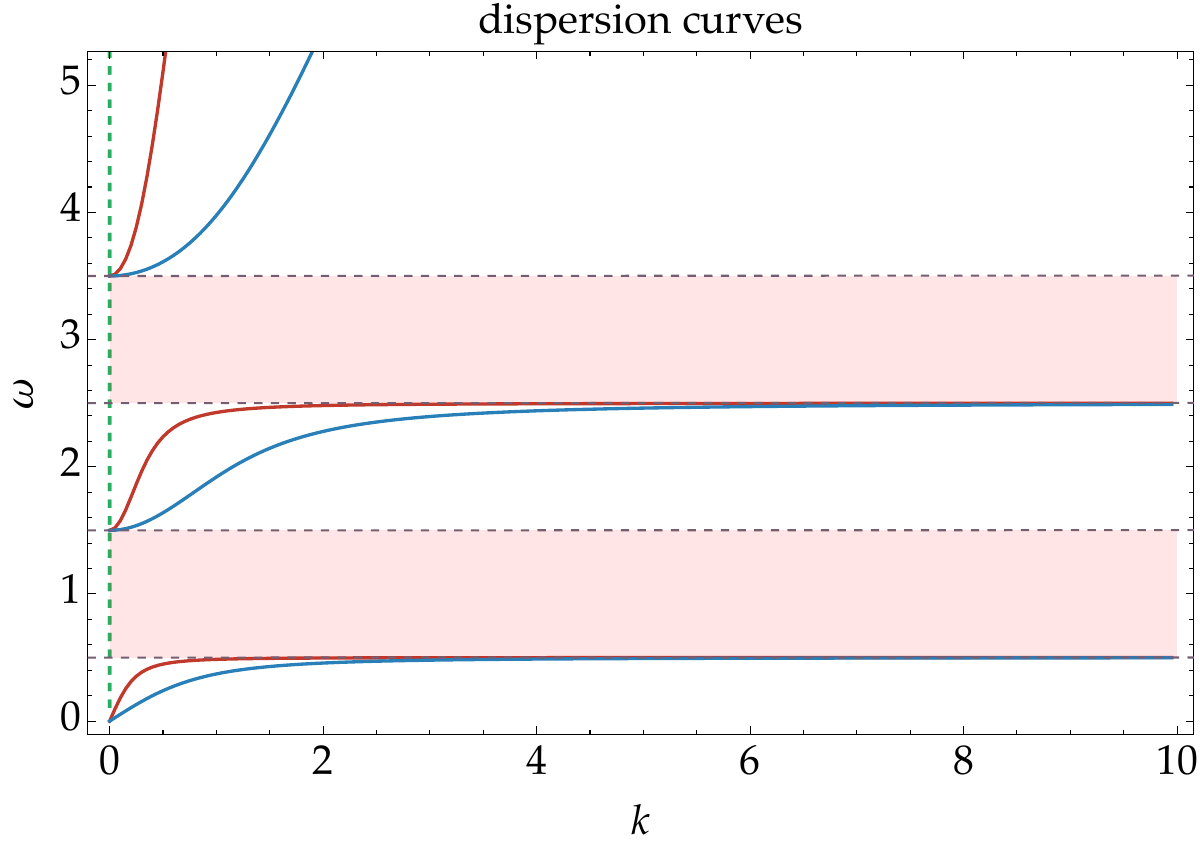}
\caption{\footnotesize{Dispersion curves in which the following values for the parameters have been used: $\rho=900$ kg/m$^3$, $\lambda=2898$ Pa, $\mu=262$ Pa, $a_1=0.079$, $b_1=0.0071\,\text{s}^2$, $a_2=47.36$, $b_2=13.97\,\text{s}^2$.
The curves for the frequency-dependent model are also reproduced by the corresponding enriched model, although the enriched model has the additional solution $k=0$.}}
\label{fig:disp_curves_CC_rho_freq_depe_z}
\end{figure}
}
%
%
%
\end{document}